\documentclass[notitlepage,prd,preprintnumbers,longbibliography,showpacs]{revtex4-1}
\usepackage[%
colorlinks=true,
urlcolor=blue,
linkcolor=blue,
citecolor=blue
]{hyperref}
\usepackage{slashed,color,amsmath,amssymb,mathrsfs}
\usepackage{graphicx}
\usepackage{hyperref}
\usepackage{longtable}
\usepackage{array}
\usepackage{accents}
\usepackage{autobreak}
\usepackage{tensor}
\usepackage{longtable}
\usepackage{booktabs}
\usepackage{multirow}
\usepackage{bm}
\usepackage{subfigure}
\usepackage{amssymb}
\usepackage{mathrsfs}
\usepackage{amsmath}
\usepackage[]{pifont}
\graphicspath{{figures/}}

\newcommand{\la}{\langle}
\newcommand{\ra}{\rangle}

\newcommand{\chip}{\chi_+}

\newcommand{\chim}{\chi_-}

\newcommand{\gamf}{\gamma_5}

\newcommand{\Bs}{B_6}

\newcommand{\Bsb}{\bar{B}_6}
\newcommand{\Bss}{{B^*_6}}

\newcommand{\Bssb}{{\bar{B}^*_6}}
\newcommand{\Bt}{B_{\bar{3}}}

\newcommand{\Btb}{\bar{B}_{\bar{3}}}

\newcommand{\psibm}{\bar{\psi}}
\newcommand{\psim}{\psi}

\newcommand{\psib}{\bar{\psi}}
\allowdisplaybreaks

\begin{document}

\title[]{Chiral Lagrangians for singly heavy baryons to $\mathcal{O}(p^{4})$ order}
\author{Yuan-He Zou$^{1}$}
\author{Hao Liu$^{1}$}
\author{Yan-Rui Liu$^2$}\email{yrliu@sdu.edu.cn}
\author{Shao-Zhou Jiang$^{1}$}
\email{jsz@gxu.edu.cn}
\affiliation{$^{1}$ Key Laboratory for Relativistic Astrophysics, School of Physical Science and Technology, Guangxi University, Nanning 530004, People's Republic of China\\
$^2$School of Physics, Shandong University, Jinan 250100, People's Republic of China}

\date{\today}
\begin{abstract}
The chiral Lagrangians for singly heavy baryons are constructed up to the $\mathcal{O}(p^{4})$ order. The involved baryons may be in a flavor antitriplet with spin-1/2, a flavor sextet with spin-$1/2$, or a flavor sextet with spin-3/2 when one considers three light flavors. For the relativistic version of Lagrangian, from $\mathcal{O}(p^{2})$ to $\mathcal{O}(p^{4})$, there exist 48, 199, and 1242 independent terms in the $SU(2)$ case and 59, 307, and 2454 independent terms in the $SU(3)$ case, respectively. For the Lagrangian in the heavy quark limit, from $\mathcal{O}(p^{2})$ to $\mathcal{O}(p^{4})$, the numbers of independent terms are reduced to 16, 64, and 412 in the $SU(2)$ case and to 17, 88, and 714 in the $SU(3)$ case, respectively. We obtain the low-energy constant relations between the relativistic case and the heavy-quark case up to the $\mathcal{O}(p^{3})$ order. The relations between low-energy constants of independent relativistic terms are also presented up to this order by using the heavy quark symmetry.
\end{abstract}

\maketitle
\section{Introduction}\label{Sec:I}

More than forty years ago, evidence for the existence of the charmed baryons $\Lambda_{c}^{+}$ and $\Sigma_{c}^{++}$ was reported by several experiments \cite{Cazzoli:1975et,Abrams:1979iu,Baltay:1979rn,Giboni:1979rm}. Later, more and more heavy baryons were observed \cite{ParticleDataGroup:2022pth}. The theoretical study of these particles directly from QCD is a formidable challenge because of its highly non-perturbative nature. Until now, various approaches compatible with QCD have been adopted in the related studies \cite{Chen:2016spr}, such as lattice QCD \cite{Padmanath:2019ybu,Na:2006qz,Can:2013tna,Can:2021ehb}, QCD sum rule \cite{Wang:2010fq,Zhang:2009iya,Zhu:1997as,Zhu:1998ih,Agamaliev:2016fou}, bag model \cite{Bose:1980vy,Bernotas:2013eia}, pion mean-field approach \cite{Kim:2021xpp},  bound-state approach \cite{Scholl:2003ip}, chiral perturbation theory (ChPT) \cite{Wise:1992hn,Yan:1992gz,Cho:1994vg,Savage:1994wa,Banuls:1999br,Tiburzi:2004mv,Jiang:2015xqa,Wang:2018gpl,Meng:2018gan}, etc.

Of the theoretical methods, ChPT as a low-energy effective field theory of QCD is a very useful tool in studying hadron properties. In this theory, Goldstone bosons generated from the spontaneous breaking of chiral symmetry are taken as the lowest pseudoscalar mesons \cite{Weinberg:1978kz,Gasser:1984gg,Gasser:1983yg}. For an effective theory, the physics is usually described by a Lagrangian focusing only on the low-energy degrees of freedom below some scale. The high-energy degrees of freedom are integrated out and their effects are usually encoded in the theory parameters. In this sense, ChPT describes only physics in the energy region below the chiral symmetry breaking scale $\Lambda_\chi$. As the first step to establish the framework of ChPT, constructing effective Lagrangian containing necessary fields is essential. What one needs to restrict the terms in the theory is just the consideration of QCD symmetries. Picking up the independent interaction operators $O_{i}$'s, one may write the Lagrangian as a sum $\mathcal{L}=\sum_{i}c_{i}O_{i}$. Here, the parameters $c_{i}$'s are the coupling constants or low-energy constants (LECs). The effects of high-energy particles and the details of short-range QCD interactions are hidden in these parameters. With the chiral Lagrangian involving singly heavy baryons, one can study a range of problems, such as the strong CP violation \cite{Unal:2020ezc}, the properties of singly heavy baryons including magnetic moment \cite{Meng:2018gan,Wang:2018gpl}, mass \cite{Guo:2008ns,Savage:1995dw,Guo:2002tg,Jiang:2014ena}, and decay \cite{Groote:2000ma,Shi:2022dhw,Wang:2018cre}, etc.

In general, the number of terms satisfying all possible QCD symmetries are infinite, but not all of them give important contributions to the considered matrix element. Since the pion momentum $p$ is a small quantity compared to $\Lambda_\chi$, $p/\Lambda_\chi$ is taken to be the expansion parameter in ChPT and the Lagrangian and matrix element are usually organized order by order with this parameter. Whether the contribution from a term is important or not relies on its chiral order. In principle, a chirally perturbative calculation involving high-order terms is better than that using only low-order terms. However, higher order implies more LECs. The determination of such LECs is one of the greatest difficulties in the application of ChPT. For ChPT with heavy quark baryons, it is possible to reduce the number of independent LECs to a certain extent by finding out LEC relations with the heavy quark symmetry.

The quark component of singly heavy baryons is $Qqq$ ($Q=c,b$; $q=u,d,s$). By treating the heavy quark as a singlet, such a baryon belongs to a $\bar{3}$ or $6$ representation in the flavor $SU(3)$ group. The spins of the light diquark in $\bar{3}$ and 6 are 0 and 1, respectively, because of the requirement from the Pauli principle. Combining with the heavy quark spin, one gets a spin-$\frac{1}{2}$ antitriplet, a spin-$\frac{1}{2}$ sextet, and a spin-$\frac{3}{2}$ sextet. In the limit that the heavy quark mass goes to infinity, $m_{Q}\to \infty$, there is a heavy quark spin symmetry \cite{Isgur:1989vq,Wise:1992hn,Isgur:1990yhj,Georgi:1990um} which says that the heavy quark spin does not affect the baryon dynamics and, thus, the two sextets are degenerate. With this approximate symmetry, the number of LECs will be reduced to a much smaller value \cite{Yan:1992gz,Cheng:1993kp,Meng:2018gan,Wang:2018gpl}.

Until now, chiral Lagrangians with nucleons \cite{Fettes:2000gb}, octet baryons \cite{Oller:2006yh,Frink:2006hx,Jiang:2016vax}, $\Delta(1232)$ \cite{Jiang:2017yda}, decuplet baryons \cite{Jiang:2018mzd,Holmberg:2018dtv}, heavy-light mesons \cite{Jiang:2019hgs}, and doubly charmed baryons \cite{Qiu:2020omj,Liu:2023lsg} have been obtained up to the $\mathcal{O}(p^{4})$ order. However, only the lowest order and a part of incomplete high-order chiral Lagrangian with singly heavy baryons have been constructed \cite{Yan:1992gz,Meng:2018gan,Wang:2018gpl}. In all these versions of ChPT, the two-loop contributions to amplitudes start to enter at the ${\cal O}(p^5)$ order. For considerations to complete the necessary ingredient of one-loop level investigations, to check convergence of chiral expansion better, to motivate future LEC studies, and so on, here we would like to construct relativistic chiral Lagrangians with singly heavy baryons up to the $\mathcal{O}(p^{4})$ order. In principle, when one determines the values of LECs from experimental data, all the operators should be independent. Otherwise overfitting would appear. The Lagrangians given in the present study will be complete and the terms will be independent. Moreover, the Lagrangian in the heavy quark limit will also be constructed. It will be used to find out LEC relations with the heavy quark symmetry.

This paper is organized as follows. In Sec. \ref{Sec:II}, the building blocks and multiplets of singly heavy baryons are introduced. In Sec. \ref{Sec:III}, the structures of chiral Lagrangians, properties of building blocks, and some linear relations are presented. Section \ref{sec:IV} shows the method to obtain LEC relations with the heavy quark symmetry. Section \ref{sec:res} is for the results and discussions. A brief summary is given in Sec.\ref{Sec:V}.

\section{Building blocks for Lagrangian construction}\label{Sec:II}

This section collects the basic building blocks in constructing the chirally invariant terms. Their basic properties are given directly. One can find more details about them in Refs. \cite{Yan:1992gz,Gasser:1983yg,Gasser:1984gg,Wise:1992hn,Georgi:1990um,Fearing:1994ga,Bijnens:1999sh,Bijnens:2001bb,Jiang:2019hgs,Jiang:2018mzd,Jiang:2017yda,Jiang:2016vax,Bijnens:2018lez,Gasser:1987rb,Wise:1993wa,Jenkins:1992zx,Casalbuoni:1996pg,Cata:2007ns}.

\subsection{Goldstone bosons dynamics and external sources}\label{sec:meson}
The QCD Lagrangian with light quarks coupling to external sources reads
\begin{align}
\mathcal{L}=\mathcal{L}_{\mathrm{QCD}}^{0}+\bar{q}\gamma^{\mu}\left(v_{\mu}+\gamma_{5}a_{\mu} \right)q+\bar{q}\left(i\gamma_{5}p-s\right)q,
\end{align}
where $\mathcal{L}_{\mathrm{QCD}}^{0}$ is the original QCD Lagrangian in the chiral limit, $q$ denotes the light quark field, and $v_{\mu}$, $a_{\mu}$, $s$, and $p$ are vector, axial-vector, scalar, and pseudoscalar sources, respectively. In order to consider the electroweak interactions, $v^\mu$ is taken to be traceable in the flavor space, i.e. $\langle v^\mu\rangle\neq 0$, in the relativistic case. When the heavy quark limit is considered, we take $\langle v^\mu\rangle\neq 0$ in the two-flavor ($N_f=2$) case and $\langle v^\mu\rangle=0$ in the three-flavor ($N_f=3$) case. For the axial-vector source $a^{\mu}$, it is always taken to be traceless. We do not consider the tensor external source \cite{Cata:2007ns} and the strong CP violation term in this paper.

The mesonic field $u$ denotes the nonlinear representation for the pseudoscalar Goldstone bosons which are generated from the spontaneous breaking of the chiral symmetry $SU(N_f)_L\times SU(N_f)_R$ into the vector symmetry $SU(N_f)_V$. Under the $g_{L}\times g_{R}$ chiral rotation, one has
\begin{align}
u\rightarrow g_{L}uh^{\dagger}=hug_{R}^{\dagger},
\end{align}
where $h$ is a compensator field related to the pion fields.

For convenience, we choose the following building blocks to construct chiral Lagrangians \cite{Fettes:2000gb,Jiang:2016vax,Jiang:2017yda,Jiang:2018mzd,Jiang:2019hgs}
\begin{align}
u^{\mu} &=i\left\{u^{\dagger}\left(\partial^{\mu}-i r^{\mu}\right) u-u\left(\partial^{\mu}-i l^{\mu}\right) u^{\dagger}\right\}, \\
\chi_{\pm} &=u^{\dagger} \chi u^{\dagger} \pm u \chi^{\dagger} u, \label{eq:dchi} \\
h^{\mu \nu} &=\nabla^{\mu} u^{\nu}+\nabla^{\nu} u^{\mu}, \\
f_{+}^{\mu \nu} &=u F_{L}^{\mu \nu} u^{\dagger}+u^{\dagger} F_{R}^{\mu \nu} u \label{eq:df},\\
f_{-}^{\mu \nu} &=u F_{L}^{\mu \nu} u^{\dagger}-u^{\dagger} F_{R}^{\mu \nu} u=-\nabla^{\mu} u^{\nu}+\nabla^{\nu} u^{\mu},
\end{align}
where $r^{\mu}=v^{\mu}+a^{\mu}$, $l^{\mu}=v^{\mu}-a^{\mu}$, $\chi=2 B_{0}(s+i p)$, $F_{R}^{\mu \nu}=\partial^{\mu} r^{\nu}-\partial^{\nu} r^{\mu}-i\left[r^{\mu}, r^{\nu}\right]$, $F_{L}^{\mu \nu}=\partial^{\mu} l^{\nu}-\partial^{\nu} l^{\mu}-i\left[l^{\mu}, l^{\nu}\right]$, and $B_{0}$ is a parameter related to the quark condensate. The advantage to use such notations is that the chiral rotation of an arbitrary building block $X$ is
\begin{align}
X\rightarrow X^{\prime}=hXh^{\dagger}.
\end{align}

The covariant derivative of $X$ reads
\begin{align}
\nabla^{\mu} X &= \partial^{\mu} X+\left[\Gamma^{\mu}, X\right] \label{eq:dcd}, \\
\Gamma^{\mu}&=\frac{1}{2}\left\{u^{\dagger}\left(\partial^{\mu}-i r^{\mu}\right) u+u\left(\partial^{\mu}-i l^{\mu}\right) u^{\dagger}\right\}.
\end{align}


\subsection{Singly heavy baryons}

For the singly heavy baryon $Qqq$, the formed antitriplet (sextet) is antisymmetric (symmetric) in flavor space for the exchange of the two light quarks. One can put all the involved ground baryon fields into three matrices \cite{Yan:1992gz,Cheng:1993kp},
\begin{align}
B_{\bar{3}}&=\left(\begin{array}{ccc}
0 & \Lambda_{Q} & \Xi_{Q}^{+1 / 2} \\
-\Lambda_{Q} & 0 & \Xi_{Q}^{-1 / 2} \\
-\Xi_{Q}^{+1 / 2} & -\Xi_{Q}^{-1 / 2} & 0
\end{array}\right),  \qquad
B_{6}=\left(\begin{array}{ccc}
\Sigma_{Q}^{+1} & \frac{1}{\sqrt{2}} \Sigma_{Q}^{0} & \frac{1}{\sqrt{2}} \Xi_{Q}^{\prime +1 / 2} \\
\frac{1}{\sqrt{2}} \Sigma_{Q}^{0} & \Sigma_{Q}^{-1} & \frac{1}{\sqrt{2}} \Xi_{Q}^{\prime -1 / 2} \\
\frac{1}{\sqrt{2}} \Xi_{Q}^{\prime +1 / 2} & \frac{1}{\sqrt{2}} \Xi_{Q}^{\prime-1 / 2} & \Omega_{Q}
\end{array}\right),\notag\\
B_{6}^{*}&=\left(\begin{array}{ccc}
\Sigma_{Q}^{*+1} & \frac{1}{\sqrt{2}} \Sigma_{Q}^{*0} & \frac{1}{\sqrt{2}} \Xi_{Q}^{\prime*+1 / 2} \\
\frac{1}{\sqrt{2}} \Sigma_{Q}^{*0} & \Sigma_{Q}^{*-1} & \frac{1}{\sqrt{2}} \Xi_{Q}^{\prime*-1 / 2} \\
\frac{1}{\sqrt{2}} \Xi_{Q}^{\prime*+1 / 2} & \frac{1}{\sqrt{2}} \Xi_{Q}^{\prime*-1 / 2} & \Omega_{Q}^{*}
\end{array}\right),\label{eq:baryon}
\end{align}
where $B_{\bar{3},6}$ ($B_6^*$) is for spin-1/2 (spin-3/2) baryons and the superscripts $\pm 1$, 0, and $\pm 1/2$ denote the $z$-components of the isospin. In the two-flavor case, only the upper left $2\times2$ matrices are considered.

The chiral transformations for these singly heavy baryon fields are
\begin{align}
B&\to B^{\prime}=h B h^{T},  \\
\bar{B}& \to	\bar{B}^{\prime}=h^{*} \bar{B} h^{\dagger},
\end{align}
where $B=B_{\bar{3}},\;B_{6}$, or $B_{6}^{*}$. The covariant derivative for $B$ is
\begin{align}
D_{\mu} B=\partial_{\mu} B+\Gamma_{\mu} B+B \Gamma_{\mu}^{T}.
\end{align}
If more than one covariant derivatives act on $B$, a totally symmetrical form is defined
\begin{align}
D^{\mu \nu \cdots \rho} &\equiv \frac{1}{n !}(\underbrace{D^{\mu} D^{\nu} \cdots D^{\rho}}_{n}+\text{full permutation of $D$’s}).
\end{align}
Any anti-symmetrical derivative is related to the higher order contributions \cite{Fettes:2000gb}.

The two sextets $B_{6}$ and $B_{6}^{*}$ have the same light degree of freedom. In the heavy quark limit $m_Q\to\infty$,  they are degenerate and one usually puts $B_6$ and $B_6^*$ baryons into a superfield $\psi^\mu$ to reflect the heavy quark spin symmetry. In the literature, there are different relative phases between $B_{6}$ and $B_{6}^{*}$ in the definition of $\psi^\mu$. Here, we choose to use \cite{Detmold:2011rb,Meng:2018gan,Kawakami:2018olq,Meng:2018gan}
\begin{align}
\begin{aligned}
&\psi^{\mu}=B_{6}^{* \mu}-\sqrt{\frac{1}{3}}\left(\gamma^{\mu}+v^{\mu}\right) \gamma_{5} B_{6}, \\
&\bar{\psi}^{\mu}=\bar{B}_{6}^{* \mu}+\sqrt{\frac{1}{3}} \bar{B}_{6} \gamma_{5}\left(\gamma^{\mu}+v^{\mu}\right), \label{eq:sf}
\end{aligned}
\end{align}
where $v^{\mu}$ with $v^{2}=1$ is the velocity of heavy baryons \footnote{From now on, $v^\mu$ will no longer denote vector external source}.  Since the superfield describes baryons in the nonrelativiestic case, it contains only the annihilation operators. Usually, the velocity-related superfield is marked by a subscript $v$, i.e. $\psi^{\mu}_v$. In this paper, we do not show it explicitly. Under the chiral rotation, the behaviors for $\psi^{\mu}$ and $\bar{\psi}^{\mu}$ are the same as the baryon field $B$ and $\bar{B}$, respectively. In the heavy quark limit, the superfield satisfies the relations
\begin{align}
\slashed{v}\psi^\mu=\psi^\mu, \qquad v^{\mu} \psi_{\mu}=0.
\end{align}

\section{Construction of the chiral Lagrangian for singly heavy baryons}\label{Sec:III}

In this section, we introduce the method of constructing the singly-heavy-baryon chiral Lagrangian. First, we show the basic structures of the considered Lagrangian. Second, we present the transformation properties of building blocks, such as parity (P), charge conjugation (C), and Hermiticity (h.c.). Finally, we give a short description for the linear relations which are used to remove dependent terms.

\subsection{Structures of chiral Lagrangians}

Since singly heavy baryons form three types of fields, $B_{\bar{3}}$, $B_{6}$, and $B_{6}^{*}$, the relativistic chiral Lagrangian has six different parts
\begin{align}
\begin{aligned}
\mathscr{L}={}& \mathscr{L}_{B_{\bar{3}} B_{\bar{3}}}+\mathscr{L}_{B_{6} B_{6}}+\mathscr{L}_{B_{6}^{*} B_{6}^{*}}+\mathscr{L}_{B_{\bar{3}} B_{6}}+\mathscr{L}_{B_{\bar{3}} B_{6}^{*}}+\mathscr{L}_{B_{6} B_{6}^{*}} \\
={}& \sum_{n} C_{n}\left(\left\langle\bar{B}_{\bar{3}} O\Gamma D B_{\bar{3}} Q^{T}\right\rangle+\text{H.c.}\right)
+\sum_{m} C_{m}\left(\left\langle\bar{B}_{6} O \Gamma D B_{6} Q^{T} \right\rangle+\text{H.c.}\right) \\
&+\sum_{p} C_{p}\left(\left\langle\bar{B}_{6}^{*} O \Gamma D B_{6}^{*} Q^{T}\right\rangle+\text{H.c.}\right)
+\sum_{q} C_{q}\left(\left\langle\bar{B}_{\bar{3}} O \Gamma D B_{6} Q^{T}\right\rangle+\text{H.c.}\right) \\
&+\sum_{r} C_{r}\left(\left\langle\bar{B}_{\bar{3}} O \Gamma D B_{6}^{*} Q^{T}\right\rangle+\text{H.c.}\right)
+\sum_{s} C_{s}\left(\left\langle\bar{B}_{6} O \Gamma D B_{6}^{*} Q^{T}\right\rangle+\text{H.c.}\right), \label{eq:rl}
\end{aligned}
\end{align}
where the Lorentz indices are omitted, $C_{n,\;m,\;p,\;q,\;r,\;s}$ are the LECs, and $\left\langle \cdots \right\rangle$ denotes the trace in the flavor space. In a Lagrangian term, $\Gamma$ denotes an element in the Clifford algebra, $O$ and $Q$ represent all possible combinations of the building blocks, and H.c. means Hermitian conjugate. In the first three parts of the Lagrangian, we use a convention that the covariant derivative $D$ (when it appears) acts only on the right baryon fields $B_{\bar{3}}$, $B_6$, and $B_6^*$. If a term is Hermitian, the H.c. of that term is not needed. In the last three parts of the Lagrangian, the H.c. of a term is always needed and we use a convention that the covariant derivative $D$ (when it appears) acts on $B_{\bar{3}}$, $B_6$, or $B_6^*$ in that term and on $\bar{B}_{\bar{3}}$, $\bar{B}_6$, or $\bar{B}_6^*$ in H.c. of that term. With such a convention, all the LECs are real. We do not show the kinetic terms in the above Lagrangian. They will be given in the leading order Lagrangian (Eq. \eqref{eq:lol}).

In the heavy quark limit, one has two mass terms $M_{\bar 3}\langle\bar{B}_{\bar 3}B_{\bar 3}\rangle$ and $M_6\langle \bar{\psi}^\mu \psi_\mu\rangle$. The first term can be removed by scaling the baryon fields with the factor $e^{-i M_{\bar 3} v \cdot x}$. Then the covariant derivatives acting on the matter fields $B_{\bar 3}$ and $\psi^\nu$ become $D^\mu B_{\bar 3}=-iM_{\bar 3}v^{\mu}B_{\bar 3}$ and $D^{\mu}\psi^\nu=-iM_6v^{\mu}\psi^\nu$, respectively. Since the heavy quark spin has no impact on the baryon dynamics and the light diquark has integer spin, the heavy quark symmetry imposes a constraint on the Lagrangian which says that the Clifford algebra element in each term can only be a unit matrix. Therefore, the structures of the chiral Lagrangian in the heavy quark limit are
\begin{align}
\begin{aligned}
\mathscr{L}&=\mathscr{L}_{B_{\bar{3}}B_{\bar{3}}}+\mathscr{L}_{B_{\bar{3}}\psi}+\mathscr{L}_{\psi\psi}\\
&=\sum_{m} D_{m}\left(\left\langle\bar{B}_{\bar{3}} O  v  B_{\bar{3}} Q^{T}\right\rangle + \text{H.c.}\right)
+\sum_{n} D_{n}\left(\left\langle\bar{B}_{\bar{3}} O  v  \psi Q^{T}\right\rangle + \text{H.c.}\right)
+\sum_{p}D_{p}\left(\langle\bar{\psi} O  v \psi Q^{T}\rangle+\text{H.c.}\right), \label{eq:sfl}
\end{aligned}
\end{align}
where $D_{n,\;m,\;p}$ are the relevant LECs. Other symbols are the same as those in Eq. \eqref{eq:rl} except for $v$ which just means possible velocities from the covariant derivatives acting on baryons. The kinetic terms are also omitted and they are given in the following Eq. \eqref{sfl1}.

\subsection{Properties of building blocks}\label{pbd}

According to the basic idea to establish an effective field theory, the Lagrangians in Eqs. \eqref{eq:rl} and \eqref{eq:sfl} should satisfy all the QCD symmetries, i.e. Lorentz symmetry, chiral symmetry, parity symmetry ($P$), charge conjugation symmetry ($C$), and time-reversal symmetry ($T$). With the building blocks discussed in Sec. \ref{sec:meson}, it is easy to obtain operators satisfying both Lorentz symmetry and chiral symmetry. One also needs to know the properties of building blocks in other symmetries. Because of the CPT theorem which says that a local theory is invariant under the combination of CPT operations if it is hermitian and Lorentz invariant, one may consider only $P$ and $C$ behaviors of the operators in ChPT. The $T$ invariance of the Lagrangian is automatically guaranteed by the CP invariance.

In Table \ref{table1}, we list the $P$ parity, charge conjugation, Hermiticity, and chiral dimensions (Dim) of the building blocks defined in Sec. \ref{sec:meson}. In Table \ref{table2}, the properties of Clifford algebra, Levi-Civita tensor, covariant derivative acting on baryons, and velocity of heavy baryons are shown. We use subscript $BB$ to denote terms in the Lagrangians $\mathscr{L}_{B_{\bar 3} B_{\bar 3}}$, $\mathscr{L}_{B_6 B_6}$, $\mathscr{L}_{B_6^* B_6^*}$, and $\mathscr{L}_{\psi \psi}$, while $BB^{\prime}$ to denote those in $\mathscr{L}_{B_{\bar 3} B_6}$, $\mathscr{L}_{B_{\bar 3} B_6^{*}}$, $\mathscr{L}_{B_6 B_6^{*}}$, and $\mathscr{L}_{B_{\bar 3} \psi}$. Only signs after  transformations are displayed. We give some explanations about the sign assignment by noting that the parity transformations for spin-$\frac{1}{2}$ and spin-$\frac{3}{2}$ fields are different \cite{Hemmert:1997ye} and the final Lagrangian should have the structures shown in Eq. \eqref{eq:rl} or \eqref{eq:sfl}. For convenience, we use the same sign for $P_{BB}$ and $P_{BB^\prime}$ for $D^\mu$ and $v^\mu$, but compensate a minus sign in $P_{BB^\prime}$ for Clifford algebra and $\varepsilon^{\mu\nu\lambda\rho}$. An exception is the sign assignment for parity transformation for $\mathscr{L}_{B_{\bar 3} B_6}$. It is taken to be the same as $\mathscr{L}_{BB}$, but not $\mathscr{L}_{BB^{\prime}}$, since both $B_{\bar 3}$ and $B_6$ contain only spin-$\frac{1}{2}$ fields. For charge conjugation and Hermiticity, the signs for $D^\mu$ are different between $\mathscr{L}_{B B}$ and $\mathscr{L}_{B B'}$ because of the convention adopted in Eqs. \eqref{eq:rl} and \eqref{eq:sfl}.

\begin{table}[h]
\caption{Chiral dimension (Dim), parity ($P$), charge conjugation ($C$), and Hermiticity (h.c.) of the building blocks.}\label{table1}
\begin{tabular}{ccccc}
\hline \hline & \text { Dim } & P & C & \text{h.c.} \\
\hline $u^{\mu}$ & 1 & $-u_{\mu}$ & $\left(u^{\mu}\right)^{T}$ & $u^{\mu}$ \\
$h^{\mu \nu}$ & 2 & $-h_{\mu \nu}$ & $\left(h^{\mu \nu}\right)^{T}$ & $h^{\mu \nu}$ \\
$\chi_{\pm}$ & 2 & $\pm \chi_{\pm}$ & $\left(\chi_{\pm}\right)^{T}$ & $\pm \chi_{\pm}$ \\
$f_{\pm}^{\mu \nu}$ & 2 & $\pm f_{\pm \mu \nu}$ & $\mp\left(f_{\pm}^{\mu \nu}\right)^{T}$ & $f_{\pm}^{\mu \nu}$ \\
\hline \hline
\end{tabular}
\end{table}

\begin{table}[h]
\caption{Chiral dimension (Dim), parity ($P$), charge conjugation ($C$), and Hermiticity (h.c.) of the Clifford algebra elements, Levi-Civita tensor, covariant derivative acting on baryons and velocity of baryons. The subscripts $BB$ and $BB^\prime$ are explained in the text.}\label{table2}
\begin{tabular}{cccccccc}
\hline \hline & $\operatorname{Dim}$ & $P_{B B}$ & $C_{B B}$  & h.c.$_{B B}$ & $P_{B B^{\prime}}$ & $C_{B B^{\prime}}$ & h.c.$_{B B^{\prime}}$ \\
\hline  1 & 0 & $+$ & $+$ & $+$ & $-$ & $+$ & $+$ \\
$\gamma_5$ & 1 & $-$ & $+$ & $-$ & $+$ & $+$ & $-$ \\
$\gamma^\mu$ & 0 & $+$ & $-$ & $+$ & $-$ & $-$ & $+$ \\
$\gamma_5 \gamma^\mu$ & 0 & $-$ & $+$ & $+$ & $+$ & $+$ & $+$ \\
$\sigma^{\mu \nu}$ & 0 & $+$ & $-$ & $+$ & $-$ & $-$ & $+$ \\
$\varepsilon^{\mu \nu \lambda \rho}$ & 0 & $-$ & $+$ & $+$ & $+$ & $+$ & $+$ \\
$ D^{\mu} $ & 0 & $+$ &  $-$ & $-$ &  $+$ & $+$ & $+$    \\
$ v^{\mu} $ & 0 & $+$ &  $-$ & $+$ &  $+$ & $-$ & $+$    \\
\hline \hline
\end{tabular}
\end{table}

\subsection{Linear relations}\label{secIIIC}

With the above recipes, terms satisfying all the symmetries can be obtained. However, these terms are not completely independent. There are linear relations restricting the constructed terms to a minimal set. They include equations of motions, Schouten identity, $SU(2)$ and $SU(3)$ Cayley-Hamilton relations, Bianchi identity, and relations about derivatives. Such relations are very similar to those for the doubly charmed baryons \cite{Liu:2023lsg} and we do not repeat them here. The details about them have been discussed widely \cite{Gasser:1983yg,Gasser:1984gg,Yan:1992gz,Fearing:1994ga,Bijnens:1999sh,Fettes:2000gb,Bijnens:2001bb,Ebertshauser:2001nj,Jiang:2014via,Jiang:2016vax,Bijnens:2018lez,Jiang:2017yda,Jiang:2017yda,Jiang:2018mzd}. However, now we need to consider one more relation because the flavor multiplets of baryons are different.

The baryon antitriplet is represented by an antisymmetric matrix $B_{\bar{3}}$ and the two baryon sextets are represented by two symmetric matrices $B_6$ and $B_6^*$. In an equation form, one has
\begin{align}
B_{\bar{3}}^T=-B_{\bar{3}},\quad(B_6^{(*)})^T=B_6^{(*)}.\label{eq:sym}
\end{align}
Noticing further that the trace of a matrix is equal to the trace of its transpose, one may get a new relation
\begin{align}
\left\langle\bar{B} O \Gamma B^{\prime} Q^T\right\rangle=\delta_B \delta_{B^{\prime}}\left\langle\bar{B} Q \Gamma B^{\prime} O^T\right\rangle,
\end{align}
where $B^{(\prime)}$ can be $B_{\bar{3}}$, $B_{6}$, or $B_{6}^{*}$, $\delta_{B^{(\prime)}}=-1$ for $B_{\bar{3}}$, and $\delta_{B^{(\prime)}}=1$ for $B_{6}$ and $ B_{6}^{*}$.

All the non-independent chiral-invariant terms can be eliminated with the linear relations. To do that, we adopt a programmatic approach which is almost the same as those used in Refs. \cite{Jiang:2014via,Jiang:2016vax,Jiang:2017yda}. The details can be found there.

\section{LEC relations in the heavy quark limit}\label{sec:IV}

In the heavy quark limit, not all terms in the relativistic chiral Lagrangian are independent. The heavy quark symmetry leads to some relations between the relativistic LECs. Considering the fact that one covariant derivative acting on the baryon field gives one mass scale, we revise the dimensions of operators and LECs in Eq. (\ref{eq:rl}) with the new definitions
\begin{align}
\tilde{O}_n=O_n / M^m, \qquad \tilde{C}_n=C_n M^m,
\end{align}
where $m$ is the number of covariant derivatives acting on the baryon field and $M=M_{\bar 3}$. Similar to the chiral perturbation theory with explicit $\Delta(1232)$ baryons \cite{Hemmert:1997ye} where the mass difference $(m_\Delta-m_N)$ is treated as the same order as the pion momentum, we here also take $(M_6-M_{\bar 3})\sim {\cal O}(p^1)$. With the redefined symbols, the dimensions of LECs $\tilde{C}_n$ at the $O(p^{n})$ order are all ($1-n$). The Lagrangians at the order $O(p^{m})$ can then be written as
\begin{align}
&\mathscr{L}^{(m)}=\sum_n C_n^{(m)} O_n^{(m)}=\sum_n \tilde{C}_n^{(m)} \tilde{O}_n^{(m)}, N_f=3, \\
&\mathscr{L}^{(m)}=\sum_n c_n^{(m)} o_n^{(m)}=\sum_n \tilde{c}_n^{(m)} \tilde{o}_n^{(m)}, N_f=2, \\
&\mathscr{L}_{\mathrm{HQ}}^{(m)}=\sum_n D_n^{(m)} P_n^{(m)}, N_f=3, \\
&\mathscr{L}_{\mathrm{HQ}}^{(m)}=\sum_n d_n^{(m)} p_n^{(m)}, N_f=2.
\end{align}

We discussed two methods to get the relations between LECs for heavy-light mesons in Ref. \cite{Jiang:2019hgs}. Here, we adopt a method similar to the second one there, i.e. changing the heavy-quark form to the relativistic form. Substituting Eq. \eqref{eq:sf} into Eq. \eqref{eq:sfl}, one has
\begin{align}
\psibm^{\mu}\psim_{\nu}\to&-\frac{i}{\sqrt{3}}\Bsb\gamf D^{\mu}\Bss_{\nu}
+\frac{1}{\sqrt{3}}\Bsb\gamf\gamma^{\mu}\Bss_{\nu}
+\frac{i}{\sqrt{3}}\Bssb^{\mu}\gamf D_{\nu}\Bs
+\frac{1}{\sqrt{3}}\Bssb^{\mu}\gamf\gamma_{\nu}\Bs
\notag\\
&-\frac{1}{3}{g^{\mu}}_{\nu}\Bsb\Bs
-\frac{1}{3}\Bsb {D^{\mu}}_{\nu}\Bs
+\frac{i}{3}\Bsb{\sigma^{\mu}}_{\nu}\Bs
+\Bssb^{\mu}\Bss_{\nu},\\
\Btb\psim^{\mu}\to&\frac{i}{\sqrt{3}}\Btb\gamf D^{\mu}\Bs
+\frac{1}{\sqrt{3}}\Btb\gamf\gamma^{\mu}\Bs
+\Btb\Bss^{\mu}.
\end{align}
To find LEC relations, we change the forms of $O(p^{m})$ order chiral Lagrangians to be
\begin{align}
\begin{aligned}
&\mathscr{L}^{(m)} \sim \sum_{k} \tilde{C}_k \tilde{O}_k=\sum_{k, i} \tilde{C}_k X_{k i} A_i, \\
&\mathscr{L}_{\mathrm{HQ}}^{(m)} \sim \sum_{l} D_l P_l=\sum_{l, i} D_l Y_{l i} A_i,
\end{aligned}
\end{align}
where $A_i$'s are possible terms maintaining all the QCD symmetries, but they do not need to be independent. The number of $A_i$'s is larger than those of $\tilde{O}_k$'s and $P_l$'s and the number of $\tilde{O}_k$'s is larger than that of $P_l$'s. $X$ and $Y$ are the coefficient matrices. Using all linear relations mentioned in Sec. \ref{secIIIC}, one may pick up the independent $\tilde{O}_k$'s in the heavy quark limit via linear algebra. Then the remaining number of $\tilde{O}_k$'s is equal to that of $P_l$'s. The relations between $\tilde{C}_k$'s and $D_l$'s and thus the relations among $\tilde{C}_k$'s can be obtained.

\section{Results and Discussions}\label{sec:res}

\subsection{$\mathcal{O}(p^1)$ and $\mathcal{O}(p^2)$ orders}
The lowest-order relativistic chiral Lagrangian is \cite{Yan:1992gz}
\begin{align}
\begin{aligned}
\mathscr{L}^{(1)}={}&\left\langle\bar{B}_{6}\left(i \slashed{D}-m_{6}\right) B_{6}\right\rangle  +\frac{1}{2} \left\langle\bar{B}_{\bar{3}}\left(i \slashed{D}-m_{\bar{3}}\right) B_{\bar{3}}\right \rangle\\
&+\left\langle\bar{B}_{6}^{* \mu}\left[-g_{\mu \nu}\left(i \slashed{D}-m_{6}^*\right) +i\left(\gamma_\mu D_\nu+\gamma_\nu D_\mu\right)-\gamma_\mu\left(i D+m_{6}^*\right) \gamma_\nu\right] B_{6}^{*\nu}\right\rangle\\
&+g_1(\langle\bar{B}_6 u^\mu \gamma_5 \gamma_\mu B_6\rangle+\text{H.c.})+g_2(\langle\bar{B}_{\bar{3}} u^\mu \gamma_5 \gamma_\mu B_6\rangle+\text{H.c.})+g_3 (\langle\bar{B}_6 u^\mu B_{6 \mu}^*\rangle+\text{H.c.})\\
&+g_4 (\langle\bar{B}_{\bar{3}} u^\mu B_{6 \mu}^*\rangle+\text{H.c.})+g_5\langle\bar{B}_6^{* \mu} u^\nu \gamma_5 \gamma_\nu B_{6 \mu}^*\rangle+g_6 \langle\bar{B}_{\bar{3}} u^\mu \gamma_5 \gamma_\mu B_{\bar{3}}\rangle. \label{eq:lol}
\end{aligned}
\end{align}
In the heavy quark limit, the Lagrangian is \cite{Cheng:1993kp,Cho:1992nt}
\begin{align}
\begin{aligned}
\mathscr{L}_{\mathrm{HQ}}^{(1)}=&\frac{i}{2} \left\langle\bar{B}_{\bar3} v \cdot D B_{\bar 3}\right\rangle -i \left\langle\bar{\psi}^\mu v \cdot D \psi_\mu\right\rangle+(M_6-M)\langle \bar{\psi}^\mu\psi_\mu\rangle
+i d_1 \epsilon_{\mu \nu\lambda \rho} \left\langle\bar{\psi}^\mu v^\nu u^\lambda \psi^\rho\right\rangle+d_2 (\langle\bar{\psi}^\mu u_\mu B_3\rangle+\text{H.c.}) \label{sfl1}
 \end{aligned}
\end{align}
With the heavy quark symmetry, the relations between LECs in $\mathscr{L}^{(1)}$ and those in $\mathscr{L}_{\mathrm{HQ}}^{(1)}$ are
\begin{align}
g_1=\frac{2}{3} d_1, \quad g_3=\frac{1}{\sqrt{3}} d_1, \quad g_5=-d_1, \quad g_2=-\frac{1}{\sqrt{3}} d_2, \quad g_4=d_2, \quad g_6=0.
\end{align}
These results are valid in both two- and three-flavor cases.

The relativistic chirally invariant terms at the $\mathcal{O}(p^2)$ order are shown in Table \ref{tab:table3}. The second and fifth columns label the number for terms in the two- and three-flavor cases, respectively. The third and sixth columns give the relations between relativistic LECs and the LECs in the heavy quark limit. The fourth and seventh columns give the constraints for the relativistic LECs with the heavy quark symmetry, where the independent ones are marked by ``I''. Long LEC relations are marked by ``$*$'' and those in the columns 3, 4, 6, and 7 are displayed explicitly in the following Eqs. \eqref{relation-p2-1}, \eqref{relation-p2-2}, \eqref{relation-p2-3}, and \eqref{relation-p2-4}, respectively. Table \ref{tab:table2} lists the Lagrangian terms in the heavy quark limit. The $\mathscr{L}_{B_{\bar{3}}B_{\bar{3}}}$ terms are almost the same as those in Table \ref{tab:table3}, except that terms containing $\gamma$ matrices vanish now. Therefore, we do not show $\mathscr{L}_{B_{\bar{3}}B_{\bar{3}}}$ terms explicitly in this table. Table \ref{tab:table3} does not give the LECs relations about $\mathscr{L}_{B_{\bar{3}}B_{\bar{3}}}$, either. A part of  $\mathcal{O}(p^{2})$ Lagrangians are given in Refs. \cite{Meng:2018gan,Yan:1992gz,Meng:2022ozq}, which is consistent with ours.

\begin{longtable}{lcrrcrr}
\caption{\label{tab:table3}Independent terms in the relativistic Lagrangian at the $\mathcal{O}(p^{2})$ order. The columns 3 and 6 list the LEC relations between the relativistic case and the heavy quark case. The columns 4 and 7 show the LEC relations between different terms in the relativistic case by using the heavy quark symmetry. In these two columns, ``I'' means that we choose the term as an independent one when employing the heavy quark symmetry. In the columns 3, 4, 6, and 7, the LEC relations marked by ``*'' are given in Eqs. \eqref{relation-p2-1}, \eqref{relation-p2-2}, \eqref{relation-p2-3}, and \eqref{relation-p2-4}, respectively. ``0'' means that the LECs in the heavy quark limit vanish. ``$-$'' means that the LEC in the heavy quark case vanishes or is equal to that in the relativistic case.}\\
\hline\hline $O_n$ & $SU(2)$ & $\tilde{c}^{(2)}_n$ & $\tilde{c}^{(2)}_n$ & $SU(3)$ & $\tilde{C}^{(3)}_n$ & $\tilde{C}^{(3)}_n$\\
\hline\endfirsthead
\hline\hline $O_n$ & $SU(2)$ & $\tilde{c}^{(2)}_n$ & $\tilde{c}^{(2)}_n$ & $SU(3)$ & $\tilde{C}^{(3)}_n$ & $\tilde{C}^{(3)}_n$\\
\hline\endhead
\hline\hline 
\endfoot
\hline\endlastfoot
$\la\Btb u^{\mu}u_{\mu}\Bt\ra$ & 1 & $-\quad$ & $-\quad$ & 1 & $-\quad$ & $-\quad$  \\
$\la\Btb u^{\mu}u^{\nu}D_{\mu\nu}\Bt\ra$ & 2 & $-\quad$ & $-\quad$ & 2 & $-\quad$ & $-\quad$  \\
$i\la\Btb u^{\mu}u^{\nu}\sigma_{\mu\nu}\Bt\ra$ & 3 & $-\quad$ & $-\quad$ & 3 & $-\quad$ & $-\quad$  \\
$\la\Btb u^{\mu}\Bt{u^{T}}_{\mu}\ra$ & 4 & $-\quad$ & $-\quad$ & 4 & $-\quad$ & $-\quad$  \\
$\la\Btb u^{\mu}{D_{\mu}}^{\nu}\Bt{u^{T}}_{\nu}\ra$ & 5 & $-\quad$ & $-\quad$ & 5 & $-\quad$ & $-\quad$  \\
$\la\Btb\Bt\ra\la u^{\mu}u_{\mu}\ra$ &  &  &  & 6 & $-\quad$ & $-\quad$  \\
$\la\Btb D^{\mu\nu}\Bt\ra\la u_{\mu}u_{\nu}\ra$ &  &  &  & 7 & $-\quad$ & $-\quad$  \\
$\la\Btb{f_{+}}^{\mu\nu}\sigma_{\mu\nu}\Bt\ra$ & 6 & $-\quad$ & $-\quad$ & 8 & $-\quad$ & $-\quad$  \\
$\la{f_{+}}^{\mu\nu}\ra\la\Btb\sigma_{\mu\nu}\Bt\ra$ & 7 & $-\quad$ & $-\quad$ & 9 & $-\quad$ & $-\quad$  \\
$\la\Btb\chip\Bt\ra$ & 8 & $-\quad$ & $-\quad$ & 10 & $-\quad$ & $-\quad$  \\
$\la\chip\ra\la\Btb\Bt\ra$ & 9 & $-\quad$ & $-\quad$ & 11 & $-\quad$ & $-\quad$  \\
$\la\Bsb u^{\mu}u_{\mu}\Bs\ra$ & 10 & $*\quad$ & $\mathrm{I}\quad$ & 12 & $*\quad$  & $\mathrm{I}\quad$  \\
$\la\Bsb u^{\mu}u^{\nu}D_{\mu\nu}\Bs\ra$ & 11 & $*\quad$ & $\mathrm{I}\quad$ & 13 & $*\quad$  & $\mathrm{I}\quad$  \\
$i\la\Bsb u^{\mu}u^{\nu}\sigma_{\mu\nu}\Bs\ra$ & 12 & $*\quad$ & $\mathrm{I}\quad$ & 14 & $*\quad$  & $\mathrm{I}\quad$  \\
$\la\Bsb u^{\mu}\Bs{u^{T}}_{\mu}\ra$ & 13 & $*\quad$ & $\mathrm{I}\quad$ & 15 & $*\quad$  & $\mathrm{I}\quad$  \\
$\la\Bsb u^{\mu}{D_{\mu}}^{\nu}\Bs{u^{T}}_{\nu}\ra$ & 14 & $*\quad$ & $\mathrm{I}\quad$ & 16 & $*\quad$  & $\mathrm{I}\quad$  \\
$\la\Bsb\Bs\ra\la u^{\mu}u_{\mu}\ra$ &  &  &  & 17 & $*\quad$  & $\mathrm{I}\quad$  \\
$\la\Bsb D^{\mu\nu}\Bs\ra\la u_{\mu}u_{\nu}\ra$ &  &  &  & 18 & $*\quad$  & $\mathrm{I}\quad$  \\
$\la\Bsb{f_{+}}^{\mu\nu}\sigma_{\mu\nu}\Bs\ra$ & 15 & $-\frac{1}{3}d^{(2)}_{13}$ & $\mathrm{I}\quad$ & 19 & $-\frac{1}{3}D^{(2)}_{15}$  & $\mathrm{I}\quad$  \\
$\la{f_{+}}^{\mu\nu}\ra\la\Bsb\sigma_{\mu\nu}\Bs\ra$ & 16 & $-\frac{1}{3}d^{(2)}_{14}$ & $\mathrm{I}\quad$ & 20 & $0\quad$  & $0\quad$  \\
$\la\Bsb\chip\Bs\ra$ & 17 & $-d^{(2)}_{15}$ & $\mathrm{I}\quad$ & 21 & $-D^{(2)}_{16}$  & $\mathrm{I}\quad$  \\
$\la\chip\ra\la\Bsb\Bs\ra$ & 18 & $-d^{(2)}_{16}$ & $\mathrm{I}\quad$ & 22 & $-D^{(2)}_{17}$  & $\mathrm{I}\quad$  \\
$\la\Bssb^{\mu}u_{\mu}u^{\nu}\Bss_{\nu}\ra$ & 19 & $d^{(2)}_{6}$ & $\mathrm{I}\quad$ & 23 & $D^{(2)}_{5}$  & $\mathrm{I}\quad$  \\
$\la\Bssb^{\mu}u^{\nu}u_{\mu}\Bss_{\nu}\ra$ & 20 & $d^{(2)}_{7}$ & $*\quad$ & 24 & $D^{(2)}_{6}$  & $*\quad$  \\
$\la\Bssb^{\mu}u^{\nu}u_{\nu}\Bss_{\mu}\ra$ & 21 & $d^{(2)}_{8}$ & $*\quad$ & 25 & $D^{(2)}_{7}$  & $*\quad$  \\
$\la\Bssb^{\mu}u^{\nu}u^{\lambda}D_{\nu\lambda}\Bss_{\mu}\ra$ & 22 & $-d^{(2)}_{9}$ & $*\quad$ & 26 & $-D^{(2)}_{8}$  & $*\quad$  \\
$\la\Bssb^{\mu}u_{\mu}\Bss^{\nu}{u^{T}}_{\nu}\ra+\mathrm{H.c.}$ & 23 & $d^{(2)}_{10}$ & $\mathrm{I}\quad$ & 27 & $D^{(2)}_{9}$  & $\mathrm{I}\quad$  \\
$\la\Bssb^{\mu}u^{\nu}\Bss_{\mu}{u^{T}}_{\nu}\ra$ & 24 & $d^{(2)}_{11}$ & $*\quad$ & 28 & $D^{(2)}_{10}$  & $*\quad$  \\
$\la\Bssb^{\mu}u^{\nu}{D_{\nu}}^{\lambda}\Bss_{\mu}{u^{T}}_{\lambda}\ra$ & 25 & $-d^{(2)}_{12}$ & $*\quad$ & 29 & $-D^{(2)}_{11}$  & $*\quad$  \\
$\la\Bssb^{\mu}\Bss_{\mu}\ra\la u^{\nu}u_{\nu}\ra+\mathrm{H.c.}$ &  &  &  & 30 & $D^{(2)}_{12}$  & $\mathrm{I}\quad$  \\
$\la\Bssb^{\mu}\Bss^{\nu}\ra\la u_{\mu}u_{\nu}\ra+\mathrm{H.c.}$ &  &  &  & 31 & $D^{(2)}_{13}$  & $*\quad$  \\
$\la\Bssb^{\mu}D^{\nu\lambda}\Bss_{\mu}\ra\la u_{\nu}u_{\lambda}\ra+\mathrm{H.c.}$ &  &  &  & 32 & $-D^{(2)}_{14}$  & $*\quad$  \\
$i\la\Bssb^{\mu}{f_{+\mu}}^{\nu}\Bss_{\nu}\ra$ & 26 & $d^{(2)}_{13}$ & $-3\tilde{c}^{(2)}_{15}$ & 33 & $D^{(2)}_{15}$  & $-3\tilde{C}^{(2)}_{19}$  \\
$i\la{f_{+}}^{\mu\nu}\ra\la\Bssb_{\mu}\Bss_{\nu}\ra$ & 27 & $d^{(2)}_{14}$ & $-3\tilde{c}^{(2)}_{16}$ & 34 & $0\quad$  & $0\quad$  \\
$\la\Bssb^{\mu}\chip\Bss_{\mu}\ra$ & 28 & $d^{(2)}_{15}$ & $-\tilde{c}^{(2)}_{17}$ & 35 & $D^{(2)}_{16}$  & $-\tilde{C}^{(2)}_{21}$  \\
$\la\chip\ra\la\Bssb^{\mu}\Bss_{\mu}\ra$ & 29 & $d^{(2)}_{16}$ & $-\tilde{c}^{(2)}_{18}$ & 36 & $D^{(2)}_{17}$  & $-\tilde{C}^{(2)}_{22}$  \\
$\la\Btb u^{\mu}u_{\mu}\Bs\ra+\mathrm{H.c.}$ & 30 & $0\quad$ & $0\quad$ & 37 & $0\quad$  & $0\quad$  \\
$\la\Btb u^{\mu}u^{\nu}D_{\mu\nu}\Bs\ra+\mathrm{H.c.}$ & 31 & $0\quad$ & $0\quad$ & 38 & $0\quad$  & $0\quad$  \\
$i\la\Btb u^{\mu}u^{\nu}\sigma_{\mu\nu}\Bs\ra+\mathrm{H.c.}$ & 32 & $-\frac{\sqrt{3}}{3}d^{(2)}_{1}$ & $\mathrm{I}\quad$ & 39 & $-\frac{\sqrt{3}}{3}D^{(2)}_{1}$  & $\mathrm{I}\quad$  \\
$i\la\Btb u^{\mu}{\sigma_{\mu}}^{\nu}\Bs{u^{T}}_{\nu}\ra+\mathrm{H.c.}$ & 33 & $-\frac{\sqrt{3}}{3}d^{(2)}_{2}$ & $\mathrm{I}\quad$ & 40 & $-\frac{\sqrt{3}}{3}D^{(2)}_{2}$  & $\mathrm{I}\quad$  \\
$\la\Btb\Bs\ra\la u^{\mu}u_{\mu}\ra+\mathrm{H.c.}$ &  &  &  & 41 & $0\quad$  & $0\quad$  \\
$\la\Btb D^{\mu\nu}\Bs\ra\la u_{\mu}u_{\nu}\ra+\mathrm{H.c.}$ &  &  &  & 42 & $0\quad$  & $0\quad$  \\
$\la\Btb{f_{-}}^{\mu\nu}\gamf\gamma_{\mu}D_{\nu}\Bs\ra+\mathrm{H.c.}$ & 34 & $\frac{\sqrt{3}}{3}d^{(2)}_{3}$ & $\mathrm{I}\quad$ & 43 & $\frac{\sqrt{3}}{3}D^{(2)}_{3}$  & $\mathrm{I}\quad$  \\
$\la\Btb{f_{+}}^{\mu\nu}\sigma_{\mu\nu}\Bs\ra+\mathrm{H.c.}$ & 35 & $-\frac{\sqrt{3}}{3}d^{(2)}_{4}$ & $\mathrm{I}\quad$ & 44 & $-\frac{\sqrt{3}}{3}D^{(2)}_{4}$  & $\mathrm{I}\quad$  \\
$\la{f_{+}}^{\mu\nu}\ra\la\Btb\sigma_{\mu\nu}\Bs\ra+\mathrm{H.c.}$ & 36 & $-\frac{\sqrt{3}}{3}d^{(2)}_{5}$ & $\mathrm{I}\quad$ & 45 & $0\quad$  & $0\quad$  \\
$\la\Btb\chip\Bs\ra+\mathrm{H.c.}$ & 37 & $0\quad$ & $0\quad$ & 46 & $0\quad$  & $0\quad$  \\
$\la\chip\ra\la\Btb\Bs\ra+\mathrm{H.c.}$ & 38 & $0\quad$ & $0\quad$ & 47 & $0\quad$  & $0\quad$  \\
$\la\Btb u^{\mu}u^{\nu}\gamf\gamma_{\mu}\Bss_{\nu}\ra+\mathrm{H.c.}$ & 39 & $d^{(2)}_{1}$ & $-\sqrt{3}\tilde{c}^{(2)}_{32}$ & 48 & $D^{(2)}_{1}$  & $-\sqrt{3}\tilde{C}^{(2)}_{39}$  \\
$\la\Btb u^{\mu}u^{\nu}\gamf\gamma_{\nu}\Bss_{\mu}\ra+\mathrm{H.c.}$ & 40 & $-d^{(2)}_{1}$ & $\sqrt{3}\tilde{c}^{(2)}_{32}$ & 49 & $-D^{(2)}_{1}$  & $\sqrt{3}\tilde{C}^{(2)}_{39}$  \\
$\la\Btb u^{\mu}\gamf\gamma_{\mu}\Bss^{\nu}{u^{T}}_{\nu}\ra+\mathrm{H.c.}$ & 41 & $2d^{(2)}_{2}$ & $-2\,\sqrt{3}\tilde{c}^{(2)}_{33}$ & 50 & $2D^{(2)}_{2}$  & $-2\,\sqrt{3}\tilde{C}^{(2)}_{40}$  \\
$\la u^{\mu}u^{\nu}\ra\la\Btb\gamf\gamma_{\mu}\Bss_{\nu}\ra+\mathrm{H.c.}$ &  &  &  & 51 & $0\quad$  & $0\quad$  \\
$i\la\Btb{f_{+}}^{\mu\nu}\gamf\gamma_{\mu}\Bss_{\nu}\ra+\mathrm{H.c.}$ & 42 & $-2d^{(2)}_{4}$ & $2\,\sqrt{3}\tilde{c}^{(2)}_{35}$ & 52 & $-2D^{(2)}_{4}$  & $2\,\sqrt{3}\tilde{C}^{(2)}_{44}$  \\
$i\la{f_{+}}^{\mu\nu}\ra\la\Btb\gamf\gamma_{\mu}\Bss_{\nu}\ra+\mathrm{H.c.}$ & 43 & $-2d^{(2)}_{5}$ & $2\,\sqrt{3}\tilde{c}^{(2)}_{36}$ & 53 & $0\quad$  & $0\quad$  \\
$\la\Bsb u^{\mu}u^{\nu}\gamf\gamma_{\mu}\Bss_{\nu}\ra+\mathrm{H.c.}$ & 44 & $\frac{\sqrt{3}}{3}d^{(2)}_{6}$ & $\frac{\sqrt{3}}{3}\tilde{c}^{(2)}_{19}$ & 54 & $\frac{\sqrt{3}}{3}D^{(2)}_{5}$  & $\frac{\sqrt{3}}{3}\tilde{C}^{(2)}_{23}$  \\
$\la\Bsb u^{\mu}u^{\nu}\gamf\gamma_{\nu}\Bss_{\mu}\ra+\mathrm{H.c.}$ & 45 & $\frac{\sqrt{3}}{3}d^{(2)}_{7}$ & $*\quad$ & 55 & $\frac{\sqrt{3}}{3}D^{(2)}_{6}$  & $*\quad$  \\
$\la\Bsb u^{\mu}\gamf\gamma_{\mu}\Bss^{\nu}{u^{T}}_{\nu}\ra+\mathrm{H.c.}$ & 46 & $\frac{\sqrt{3}}{3}d^{(2)}_{10}$ & $\frac{\sqrt{3}}{3}\tilde{c}^{(2)}_{23}$ & 56 & $\frac{\sqrt{3}}{3}D^{(2)}_{9}$  & $\frac{\sqrt{3}}{3}\tilde{C}^{(2)}_{27}$  \\
$\la u^{\mu}u^{\nu}\ra\la\Bsb\gamf\gamma_{\mu}\Bss_{\nu}\ra+\mathrm{H.c.}$ &  &  &  & 57 & $\frac{\sqrt{3}}{3}D^{(2)}_{13}$  & $*\quad$  \\
$i\la\Bsb{f_{+}}^{\mu\nu}\gamf\gamma_{\mu}\Bss_{\nu}\ra+\mathrm{H.c.}$ & 47 & $\frac{\sqrt{3}}{3}d^{(2)}_{13}$ & $-\sqrt{3}\tilde{c}^{(2)}_{15}$ & 58 & $\frac{\sqrt{3}}{3}D^{(2)}_{15}$  & $-\sqrt{3}\tilde{C}^{(2)}_{19}$  \\
$i\la{f_{+}}^{\mu\nu}\ra\la\Bsb\gamf\gamma_{\mu}\Bss_{\nu}\ra+\mathrm{H.c.}$ & 48 & $\frac{\sqrt{3}}{3}d^{(2)}_{14}$ & $-\sqrt{3}\tilde{c}^{(2)}_{16}$ & 59 & $0\quad$  & $0\quad$  \\
\hline
\end{longtable}

\begin{align}\label{relation-p2-1}
\begin{autobreak}
~
\tilde{c}^{(2)}_{10}=-\frac{1}{3}d^{(2)}_{6}-\frac{1}{3}d^{(2)}_{7}-d^{(2)}_{8},\;
\tilde{c}^{(2)}_{11}=-\frac{1}{3}d^{(2)}_{6}-\frac{1}{3}d^{(2)}_{7}+d^{(2)}_{9},\;
\tilde{c}^{(2)}_{12}=\frac{1}{3}d^{(2)}_{6}-\frac{1}{3}d^{(2)}_{7},\;
\tilde{c}^{(2)}_{13}=-\frac{2}{3}d^{(2)}_{10}-d^{(2)}_{11},\;
\tilde{c}^{(2)}_{14}=-\frac{2}{3}d^{(2)}_{10}+d^{(2)}_{12}.
\end{autobreak}
\end{align}
\begin{align}\label{relation-p2-2}
\begin{autobreak}
~
\tilde{c}^{(2)}_{20}=-3\tilde{c}^{(2)}_{12}+\tilde{c}^{(2)}_{19},\;
\tilde{c}^{(2)}_{21}=-\tilde{c}^{(2)}_{10}+\tilde{c}^{(2)}_{12}-\frac{2}{3}\tilde{c}^{(2)}_{19},\;
\tilde{c}^{(2)}_{22}=-\tilde{c}^{(2)}_{11}+\tilde{c}^{(2)}_{12}-\frac{2}{3}\tilde{c}^{(2)}_{19},\;
\tilde{c}^{(2)}_{24}=-\tilde{c}^{(2)}_{13}-\frac{2}{3}\tilde{c}^{(2)}_{23},\;
\tilde{c}^{(2)}_{25}=-\tilde{c}^{(2)}_{14}-\frac{2}{3}\tilde{c}^{(2)}_{23},\;
\tilde{c}^{(2)}_{45}=-\sqrt{3}\tilde{c}^{(2)}_{12}+\frac{\sqrt{3}}{3}\tilde{c}^{(2)}_{19}.
\end{autobreak}
\end{align}
\begin{align}\label{relation-p2-3}
\begin{autobreak}
~
\tilde{C}^{(2)}_{12}=-\frac{1}{3}D^{(2)}_{5}-\frac{1}{3}D^{(2)}_{6}-D^{(2)}_{7},\;
\tilde{C}^{(2)}_{13}=-\frac{1}{3}D^{(2)}_{5}-\frac{1}{3}D^{(2)}_{6}+D^{(2)}_{8},\;
\tilde{C}^{(2)}_{14}=\frac{1}{3}D^{(2)}_{5}-\frac{1}{3}D^{(2)}_{6},\;
\tilde{C}^{(2)}_{15}=-\frac{2}{3}D^{(2)}_{9}-D^{(2)}_{10},\;
\tilde{C}^{(2)}_{16}=-\frac{2}{3}D^{(2)}_{9}+D^{(2)}_{11},\;
\tilde{C}^{(2)}_{17}=-D^{(2)}_{12}-\frac{1}{3}D^{(2)}_{13},\;
\tilde{C}^{(2)}_{18}=-\frac{1}{3}D^{(2)}_{13}+D^{(2)}_{14}.
\end{autobreak}
\end{align}
\begin{align}\label{relation-p2-4}
\begin{autobreak}
~
\tilde{C}^{(2)}_{24}=-3\tilde{C}^{(2)}_{14}+\tilde{C}^{(2)}_{23},\;
\tilde{C}^{(2)}_{25}=-\tilde{C}^{(2)}_{12}+\tilde{C}^{(2)}_{14}-\frac{2}{3}\tilde{C}^{(2)}_{23},\;
\tilde{C}^{(2)}_{26}=-\tilde{C}^{(2)}_{13}+\tilde{C}^{(2)}_{14}-\frac{2}{3}\tilde{C}^{(2)}_{23},\;
\tilde{C}^{(2)}_{28}=-\tilde{C}^{(2)}_{15}-\frac{2}{3}\tilde{C}^{(2)}_{27},\;
\tilde{C}^{(2)}_{29}=-\tilde{C}^{(2)}_{16}-\frac{2}{3}\tilde{C}^{(2)}_{27},\;
\tilde{C}^{(2)}_{31}=-3\tilde{C}^{(2)}_{17}-3\tilde{C}^{(2)}_{30},\;
\tilde{C}^{(2)}_{32}=\tilde{C}^{(2)}_{17}-\tilde{C}^{(2)}_{18}+\tilde{C}^{(2)}_{30},\;
\tilde{C}^{(2)}_{55}=-\sqrt{3}\tilde{C}^{(2)}_{14}+\frac{\sqrt{3}}{3}\tilde{C}^{(2)}_{23},\;
\tilde{C}^{(2)}_{57}=-\sqrt{3}\tilde{C}^{(2)}_{17}-\sqrt{3}\tilde{C}^{(2)}_{30}.
\end{autobreak}
\end{align}

\begin{table}[htbp]
\caption{Independent Lagrangian terms in the heavy quark limit at the $\mathcal{O}(p^{2})$ order.}\label{tab:table2}
\begin{tabular}{lcclcc}
\hline $P_n$ & $SU(2)$ & $SU(3)$ & $P_n$ & $SU(2)$ & $SU(3)$\\
\hline
$i\epsilon^{\mu\nu\lambda\rho}\la\Btb u_{\mu}u_{\nu}v_{\lambda}\psi_{\rho}\ra$ & 1 & 1 & $\la\psib^{\mu}u^{\nu}\psi_{\mu}{u^{T}}_{\nu}\ra$ & 11 & 10  \\
$i\epsilon^{\mu\nu\lambda\rho}\la\Btb u_{\mu}v_{\nu}\psi_{\lambda}{u^{T}}_{\rho}\ra$ & 2 & 2 & $\la\psib^{\mu}u^{\nu}v_{\nu}v^{\lambda}\psi_{\mu}{u^{T}}_{\lambda}\ra$ & 12 & 11  \\
$i\la\Btb{f_{-}}^{\mu\nu}v_{\mu}\psi_{\nu}\ra$ & 3 & 3 & $\la\psib^{\mu}\psi_{\mu}\ra\la u^{\nu}u_{\nu}\ra$ &  & 12  \\
$\epsilon^{\mu\nu\lambda\rho}\la\Btb f_{+\mu\nu}v_{\lambda}\psi_{\rho}\ra$ & 4 & 4 & $\la\psib^{\mu}\psi^{\nu}\ra\la u_{\mu}u_{\nu}\ra$ &  & 13  \\
$\epsilon^{\mu\nu\lambda\rho}\la f_{+\mu\nu}\ra\la\Btb v_{\lambda}\psi_{\rho}\ra$ & 5 & & $\la\psib^{\mu}v^{\nu}v^{\lambda}\psi_{\mu}\ra\la u_{\nu}u_{\lambda}\ra$ &  & 14  \\
$\la\psib^{\mu}u_{\mu}u^{\nu}\psi_{\nu}\ra$ & 6 & 5 & $i\la\psib^{\mu}{f_{+\mu}}^{\nu}\psi_{\nu}\ra$ & 13 & 15  \\
$\la\psib^{\mu}u^{\nu}u_{\mu}\psi_{\nu}\ra$ & 7 & 6 & $i\la{f_{+}}^{\mu\nu}\ra\la\psib_{\mu}\psi_{\nu}\ra$ & 14 &  \\
$\la\psib^{\mu}u^{\nu}u_{\nu}\psi_{\mu}\ra$ & 8 & 7 & $\la\psib^{\mu}\chip\psi_{\mu}\ra$ & 15 & 16  \\
$\la\psib^{\mu}u^{\nu}u^{\lambda}v_{\nu}v_{\lambda}\psi_{\mu}\ra$ & 9 & 8 & $\la\chip\ra\la\psib^{\mu}\psi_{\mu}\ra$ & 16 & 17  \\
$\la\psib^{\mu}u_{\mu}\psi^{\nu}{u^{T}}_{\nu}\ra$ & 10 & 9  \\
\hline
\end{tabular}
\end{table}

\subsection{$\mathcal{O}(p^3)$ and $\mathcal{O}(p^4)$ orders}

The $\mathcal{O}(p^{3})$ and $\mathcal{O}(p^4)$ order results are very long. We give them in Appendices. Table \ref{tab:table5} of Appendix \ref{app:A} list the relativistic terms at the $\mathcal{O}(p^{3})$ order. The contents are similar to Table \ref{tab:table3}. When the obtained LEC relations are not simple, we show them in Eqs. \eqref{relation-p3-1}-\eqref{relation-p3-4}. In Table \ref{tab:table4}, we collect the heavy-quark Lagrangian terms at the order $\mathcal{O}(p^{3})$. The $\mathcal{O}(p^{4})$ relativistic terms are given in Table \ref{tab:table6} of Appendix \ref{app:B}. For the Lagrangian in the heavy quark limit, Table \ref{tab:table7} display its independent terms. In the present study, we do not consider the LEC relations at the $\mathcal{O}(p^4)$ order because of the large number of terms and the computational conditions. We only mark the independent relativistic terms in the columns 3, 5, 8, and 10 of Table \ref{tab:table6} when employing the heavy quark symmetry. The contact terms appear at this order and they are placed at the end of each part of interaction terms in Eqs. \eqref{eq:rl} and \eqref{eq:sfl}.

\subsection{Discussions}

Comparing chiral Lagrangians for different baryons, one can find some similar structures. This may help us to check the results from some sides. $B_{\bar 3}$ and $B_6$ differ only in the transposition property of matrices. The structures of $\mathscr{L}_{B_{\bar{3}} B_{\bar{3}}}$ and $\mathscr{L}_{B_{6} B_{6}}$ should be similar for both two- and three-flavor cases. One may confirm this from the obtained interaction terms. The existence of transposed building blocks in part of interaction terms is a key characteristic of the singly-heavy-baryon Lagrangians. Excluding such terms, one finds that the two-flavor Lagrangians $\mathscr{L}_{B_{\bar{3}} B_{\bar{3}}, B_6B_6}$ and the pion-nucleon Lagrangian \cite{Fettes:2000gb} have the same number of independent structures. If appropriate combinations of terms are adopted, similar structures in these two cases may be further observed. In Ref. \cite{Liu:2023lsg}, we obtained chiral Lagrangians with doubly charmed baryons. Up to the $\mathcal{O}(p^3)$ order (for both two and three flavor cases), still excluding the terms with transposed building blocks, one finds that the structures in $\mathscr{L}_{B_{\bar{3}} B_{\bar{3}},B_6B_6}$, $\mathscr{L}_{B_{6}^{*} B_{6}^{*}}$, and $\mathscr{L}_{B_{\bar{3}} B_{6}^{*},B_6B_6^*}$ are similar to those in $\mathcal{L}_{B B}$, $\mathcal{L}_{TT}$, and $\mathcal{L}_{B T}$, respectively, where $B$ ($T$) denotes spin-$\frac{1}{2}$ (spin-$\frac{3}{2}$) doubly charmed baryon fields. For the $SU(2)$ or $SU(3)$ Lagrangians $\mathscr{L}_{B_{\bar{3}} B_{6}^{*}}$ and $\mathscr{L}_{B_{6} B_{6}^{*}}$, same structures are found up to the $\mathcal{O}(p^3)$ order. At the $\mathcal{O}(p^4)$ order, however, the feature of same structure is not always true, because the symmetries of baryons are different [see Eq. \eqref{eq:sym}].

We chose the definition of the superfield [Eq. \eqref{eq:sf}] in Refs. \cite{Detmold:2011rb,Meng:2018gan,Kawakami:2018olq,Meng:2018gan}. A different choice of phase between $B_6$ and $B_6^*$ may be found in the literature. It affects LEC relations involving terms in ${\cal L}_{B_{\bar 3}B_6}$, ${\cal L}_{B_{\bar 3}B_6^*}$, and ${\cal L}_{B_6B_6^*}$, but does not affect forms of obtained results and any physics.

\section{Summary and outlook}\label{Sec:V}

In this paper, we construct the relativistic chiral Lagrangians for singly heavy baryons to the ${\cal O}(p^4)$ order, including both $SU(2)$ and $SU(3)$ cases. The chiral Lagrangians in the heavy quark limit are also obtained to this order. We collect the numbers of independent terms at each order in Table \ref{numberofterms}. Using the heavy quark symmetry, one finds that the number of independent LECs could be reduced to about $1/3$ of its original value. This symmetry may partly solve the problem that unknown LECs at high orders would affect the applications of ChPT. In the present study, we obtain LEC relations between the relativistic case and the heavy-quark-limit case up to the third chiral order. The relations among independent and non-independent relativistic LECs are also obtained.

The studies of magnetic moments \cite{Meng:2018gan,Wang:2018gpl} and radiative decays \cite{Wang:2018cre} of singly heavy baryons have involved the ${\cal O}(p^4)$ order chiral Lagrangians. Now, with this order Lagrangians constructed here, one may check whether the adopted operators in these references are complete. Of course, the newly constructed Lagrangians can also serve for future studies of other quantities to the fourth chiral order which can help us to understand the convergence and power counting problem better. A natural question about LEC determination follows such studies. Although the number of independent ${\cal O}(p^4)$ order terms and LECs are very large, not all of them are needed in the study of a special problem. From the experience of high order chiral studies, only several LECs or several LEC combinations are usually involved. One may turn to various theoretical methods (e.g. quark model, large $N_c$, and lattice QCD) to determine or constrain their values in the case that available experimental data are not enough.

In fact, the complete Lagrangians may motivate future studies of LEC determinations or LEC relations. At the leading order, the LECs can be determined with the quark model or other methods \cite{Yan:1992gz,Meng:2018gan,Liu:2011xc,Meguro:2011nr}. At higher orders, however, only few results have been obtained \cite{Meng:2018gan,Wang:2018cre,Jiang:2014ena}. Since the numbers of LECs are significantly increased for high-order terms, theoretical methods to study LECs in a systematic way are welcome. In Ref. \cite{Jiang:2015dba}, an analytical method was adopted to calculate the LECs for mesonic chiral Lagrangians. A possible method in the present case would also be developed in future with complete and independent Lagrangians. In Refs. \cite{Jiang:2022gjy,Jiang:2023zqq}, we preliminarily studied the LEC relations with the help of chiral quark model up to the third chiral order, which is based on operator correspondences. In principle, such a study can be extended to any chiral order, but it needs complete Lagrangians and independent terms.

The explicit high order Lagrangians are also helpful to improve methods to analyze the structure of chiral effective theories. It is a natural problem in ChPT how many independent structures are at a given chiral order. In the mesonic case, the authors of Ref. \cite{Graf:2020yxt} obtained total numbers of independent terms in the chiral Lagrangian at different orders with the help of Hilbert series techniques. They validated the results with concrete Lagrangians. Once the study in meson-baryon case is performed, the Lagrangian constructed here would be also helpful.

\begin{table}[htbp]
\caption{Numbers of independent terms at each chiral order.}\label{numberofterms}
\begin{tabular}{ccccc|cccc}\hline\hline
&\multicolumn{4}{c|}{Relativistic case}&\multicolumn{4}{c}{Heavy quark case}\\
Chiral order&${\cal O}(p^1)$&${\cal O}(p^2)$&${\cal O}(p^3)$&${\cal O}(p^4)$&${\cal O}(p^1)$&${\cal O}(p^2)$&${\cal O}(p^3)$&${\cal O}(p^4)$\\
\hline $SU(2)$&6&48&199&1242&2&16&64&412\\
$SU(3)$&6&59&307&2454&2&17&88&714\\\hline\hline
\end{tabular}
\end{table}

\begin{acknowledgments}
This project was supported by the National Natural Science Foundation of China under Grants No. 11775132, No. 12235008, and Guangxi Science Foundation under Grants No. 2022GXNSFAA035489.
\end{acknowledgments}

\appendix
\section{$\mathcal{O}(p^{3})$ order results} \label{app:A}

\begin{longtable}{ccccccccccccc}
\caption{\label{tab:table5}Independent terms in the relativistic Lagrangian at the $\mathcal{O}(p^{3})$ order. The columns 3 and 6 list the LEC relations between the relativistic case and the heavy quark case. The columns 4 and 7 show the LEC relations between different terms in the relativistic case by using the heavy quark symmetry. In these two columns, ``I'' means that we choose the term as an independent one when employing the heavy quark symmetry. In the columns 3, 4, 6, and 7, the LEC relations marked by ``*'' are given in Eqs. \eqref{relation-p3-1}, \eqref{relation-p3-2}, \eqref{relation-p3-3}, and \eqref{relation-p3-4}, respectively, in this Appendix. ``0'' means that the LECs in the heavy quark limit vanish. ``$-$'' means that the LEC in the heavy quark case vanishes or is equal to that in the relativistic case.}\\
\hline\hline $O_n$ & $SU(2)$ & $\tilde{c}^{(2)}_n$ & $\tilde{c}^{(2)}_n$ & $SU(3)$ & $\tilde{C}^{(3)}_n$ & $\tilde{C}^{(3)}_n$\\
\hline\endfirsthead

\hline\hline $O_n$ & $SU(2)$ & $\tilde{c}^{(2)}_n$ & $\tilde{c}^{(2)}_n$ & $SU(3)$ & $\tilde{C}^{(3)}_n$ & $\tilde{C}^{(3)}_n$\\
\hline\endhead

\hline\hline 
\endfoot
\hline\endlastfoot
$\la\Btb u^{\mu}u_{\mu}u^{\nu}\gamf\gamma_{\nu}\Bt\ra+\mathrm{H.c.}$ & 1 & $-\quad$ & $-\quad$ & 1 & $-\quad$ & $-\quad$  \\
$\la\Btb u^{\mu}u^{\nu}u_{\mu}\gamf\gamma_{\nu}\Bt\ra$ & 2 & $-\quad$ & $-\quad$ & 2 & $-\quad$ & $-\quad$  \\
$\la\Btb u^{\mu}u^{\nu}u^{\lambda}\gamf\gamma_{\mu}D_{\nu\lambda}\Bt\ra+\mathrm{H.c.}$ & 3 & $-\quad$ & $-\quad$ & 3 & $-\quad$ & $-\quad$  \\
$\la\Btb u^{\mu}u^{\nu}u^{\lambda}\gamf\gamma_{\nu}D_{\mu\lambda}\Bt\ra$ & 4 & $-\quad$ & $-\quad$ & 4 & $-\quad$ & $-\quad$  \\
$\la\Btb u^{\mu}u_{\mu}\gamf\gamma^{\nu}\Bt{u^{T}}_{\nu}\ra$ &  &  &  & 5 & $-\quad$ & $-\quad$  \\
$\la\Btb u^{\mu}u^{\nu}\gamf\gamma_{\mu}\Bt{u^{T}}_{\nu}\ra+\mathrm{H.c.}$ &  &  &  & 6 & $-\quad$ & $-\quad$  \\
$\la\Btb u^{\mu}u^{\nu}\gamf\gamma_{\mu}{D_{\nu}}^{\lambda}\Bt{u^{T}}_{\lambda}\ra+\mathrm{H.c.}$ &  &  &  & 7 & $-\quad$ & $-\quad$  \\
$\la\Btb u^{\mu}u^{\nu}\gamf\gamma^{\lambda}D_{\mu\nu}\Bt{u^{T}}_{\lambda}\ra$ &  &  &  & 8 & $-\quad$ & $-\quad$  \\
\hline
\end{longtable}

\begin{align}\label{relation-p3-1}
\begin{autobreak}
~
\tilde{c}^{(3)}_{29}=-\frac{1}{3}d^{(3)}_{29}+\frac{2}{3}d^{(3)}_{32}-\frac{1}{6}d^{(3)}_{34},\;
\tilde{c}^{(3)}_{30}=\frac{2}{3}d^{(3)}_{29}+\frac{1}{3}d^{(3)}_{34},\;
\tilde{c}^{(3)}_{31}=-\frac{1}{3}d^{(3)}_{29}-\frac{2}{3}d^{(3)}_{30}-\frac{1}{6}d^{(3)}_{34},\;
\tilde{c}^{(3)}_{32}=\frac{2}{3}d^{(3)}_{29}+\frac{4}{3}d^{(3)}_{30}+\frac{1}{3}d^{(3)}_{34},\;
\tilde{c}^{(3)}_{33}=\frac{2}{3}d^{(3)}_{29}-\frac{2}{3}d^{(3)}_{31},\;
\tilde{c}^{(3)}_{34}=-\frac{2}{3}d^{(3)}_{33}-\frac{2}{3}d^{(3)}_{34},\;
\tilde{c}^{(3)}_{35}=-\frac{1}{3}d^{(3)}_{35}+\frac{1}{3}d^{(3)}_{37}+d^{(3)}_{38},\;
\tilde{c}^{(3)}_{36}=\frac{1}{3}d^{(3)}_{35}+\frac{1}{3}d^{(3)}_{37},\;
\tilde{c}^{(3)}_{38}=\frac{1}{3}d^{(3)}_{39}+\frac{1}{3}d^{(3)}_{41}+d^{(3)}_{42},\;
\tilde{c}^{(3)}_{39}=\frac{1}{3}d^{(3)}_{39}+\frac{1}{3}d^{(3)}_{40}+\frac{1}{3}d^{(3)}_{41}-d^{(3)}_{43},\;
\tilde{c}^{(3)}_{40}=-\frac{1}{3}d^{(3)}_{39}+\frac{1}{3}d^{(3)}_{41},\;
\tilde{c}^{(3)}_{44}=-\frac{2}{3}d^{(3)}_{47}-\frac{1}{3}d^{(3)}_{49}+\frac{2}{3}d^{(3)}_{50},\;
\tilde{c}^{(3)}_{45}=\frac{2}{3}d^{(3)}_{47}+\frac{4}{3}d^{(3)}_{48}+\frac{1}{3}d^{(3)}_{49},\;
\tilde{c}^{(3)}_{46}=-\frac{1}{3}d^{(3)}_{47}+\frac{1}{3}d^{(3)}_{49},\;
\tilde{c}^{(3)}_{47}=\frac{2}{3}d^{(3)}_{51}+\frac{2}{3}d^{(3)}_{52},\;
\tilde{c}^{(3)}_{48}=\frac{2}{3}d^{(3)}_{53}-\frac{2}{3}d^{(3)}_{54},\;
\tilde{c}^{(3)}_{49}=\frac{2}{3}d^{(3)}_{55}-d^{(3)}_{56},\;
\tilde{c}^{(3)}_{50}=-d^{(3)}_{57}+\frac{2}{3}d^{(3)}_{58},\;
\tilde{c}^{(3)}_{54}=\frac{1}{3}d^{(3)}_{40}-d^{(3)}_{62},\;
\tilde{c}^{(3)}_{57}=-2d^{(3)}_{29}-4d^{(3)}_{30}+2d^{(3)}_{33},\;
\tilde{c}^{(3)}_{58}=d^{(3)}_{29}-d^{(3)}_{31}+d^{(3)}_{33},\;
\tilde{c}^{(3)}_{59}=2d^{(3)}_{30}+d^{(3)}_{31},\;
\tilde{c}^{(3)}_{60}=-d^{(3)}_{30}-d^{(3)}_{32},\;
\tilde{c}^{(3)}_{63}=-2d^{(3)}_{33}-d^{(3)}_{34},\;
\tilde{c}^{(3)}_{65}=-d^{(3)}_{35}-d^{(3)}_{40}-d^{(3)}_{41},\;
\tilde{c}^{(3)}_{66}=-d^{(3)}_{36}-d^{(3)}_{40}-2d^{(3)}_{41},\;
\tilde{c}^{(3)}_{67}=-d^{(3)}_{37}-d^{(3)}_{41},\;
\tilde{c}^{(3)}_{69}=-d^{(3)}_{39}-d^{(3)}_{40}-d^{(3)}_{41},\;
\tilde{c}^{(3)}_{72}=2d^{(3)}_{47}+2d^{(3)}_{48}+2d^{(3)}_{53}-d^{(3)}_{54},\;
\tilde{c}^{(3)}_{73}=d^{(3)}_{49}+2d^{(3)}_{54},\;
\tilde{c}^{(3)}_{74}=-2d^{(3)}_{48}-d^{(3)}_{49}+2d^{(3)}_{53}-d^{(3)}_{54},\;
\tilde{c}^{(3)}_{75}=-2d^{(3)}_{48}-d^{(3)}_{50},\;
\tilde{c}^{(3)}_{76}=2d^{(3)}_{51}+d^{(3)}_{52},\;
\tilde{c}^{(3)}_{78}=-2d^{(3)}_{53}+d^{(3)}_{54},\;
\tilde{c}^{(3)}_{90}=-\frac{\sqrt{3}}{3}d^{(3)}_{3}+\frac{\sqrt{3}}{3}d^{(3)}_{6},\;
\tilde{c}^{(3)}_{91}=-\frac{\sqrt{3}}{3}d^{(3)}_{4}+\frac{\sqrt{3}}{3}d^{(3)}_{6},\;
\tilde{c}^{(3)}_{96}=-\frac{2\,\sqrt{3}}{3}d^{(3)}_{7}-\frac{\sqrt{3}}{3}d^{(3)}_{8},\;
\tilde{c}^{(3)}_{102}=-\frac{2\,\sqrt{3}}{3}d^{(3)}_{10}-\frac{\sqrt{3}}{3}d^{(3)}_{11},\;
\tilde{c}^{(3)}_{105}=-\frac{\sqrt{3}}{3}d^{(3)}_{12}-\frac{\sqrt{3}}{3}d^{(3)}_{13},\;
\tilde{c}^{(3)}_{108}=\frac{\sqrt{3}}{3}d^{(3)}_{15}-\frac{\sqrt{3}}{3}d^{(3)}_{21},\;
\tilde{c}^{(3)}_{109}=\frac{\sqrt{3}}{3}d^{(3)}_{16}+\frac{\sqrt{3}}{3}d^{(3)}_{22},\;
\tilde{c}^{(3)}_{112}=\frac{\sqrt{3}}{3}d^{(3)}_{19}-\frac{\sqrt{3}}{3}d^{(3)}_{21},\;
\tilde{c}^{(3)}_{113}=\frac{\sqrt{3}}{3}d^{(3)}_{20}+\frac{\sqrt{3}}{3}d^{(3)}_{22},\;
\tilde{c}^{(3)}_{134}=-d^{(3)}_{3}-d^{(3)}_{20},\;
\tilde{c}^{(3)}_{135}=-d^{(3)}_{4}+d^{(3)}_{20},\;
\tilde{c}^{(3)}_{140}=-2d^{(3)}_{7}-d^{(3)}_{8},\;
\tilde{c}^{(3)}_{141}=2d^{(3)}_{7}+d^{(3)}_{12},\;
\tilde{c}^{(3)}_{142}=-2d^{(3)}_{7}-d^{(3)}_{8}-d^{(3)}_{12},\;
\tilde{c}^{(3)}_{145}=-d^{(3)}_{9}-d^{(3)}_{12},\;
\tilde{c}^{(3)}_{147}=-2d^{(3)}_{10}-d^{(3)}_{11},\;
\tilde{c}^{(3)}_{149}=2d^{(3)}_{12}+2d^{(3)}_{13},\;
\tilde{c}^{(3)}_{152}=-d^{(3)}_{16}-d^{(3)}_{20},\;
\tilde{c}^{(3)}_{166}=\frac{\sqrt{3}}{3}d^{(3)}_{29}-\frac{2\,\sqrt{3}}{3}d^{(3)}_{32}+\frac{\sqrt{3}}{3}d^{(3)}_{33},\;
\tilde{c}^{(3)}_{167}=-\frac{\sqrt{3}}{3}d^{(3)}_{29}+\frac{\sqrt{3}}{3}d^{(3)}_{33}-\frac{\sqrt{3}}{3}d^{(3)}_{34},\;
\tilde{c}^{(3)}_{168}=\frac{\sqrt{3}}{3}d^{(3)}_{29}+\frac{2\,\sqrt{3}}{3}d^{(3)}_{30}-\frac{\sqrt{3}}{3}d^{(3)}_{33}+\frac{\sqrt{3}}{3}d^{(3)}_{34}-\frac{2\,\sqrt{3}}{3}d^{(3)}_{47}-\frac{2\,\sqrt{3}}{3}d^{(3)}_{48}-\frac{\sqrt{3}}{3}d^{(3)}_{49},\;
\tilde{c}^{(3)}_{169}=-\frac{\sqrt{3}}{3}d^{(3)}_{29}-\frac{2\,\sqrt{3}}{3}d^{(3)}_{30}-\frac{\sqrt{3}}{3}d^{(3)}_{33}+\frac{2\,\sqrt{3}}{3}d^{(3)}_{47}+\frac{2\,\sqrt{3}}{3}d^{(3)}_{48}+\frac{\sqrt{3}}{3}d^{(3)}_{49},\;
\tilde{c}^{(3)}_{171}=-\frac{2\,\sqrt{3}}{3}d^{(3)}_{33}+\frac{\sqrt{3}}{3}d^{(3)}_{34},\;
\tilde{c}^{(3)}_{172}=\frac{2\,\sqrt{3}}{3}d^{(3)}_{33}-\frac{\sqrt{3}}{3}d^{(3)}_{34},\;
\tilde{c}^{(3)}_{173}=\frac{2\,\sqrt{3}}{3}d^{(3)}_{33}-\frac{\sqrt{3}}{3}d^{(3)}_{34},\;
\tilde{c}^{(3)}_{174}=-\frac{\sqrt{3}}{3}d^{(3)}_{35}+\frac{\sqrt{3}}{3}d^{(3)}_{39}-\frac{\sqrt{3}}{3}d^{(3)}_{40}+\frac{\sqrt{3}}{6}d^{(3)}_{55},\;
\tilde{c}^{(3)}_{175}=-\frac{\sqrt{3}}{3}d^{(3)}_{36}+\frac{\sqrt{3}}{6}d^{(3)}_{37}-\frac{\sqrt{3}}{6}d^{(3)}_{41}-\frac{\sqrt{3}}{12}d^{(3)}_{55},\;
\tilde{c}^{(3)}_{176}=-\frac{\sqrt{3}}{2}d^{(3)}_{37}-\frac{\sqrt{3}}{6}d^{(3)}_{41}+\frac{\sqrt{3}}{12}d^{(3)}_{55},\;
\tilde{c}^{(3)}_{177}=-\frac{2\,\sqrt{3}}{3}d^{(3)}_{39}-\frac{\sqrt{3}}{3}d^{(3)}_{40},\;
\tilde{c}^{(3)}_{178}=\frac{\sqrt{3}}{6}d^{(3)}_{37}-\frac{\sqrt{3}}{3}d^{(3)}_{40}-\frac{\sqrt{3}}{6}d^{(3)}_{41}+\frac{\sqrt{3}}{12}d^{(3)}_{55},\;
\tilde{c}^{(3)}_{179}=-\frac{\sqrt{3}}{6}d^{(3)}_{37}-\frac{\sqrt{3}}{2}d^{(3)}_{41}-\frac{\sqrt{3}}{12}d^{(3)}_{55},\;
\tilde{c}^{(3)}_{182}=-\frac{\sqrt{3}}{3}d^{(3)}_{35}-\frac{\sqrt{3}}{3}d^{(3)}_{39}-\frac{\sqrt{3}}{6}d^{(3)}_{55},\;
\tilde{c}^{(3)}_{183}=-\frac{\sqrt{3}}{3}d^{(3)}_{36}-\frac{\sqrt{3}}{3}d^{(3)}_{40}+\frac{\sqrt{3}}{6}d^{(3)}_{55},\;
\tilde{c}^{(3)}_{186}=-\frac{2\,\sqrt{3}}{3}d^{(3)}_{47}-\frac{\sqrt{3}}{3}d^{(3)}_{49},\;
\tilde{c}^{(3)}_{187}=\frac{\sqrt{3}}{3}d^{(3)}_{47}+\frac{\sqrt{3}}{6}d^{(3)}_{49}+\frac{2\,\sqrt{3}}{3}d^{(3)}_{53}+\frac{\sqrt{3}}{3}d^{(3)}_{54},\;
\tilde{c}^{(3)}_{188}=\frac{2\,\sqrt{3}}{3}d^{(3)}_{51}-\frac{\sqrt{3}}{3}d^{(3)}_{52},\;
\tilde{c}^{(3)}_{189}=-\frac{2\,\sqrt{3}}{3}d^{(3)}_{51}+\frac{\sqrt{3}}{3}d^{(3)}_{52},\;
\tilde{c}^{(3)}_{190}=\frac{2\,\sqrt{3}}{3}d^{(3)}_{51}-\frac{\sqrt{3}}{3}d^{(3)}_{52},\;
\tilde{c}^{(3)}_{191}=-\frac{2\,\sqrt{3}}{3}d^{(3)}_{47}-\frac{\sqrt{3}}{3}d^{(3)}_{49}+\frac{\sqrt{3}}{3}d^{(3)}_{50},\;
\tilde{c}^{(3)}_{192}=-\frac{2\,\sqrt{3}}{3}d^{(3)}_{47}-\frac{\sqrt{3}}{3}d^{(3)}_{49}-\frac{4\,\sqrt{3}}{3}d^{(3)}_{53}-\frac{2\,\sqrt{3}}{3}d^{(3)}_{54},\;
\tilde{c}^{(3)}_{193}=\frac{2\,\sqrt{3}}{3}d^{(3)}_{53}+\frac{\sqrt{3}}{3}d^{(3)}_{54},\;
\tilde{c}^{(3)}_{194}=-\frac{2\,\sqrt{3}}{3}d^{(3)}_{53}-\frac{\sqrt{3}}{3}d^{(3)}_{54},\;
\tilde{c}^{(3)}_{195}=-\frac{2\,\sqrt{3}}{3}d^{(3)}_{53}-\frac{\sqrt{3}}{3}d^{(3)}_{54}.
\end{autobreak}
\end{align}

\begin{align}\label{relation-p3-2}
\begin{autobreak}
~
\tilde{c}^{(3)}_{58}=-6\tilde{c}^{(3)}_{31}+\frac{3}{2}\tilde{c}^{(3)}_{33}+\frac{3}{2}\tilde{c}^{(3)}_{34}+\tilde{c}^{(3)}_{57},\;
\tilde{c}^{(3)}_{59}=-6\tilde{c}^{(3)}_{31}-\frac{3}{2}\tilde{c}^{(3)}_{33}+\frac{3}{2}\tilde{c}^{(3)}_{34}+\frac{1}{2}\tilde{c}^{(3)}_{57},\;
\tilde{c}^{(3)}_{60}=-\frac{3}{2}\tilde{c}^{(3)}_{29}+\frac{3}{2}\tilde{c}^{(3)}_{31},\;
\tilde{c}^{(3)}_{61}=-\frac{3}{2}\tilde{c}^{(3)}_{30}-3\tilde{c}^{(3)}_{31},\;
\tilde{c}^{(3)}_{63}=6\tilde{c}^{(3)}_{31}-\tilde{c}^{(3)}_{57},\;
\tilde{c}^{(3)}_{64}=12\tilde{c}^{(3)}_{31}-6\tilde{c}^{(3)}_{34}-2\tilde{c}^{(3)}_{57},\;
\tilde{c}^{(3)}_{67}=-3\tilde{c}^{(3)}_{36}-3\tilde{c}^{(3)}_{37}-\tilde{c}^{(3)}_{65}+\tilde{c}^{(3)}_{66},\;
\tilde{c}^{(3)}_{69}=-3\tilde{c}^{(3)}_{37}+3\tilde{c}^{(3)}_{40}+\tilde{c}^{(3)}_{66},\;
\tilde{c}^{(3)}_{70}=-\tilde{c}^{(3)}_{35}+\tilde{c}^{(3)}_{36}+2\tilde{c}^{(3)}_{37}-\tilde{c}^{(3)}_{38}-\tilde{c}^{(3)}_{40}+\frac{2}{3}\tilde{c}^{(3)}_{65}-\frac{2}{3}\tilde{c}^{(3)}_{66}-\tilde{c}^{(3)}_{68},\;
\tilde{c}^{(3)}_{71}=\tilde{c}^{(3)}_{37}-\tilde{c}^{(3)}_{39}-\tilde{c}^{(3)}_{40}-\frac{1}{3}\tilde{c}^{(3)}_{66},\;
\tilde{c}^{(3)}_{73}=-3\tilde{c}^{(3)}_{45}+6\tilde{c}^{(3)}_{46}-6\tilde{c}^{(3)}_{48}+2\tilde{c}^{(3)}_{72},\;
\tilde{c}^{(3)}_{74}=-3\tilde{c}^{(3)}_{45}+\tilde{c}^{(3)}_{72},\;
\tilde{c}^{(3)}_{75}=-\frac{3}{2}\tilde{c}^{(3)}_{44}-\frac{3}{2}\tilde{c}^{(3)}_{45},\;
\tilde{c}^{(3)}_{77}=-6\tilde{c}^{(3)}_{47}+2\tilde{c}^{(3)}_{76},\;
\tilde{c}^{(3)}_{79}=6\tilde{c}^{(3)}_{48}+2\tilde{c}^{(3)}_{78},\;
\tilde{c}^{(3)}_{85}=\tilde{c}^{(3)}_{35}-\tilde{c}^{(3)}_{36}-\tilde{c}^{(3)}_{37}-\tilde{c}^{(3)}_{54}-\frac{2}{3}\tilde{c}^{(3)}_{65}+\frac{1}{3}\tilde{c}^{(3)}_{66}+\tilde{c}^{(3)}_{68},\;
\tilde{c}^{(3)}_{134}=\sqrt{3}\tilde{c}^{(3)}_{90}-\sqrt{3}\tilde{c}^{(3)}_{93}-\sqrt{3}\tilde{c}^{(3)}_{113}-\sqrt{3}\tilde{c}^{(3)}_{115},\;
\tilde{c}^{(3)}_{135}=\sqrt{3}\tilde{c}^{(3)}_{91}-\sqrt{3}\tilde{c}^{(3)}_{93}+\sqrt{3}\tilde{c}^{(3)}_{113}+\sqrt{3}\tilde{c}^{(3)}_{115},\;
\tilde{c}^{(3)}_{141}=-2\,\sqrt{3}\tilde{c}^{(3)}_{97}+\frac{\sqrt{3}}{2}\tilde{c}^{(3)}_{106},\;
\tilde{c}^{(3)}_{142}=\sqrt{3}\tilde{c}^{(3)}_{96}-\frac{\sqrt{3}}{2}\tilde{c}^{(3)}_{106},\;
\tilde{c}^{(3)}_{145}=\sqrt{3}\tilde{c}^{(3)}_{100}-\frac{\sqrt{3}}{2}\tilde{c}^{(3)}_{106},\;
\tilde{c}^{(3)}_{151}=\sqrt{3}\tilde{c}^{(3)}_{108}+\sqrt{3}\tilde{c}^{(3)}_{114},\;
\tilde{c}^{(3)}_{152}=-\sqrt{3}\tilde{c}^{(3)}_{109}-\sqrt{3}\tilde{c}^{(3)}_{113}-2\,\sqrt{3}\tilde{c}^{(3)}_{115},\;
\tilde{c}^{(3)}_{157}=\sqrt{3}\tilde{c}^{(3)}_{112}+\sqrt{3}\tilde{c}^{(3)}_{114},\;
\tilde{c}^{(3)}_{166}=-\sqrt{3}\tilde{c}^{(3)}_{29}-3\,\sqrt{3}\tilde{c}^{(3)}_{31}+\sqrt{3}\tilde{c}^{(3)}_{34}+\frac{\sqrt{3}}{2}\tilde{c}^{(3)}_{57},\;
\tilde{c}^{(3)}_{167}=-\frac{\sqrt{3}}{2}\tilde{c}^{(3)}_{30}-3\,\sqrt{3}\tilde{c}^{(3)}_{31}+\sqrt{3}\tilde{c}^{(3)}_{34}+\frac{\sqrt{3}}{2}\tilde{c}^{(3)}_{57},\;
\tilde{c}^{(3)}_{168}=2\,\sqrt{3}\tilde{c}^{(3)}_{31}-\sqrt{3}\tilde{c}^{(3)}_{34}+\sqrt{3}\tilde{c}^{(3)}_{45}-2\,\sqrt{3}\tilde{c}^{(3)}_{46}-\frac{\sqrt{3}}{2}\tilde{c}^{(3)}_{57}-\sqrt{3}\tilde{c}^{(3)}_{72}-\sqrt{3}\tilde{c}^{(3)}_{78},\;
\tilde{c}^{(3)}_{169}=4\,\sqrt{3}\tilde{c}^{(3)}_{31}-\sqrt{3}\tilde{c}^{(3)}_{34}-\sqrt{3}\tilde{c}^{(3)}_{45}+2\,\sqrt{3}\tilde{c}^{(3)}_{46}-\frac{\sqrt{3}}{2}\tilde{c}^{(3)}_{57}+\sqrt{3}\tilde{c}^{(3)}_{72}+\sqrt{3}\tilde{c}^{(3)}_{78},\;
\tilde{c}^{(3)}_{171}=6\,\sqrt{3}\tilde{c}^{(3)}_{31}-2\,\sqrt{3}\tilde{c}^{(3)}_{34}-\sqrt{3}\tilde{c}^{(3)}_{57},\;
\tilde{c}^{(3)}_{172}=-6\,\sqrt{3}\tilde{c}^{(3)}_{31}+2\,\sqrt{3}\tilde{c}^{(3)}_{34}+\sqrt{3}\tilde{c}^{(3)}_{57},\;
\tilde{c}^{(3)}_{173}=-6\,\sqrt{3}\tilde{c}^{(3)}_{31}+2\,\sqrt{3}\tilde{c}^{(3)}_{34}+\sqrt{3}\tilde{c}^{(3)}_{57},\;
\tilde{c}^{(3)}_{174}=-\sqrt{3}\tilde{c}^{(3)}_{35}+\sqrt{3}\tilde{c}^{(3)}_{36}+2\,\sqrt{3}\tilde{c}^{(3)}_{37}-\sqrt{3}\tilde{c}^{(3)}_{40}+\frac{\sqrt{3}}{4}\tilde{c}^{(3)}_{49}+\sqrt{3}\tilde{c}^{(3)}_{65}-\frac{2\,\sqrt{3}}{3}\tilde{c}^{(3)}_{66}-\sqrt{3}\tilde{c}^{(3)}_{68}+\frac{\sqrt{3}}{4}\tilde{c}^{(3)}_{80},\;
\tilde{c}^{(3)}_{175}=\frac{\sqrt{3}}{2}\tilde{c}^{(3)}_{35}+\frac{\sqrt{3}}{2}\tilde{c}^{(3)}_{37}-\frac{\sqrt{3}}{8}\tilde{c}^{(3)}_{49}-\frac{\sqrt{3}}{6}\tilde{c}^{(3)}_{65}+\frac{\sqrt{3}}{6}\tilde{c}^{(3)}_{66}+\frac{\sqrt{3}}{2}\tilde{c}^{(3)}_{68}-\frac{\sqrt{3}}{8}\tilde{c}^{(3)}_{80},\;
\tilde{c}^{(3)}_{176}=-\frac{\sqrt{3}}{2}\tilde{c}^{(3)}_{35}-\sqrt{3}\tilde{c}^{(3)}_{36}-\frac{\sqrt{3}}{2}\tilde{c}^{(3)}_{37}+\frac{\sqrt{3}}{8}\tilde{c}^{(3)}_{49}-\frac{\sqrt{3}}{6}\tilde{c}^{(3)}_{65}+\frac{\sqrt{3}}{6}\tilde{c}^{(3)}_{66}-\frac{\sqrt{3}}{2}\tilde{c}^{(3)}_{68}+\frac{\sqrt{3}}{8}\tilde{c}^{(3)}_{80},\;
\tilde{c}^{(3)}_{177}=-\sqrt{3}\tilde{c}^{(3)}_{37}+2\,\sqrt{3}\tilde{c}^{(3)}_{40}+\frac{\sqrt{3}}{3}\tilde{c}^{(3)}_{66},\;
\tilde{c}^{(3)}_{178}=-\frac{\sqrt{3}}{2}\tilde{c}^{(3)}_{35}+\sqrt{3}\tilde{c}^{(3)}_{36}+\frac{\sqrt{3}}{2}\tilde{c}^{(3)}_{37}+\frac{\sqrt{3}}{8}\tilde{c}^{(3)}_{49}+\frac{\sqrt{3}}{2}\tilde{c}^{(3)}_{65}-\frac{\sqrt{3}}{6}\tilde{c}^{(3)}_{66}-\frac{\sqrt{3}}{2}\tilde{c}^{(3)}_{68}+\frac{\sqrt{3}}{8}\tilde{c}^{(3)}_{80},\;
\tilde{c}^{(3)}_{179}=\frac{\sqrt{3}}{2}\tilde{c}^{(3)}_{35}-\sqrt{3}\tilde{c}^{(3)}_{36}-\frac{3\,\sqrt{3}}{2}\tilde{c}^{(3)}_{37}-\frac{\sqrt{3}}{8}\tilde{c}^{(3)}_{49}-\frac{\sqrt{3}}{2}\tilde{c}^{(3)}_{65}+\frac{\sqrt{3}}{2}\tilde{c}^{(3)}_{66}+\frac{\sqrt{3}}{2}\tilde{c}^{(3)}_{68}-\frac{\sqrt{3}}{8}\tilde{c}^{(3)}_{80},\;
\tilde{c}^{(3)}_{182}=\sqrt{3}\tilde{c}^{(3)}_{35}-\sqrt{3}\tilde{c}^{(3)}_{36}-\sqrt{3}\tilde{c}^{(3)}_{37}+\sqrt{3}\tilde{c}^{(3)}_{40}-\frac{\sqrt{3}}{4}\tilde{c}^{(3)}_{49}-\frac{\sqrt{3}}{3}\tilde{c}^{(3)}_{65}+\frac{\sqrt{3}}{3}\tilde{c}^{(3)}_{66}+\sqrt{3}\tilde{c}^{(3)}_{68}-\frac{\sqrt{3}}{4}\tilde{c}^{(3)}_{80},\;
\tilde{c}^{(3)}_{183}=-\sqrt{3}\tilde{c}^{(3)}_{35}+\sqrt{3}\tilde{c}^{(3)}_{36}+2\,\sqrt{3}\tilde{c}^{(3)}_{37}+\frac{\sqrt{3}}{4}\tilde{c}^{(3)}_{49}+\frac{2\,\sqrt{3}}{3}\tilde{c}^{(3)}_{65}-\frac{\sqrt{3}}{3}\tilde{c}^{(3)}_{66}-\sqrt{3}\tilde{c}^{(3)}_{68}+\frac{\sqrt{3}}{4}\tilde{c}^{(3)}_{80},\;
\tilde{c}^{(3)}_{185}=-\frac{\sqrt{3}}{2}\tilde{c}^{(3)}_{44}+\frac{3\,\sqrt{3}}{2}\tilde{c}^{(3)}_{45}-2\,\sqrt{3}\tilde{c}^{(3)}_{46}-\sqrt{3}\tilde{c}^{(3)}_{72}-\sqrt{3}\tilde{c}^{(3)}_{78},\;
\tilde{c}^{(3)}_{186}=3\,\sqrt{3}\tilde{c}^{(3)}_{45}-4\,\sqrt{3}\tilde{c}^{(3)}_{46}-2\,\sqrt{3}\tilde{c}^{(3)}_{72}-2\,\sqrt{3}\tilde{c}^{(3)}_{78},\;
\tilde{c}^{(3)}_{187}=-\frac{3\,\sqrt{3}}{2}\tilde{c}^{(3)}_{45}+2\,\sqrt{3}\tilde{c}^{(3)}_{46}-2\,\sqrt{3}\tilde{c}^{(3)}_{48}+\sqrt{3}\tilde{c}^{(3)}_{72},\;
\tilde{c}^{(3)}_{188}=-2\,\sqrt{3}\tilde{c}^{(3)}_{47}+\sqrt{3}\tilde{c}^{(3)}_{76},\;
\tilde{c}^{(3)}_{189}=2\,\sqrt{3}\tilde{c}^{(3)}_{47}-\sqrt{3}\tilde{c}^{(3)}_{76},\;
\tilde{c}^{(3)}_{190}=-2\,\sqrt{3}\tilde{c}^{(3)}_{47}+\sqrt{3}\tilde{c}^{(3)}_{76},\;
\tilde{c}^{(3)}_{191}=\frac{\sqrt{3}}{2}\tilde{c}^{(3)}_{44}+\frac{3\,\sqrt{3}}{2}\tilde{c}^{(3)}_{45}-2\,\sqrt{3}\tilde{c}^{(3)}_{46}-\sqrt{3}\tilde{c}^{(3)}_{72}-\sqrt{3}\tilde{c}^{(3)}_{78},\;
\tilde{c}^{(3)}_{192}=3\,\sqrt{3}\tilde{c}^{(3)}_{45}-4\,\sqrt{3}\tilde{c}^{(3)}_{46}+4\,\sqrt{3}\tilde{c}^{(3)}_{48}-2\,\sqrt{3}\tilde{c}^{(3)}_{72},\;
\tilde{c}^{(3)}_{193}=-2\,\sqrt{3}\tilde{c}^{(3)}_{48}-\sqrt{3}\tilde{c}^{(3)}_{78},\;
\tilde{c}^{(3)}_{194}=2\,\sqrt{3}\tilde{c}^{(3)}_{48}+\sqrt{3}\tilde{c}^{(3)}_{78},\;
\tilde{c}^{(3)}_{195}=2\,\sqrt{3}\tilde{c}^{(3)}_{48}+\sqrt{3}\tilde{c}^{(3)}_{78}.
\end{autobreak}
\end{align}

\begin{align}\label{relation-p3-3}
\begin{autobreak}
~
\tilde{C}^{(3)}_{43}=-\frac{1}{3}D^{(3)}_{43}+\frac{2}{3}D^{(3)}_{46}+\frac{1}{3}D^{(3)}_{47}+\frac{2}{3}D^{(3)}_{55},\;
\tilde{C}^{(3)}_{44}=\frac{2}{3}D^{(3)}_{43}-\frac{2}{3}D^{(3)}_{47}+\frac{2}{3}D^{(3)}_{55},\;
\tilde{C}^{(3)}_{45}=-\frac{1}{3}D^{(3)}_{43}-\frac{2}{3}D^{(3)}_{44}+\frac{1}{3}D^{(3)}_{47}-\frac{4}{3}D^{(3)}_{55}-\frac{4}{3}D^{(3)}_{56},\;
\tilde{C}^{(3)}_{46}=\frac{2}{3}D^{(3)}_{43}+\frac{4}{3}D^{(3)}_{44}-\frac{2}{3}D^{(3)}_{47}-\frac{4}{3}D^{(3)}_{55}-\frac{4}{3}D^{(3)}_{56},\;
\tilde{C}^{(3)}_{47}=-\frac{2}{3}D^{(3)}_{50}-\frac{2}{3}D^{(3)}_{52},\;
\tilde{C}^{(3)}_{48}=\frac{2}{3}D^{(3)}_{49}+\frac{1}{3}D^{(3)}_{50}+\frac{1}{3}D^{(3)}_{52},\;
\tilde{C}^{(3)}_{49}=\frac{1}{3}D^{(3)}_{50}+\frac{2}{3}D^{(3)}_{51}+\frac{1}{3}D^{(3)}_{52},\;
\tilde{C}^{(3)}_{50}=-\frac{2}{3}D^{(3)}_{50}-\frac{4}{3}D^{(3)}_{51}-\frac{2}{3}D^{(3)}_{52},\;
\tilde{C}^{(3)}_{51}=\frac{2}{3}D^{(3)}_{53}-\frac{2}{3}D^{(3)}_{55},\;
\tilde{C}^{(3)}_{52}=\frac{4}{3}D^{(3)}_{55}+\frac{4}{3}D^{(3)}_{56},\;
\tilde{C}^{(3)}_{54}=2D^{(3)}_{55}+2D^{(3)}_{56},\;
\tilde{C}^{(3)}_{55}=\frac{2}{3}D^{(3)}_{43}-\frac{2}{3}D^{(3)}_{45}+\frac{2}{3}D^{(3)}_{47},\;
\tilde{C}^{(3)}_{56}=-\frac{2}{3}D^{(3)}_{48}-\frac{2}{3}D^{(3)}_{50}+\frac{2}{3}D^{(3)}_{52},\;
\tilde{C}^{(3)}_{58}=-\frac{1}{3}D^{(3)}_{57}+\frac{1}{3}D^{(3)}_{59}+D^{(3)}_{60},\;
\tilde{C}^{(3)}_{59}=\frac{1}{3}D^{(3)}_{57}+\frac{1}{3}D^{(3)}_{59},\;
\tilde{C}^{(3)}_{61}=\frac{1}{3}D^{(3)}_{61}+\frac{1}{3}D^{(3)}_{63}+D^{(3)}_{64},\;
\tilde{C}^{(3)}_{62}=\frac{1}{3}D^{(3)}_{61}+\frac{1}{3}D^{(3)}_{62}+\frac{1}{3}D^{(3)}_{63}-D^{(3)}_{65},\;
\tilde{C}^{(3)}_{63}=-\frac{1}{3}D^{(3)}_{61}+\frac{1}{3}D^{(3)}_{63},\;
\tilde{C}^{(3)}_{70}=-\frac{2}{3}D^{(3)}_{72}-\frac{1}{3}D^{(3)}_{74}+\frac{2}{3}D^{(3)}_{75},\;
\tilde{C}^{(3)}_{71}=\frac{2}{3}D^{(3)}_{72}+\frac{4}{3}D^{(3)}_{73}+\frac{1}{3}D^{(3)}_{74},\;
\tilde{C}^{(3)}_{72}=-\frac{1}{3}D^{(3)}_{72}+\frac{1}{3}D^{(3)}_{74},\;
\tilde{C}^{(3)}_{73}=\frac{2}{3}D^{(3)}_{76}+\frac{2}{3}D^{(3)}_{77},\;
\tilde{C}^{(3)}_{74}=\frac{2}{3}D^{(3)}_{78}-\frac{2}{3}D^{(3)}_{79},\;
\tilde{C}^{(3)}_{76}=\frac{2}{3}D^{(3)}_{80}-D^{(3)}_{81},\;
\tilde{C}^{(3)}_{82}=\frac{1}{3}D^{(3)}_{62}-D^{(3)}_{86},\;
\tilde{C}^{(3)}_{85}=-2D^{(3)}_{43}-2D^{(3)}_{44}+2D^{(3)}_{55}+2D^{(3)}_{56},\;
\tilde{C}^{(3)}_{86}=D^{(3)}_{43}+D^{(3)}_{44}-D^{(3)}_{45}+2D^{(3)}_{55}+2D^{(3)}_{56},\;
\tilde{C}^{(3)}_{87}=D^{(3)}_{44}+D^{(3)}_{45}-D^{(3)}_{47}+2D^{(3)}_{55}+2D^{(3)}_{56},\;
\tilde{C}^{(3)}_{88}=-D^{(3)}_{44}-D^{(3)}_{46}+D^{(3)}_{56},\;
\tilde{C}^{(3)}_{89}=-2D^{(3)}_{44}+2D^{(3)}_{47}+2D^{(3)}_{55}+2D^{(3)}_{56},\;
\tilde{C}^{(3)}_{90}=2D^{(3)}_{44}+D^{(3)}_{56},\;
\tilde{C}^{(3)}_{92}=-D^{(3)}_{48}-D^{(3)}_{50}-D^{(3)}_{51},\;
\tilde{C}^{(3)}_{93}=2D^{(3)}_{50}+2D^{(3)}_{51},\;
\tilde{C}^{(3)}_{94}=D^{(3)}_{48}-D^{(3)}_{51}-D^{(3)}_{52},\;
\tilde{C}^{(3)}_{95}=2D^{(3)}_{51}+2D^{(3)}_{52},\;
\tilde{C}^{(3)}_{97}=-D^{(3)}_{49}+D^{(3)}_{51},\;
\tilde{C}^{(3)}_{99}=-D^{(3)}_{53}-D^{(3)}_{56},\;
\tilde{C}^{(3)}_{100}=-2D^{(3)}_{54}-2D^{(3)}_{55}-2D^{(3)}_{56},\;
\tilde{C}^{(3)}_{101}=2D^{(3)}_{54}-2D^{(3)}_{55}-2D^{(3)}_{56},\;
\tilde{C}^{(3)}_{104}=-3D^{(3)}_{55}-3D^{(3)}_{56},\;
\tilde{C}^{(3)}_{105}=-D^{(3)}_{57}-D^{(3)}_{62}-D^{(3)}_{63},\;
\tilde{C}^{(3)}_{106}=-D^{(3)}_{58}-D^{(3)}_{62}-2D^{(3)}_{63},\;
\tilde{C}^{(3)}_{107}=-D^{(3)}_{59}-D^{(3)}_{63},\;
\tilde{C}^{(3)}_{109}=-D^{(3)}_{61}-D^{(3)}_{62}-D^{(3)}_{63},\;
\tilde{C}^{(3)}_{112}=-D^{(3)}_{68}+D^{(3)}_{70},\;
\tilde{C}^{(3)}_{113}=-D^{(3)}_{69}+2D^{(3)}_{70},\;
\tilde{C}^{(3)}_{114}=2D^{(3)}_{72}+2D^{(3)}_{73},\;
\tilde{C}^{(3)}_{116}=-2D^{(3)}_{73}-D^{(3)}_{74},\;
\tilde{C}^{(3)}_{117}=-2D^{(3)}_{73}-D^{(3)}_{75},\;
\tilde{C}^{(3)}_{118}=2D^{(3)}_{76}+D^{(3)}_{77},\;
\tilde{C}^{(3)}_{122}=2D^{(3)}_{78}-D^{(3)}_{79},\;
\tilde{C}^{(3)}_{136}=-\frac{\sqrt{3}}{3}D^{(3)}_{6}-\frac{2\,\sqrt{3}}{3}D^{(3)}_{16},\;
\tilde{C}^{(3)}_{137}=-\frac{\sqrt{3}}{3}D^{(3)}_{5}-\frac{2\,\sqrt{3}}{3}D^{(3)}_{16},\;
\tilde{C}^{(3)}_{138}=-\frac{\sqrt{3}}{3}D^{(3)}_{4}-\frac{2\,\sqrt{3}}{3}D^{(3)}_{16},\;
\tilde{C}^{(3)}_{146}=-\frac{\sqrt{3}}{3}D^{(3)}_{14}+\frac{2\,\sqrt{3}}{3}D^{(3)}_{16},\;
\tilde{C}^{(3)}_{153}=-\frac{2\,\sqrt{3}}{3}D^{(3)}_{17}-\frac{\sqrt{3}}{3}D^{(3)}_{18},\;
\tilde{C}^{(3)}_{159}=-\frac{2\,\sqrt{3}}{3}D^{(3)}_{20}-\frac{\sqrt{3}}{3}D^{(3)}_{21},\;
\tilde{C}^{(3)}_{162}=-\frac{\sqrt{3}}{3}D^{(3)}_{22}-\frac{\sqrt{3}}{3}D^{(3)}_{23},\;
\tilde{C}^{(3)}_{165}=\frac{2\,\sqrt{3}}{3}D^{(3)}_{24}-\frac{\sqrt{3}}{3}D^{(3)}_{25},\;
\tilde{C}^{(3)}_{201}=-D^{(3)}_{4}-\frac{1}{2}D^{(3)}_{33},\;
\tilde{C}^{(3)}_{202}=-D^{(3)}_{5}+D^{(3)}_{33},\;
\tilde{C}^{(3)}_{203}=-D^{(3)}_{6}-\frac{1}{2}D^{(3)}_{33},\;
\tilde{C}^{(3)}_{219}=-2D^{(3)}_{17}-D^{(3)}_{18},\;
\tilde{C}^{(3)}_{220}=2D^{(3)}_{17}+D^{(3)}_{22},\;
\tilde{C}^{(3)}_{221}=-2D^{(3)}_{17}-D^{(3)}_{18}-D^{(3)}_{22},\;
\tilde{C}^{(3)}_{224}=-D^{(3)}_{19}-D^{(3)}_{22},\;
\tilde{C}^{(3)}_{226}=-2D^{(3)}_{20}-D^{(3)}_{21},\;
\tilde{C}^{(3)}_{228}=2D^{(3)}_{22}+2D^{(3)}_{23},\;
\tilde{C}^{(3)}_{229}=2D^{(3)}_{24}-D^{(3)}_{25}-D^{(3)}_{26},\;
\tilde{C}^{(3)}_{231}=2D^{(3)}_{24}-D^{(3)}_{25}+D^{(3)}_{26},\;
\tilde{C}^{(3)}_{235}=-D^{(3)}_{29}-D^{(3)}_{33},\;
\tilde{C}^{(3)}_{253}=\frac{\sqrt{3}}{3}D^{(3)}_{43}+\frac{\sqrt{3}}{3}D^{(3)}_{45}-\frac{\sqrt{3}}{3}D^{(3)}_{46}-\frac{\sqrt{3}}{3}D^{(3)}_{55},\;
\tilde{C}^{(3)}_{254}=-\frac{\sqrt{3}}{3}D^{(3)}_{43}+\frac{\sqrt{3}}{3}D^{(3)}_{47}-\frac{\sqrt{3}}{3}D^{(3)}_{55},\;
\tilde{C}^{(3)}_{255}=-\frac{\sqrt{3}}{3}D^{(3)}_{45}-\frac{\sqrt{3}}{3}D^{(3)}_{46}-\frac{\sqrt{3}}{3}D^{(3)}_{47}-\frac{\sqrt{3}}{3}D^{(3)}_{55},\;
\tilde{C}^{(3)}_{256}=\frac{\sqrt{3}}{3}D^{(3)}_{43}+\frac{\sqrt{3}}{3}D^{(3)}_{44}+\frac{\sqrt{3}}{3}D^{(3)}_{45}-\frac{\sqrt{3}}{3}D^{(3)}_{55}-\frac{\sqrt{3}}{3}D^{(3)}_{56}-\frac{\sqrt{3}}{3}D^{(3)}_{72}-\frac{\sqrt{3}}{3}D^{(3)}_{73}-\frac{\sqrt{3}}{6}D^{(3)}_{74},\;
\tilde{C}^{(3)}_{257}=-\frac{\sqrt{3}}{3}D^{(3)}_{43}-\frac{2\,\sqrt{3}}{3}D^{(3)}_{44}+\frac{\sqrt{3}}{3}D^{(3)}_{47}-\frac{\sqrt{3}}{3}D^{(3)}_{55}-\frac{\sqrt{3}}{3}D^{(3)}_{56}+\frac{2\,\sqrt{3}}{3}D^{(3)}_{72}+\frac{2\,\sqrt{3}}{3}D^{(3)}_{73}+\frac{\sqrt{3}}{3}D^{(3)}_{74},\;
\tilde{C}^{(3)}_{258}=\frac{\sqrt{3}}{3}D^{(3)}_{44}-\frac{\sqrt{3}}{3}D^{(3)}_{45}-\frac{\sqrt{3}}{3}D^{(3)}_{47}-\frac{\sqrt{3}}{3}D^{(3)}_{55}-\frac{\sqrt{3}}{3}D^{(3)}_{56}-\frac{\sqrt{3}}{3}D^{(3)}_{72}-\frac{\sqrt{3}}{3}D^{(3)}_{73}-\frac{\sqrt{3}}{6}D^{(3)}_{74},\;
\tilde{C}^{(3)}_{260}=\frac{\sqrt{3}}{3}D^{(3)}_{43}+\frac{2\,\sqrt{3}}{3}D^{(3)}_{45}+\frac{\sqrt{3}}{3}D^{(3)}_{47},\;
\tilde{C}^{(3)}_{261}=\frac{\sqrt{3}}{3}D^{(3)}_{50}+\frac{\sqrt{3}}{3}D^{(3)}_{52},\;
\tilde{C}^{(3)}_{262}=-\frac{\sqrt{3}}{3}D^{(3)}_{48}-\frac{\sqrt{3}}{3}D^{(3)}_{49}-\frac{\sqrt{3}}{3}D^{(3)}_{52},\;
\tilde{C}^{(3)}_{263}=\frac{\sqrt{3}}{3}D^{(3)}_{48}-\frac{\sqrt{3}}{3}D^{(3)}_{49}-\frac{\sqrt{3}}{3}D^{(3)}_{50},\;
\tilde{C}^{(3)}_{264}=\frac{\sqrt{3}}{3}D^{(3)}_{50}+\frac{2\,\sqrt{3}}{3}D^{(3)}_{51}+\frac{\sqrt{3}}{3}D^{(3)}_{52},\;
\tilde{C}^{(3)}_{265}=\frac{\sqrt{3}}{3}D^{(3)}_{48}-\frac{\sqrt{3}}{3}D^{(3)}_{50}-\frac{\sqrt{3}}{3}D^{(3)}_{51},\;
\tilde{C}^{(3)}_{266}=-\frac{\sqrt{3}}{3}D^{(3)}_{48}-\frac{\sqrt{3}}{3}D^{(3)}_{51}-\frac{\sqrt{3}}{3}D^{(3)}_{52},\;
\tilde{C}^{(3)}_{267}=\frac{2\,\sqrt{3}}{3}D^{(3)}_{48}-\frac{\sqrt{3}}{3}D^{(3)}_{50}+\frac{\sqrt{3}}{3}D^{(3)}_{52},\;
\tilde{C}^{(3)}_{269}=-\frac{\sqrt{3}}{3}D^{(3)}_{53}+\frac{\sqrt{3}}{3}D^{(3)}_{55},\;
\tilde{C}^{(3)}_{270}=\frac{\sqrt{3}}{3}D^{(3)}_{55}+\frac{\sqrt{3}}{3}D^{(3)}_{56},\;
\tilde{C}^{(3)}_{272}=\frac{\sqrt{3}}{2}D^{(3)}_{55}+\frac{\sqrt{3}}{2}D^{(3)}_{56},\;
\tilde{C}^{(3)}_{274}=-\frac{\sqrt{3}}{3}D^{(3)}_{57}+\frac{\sqrt{3}}{3}D^{(3)}_{61}-\frac{\sqrt{3}}{3}D^{(3)}_{62}+\frac{\sqrt{3}}{6}D^{(3)}_{80},\;
\tilde{C}^{(3)}_{275}=-\frac{\sqrt{3}}{3}D^{(3)}_{58}+\frac{\sqrt{3}}{6}D^{(3)}_{59}-\frac{\sqrt{3}}{6}D^{(3)}_{63}-\frac{\sqrt{3}}{12}D^{(3)}_{80},\;
\tilde{C}^{(3)}_{276}=-\frac{\sqrt{3}}{2}D^{(3)}_{59}-\frac{\sqrt{3}}{6}D^{(3)}_{63}+\frac{\sqrt{3}}{12}D^{(3)}_{80},\;
\tilde{C}^{(3)}_{277}=-\frac{2\,\sqrt{3}}{3}D^{(3)}_{61}-\frac{\sqrt{3}}{3}D^{(3)}_{62},\;
\tilde{C}^{(3)}_{278}=\frac{\sqrt{3}}{6}D^{(3)}_{59}-\frac{\sqrt{3}}{3}D^{(3)}_{62}-\frac{\sqrt{3}}{6}D^{(3)}_{63}+\frac{\sqrt{3}}{12}D^{(3)}_{80},\;
\tilde{C}^{(3)}_{279}=-\frac{\sqrt{3}}{6}D^{(3)}_{59}-\frac{\sqrt{3}}{2}D^{(3)}_{63}-\frac{\sqrt{3}}{12}D^{(3)}_{80},\;
\tilde{C}^{(3)}_{282}=-\frac{\sqrt{3}}{3}D^{(3)}_{57}-\frac{\sqrt{3}}{3}D^{(3)}_{61}-\frac{\sqrt{3}}{6}D^{(3)}_{80},\;
\tilde{C}^{(3)}_{283}=-\frac{\sqrt{3}}{3}D^{(3)}_{58}-\frac{\sqrt{3}}{3}D^{(3)}_{62}+\frac{\sqrt{3}}{6}D^{(3)}_{80},\;
\tilde{C}^{(3)}_{284}=-\frac{\sqrt{3}}{3}D^{(3)}_{68}+\frac{\sqrt{3}}{3}D^{(3)}_{70},\;
\tilde{C}^{(3)}_{286}=-\frac{\sqrt{3}}{3}D^{(3)}_{68}-\frac{\sqrt{3}}{3}D^{(3)}_{70},\;
\tilde{C}^{(3)}_{290}=-\frac{2\,\sqrt{3}}{3}D^{(3)}_{72}-\frac{\sqrt{3}}{3}D^{(3)}_{74},\;
\tilde{C}^{(3)}_{291}=\frac{\sqrt{3}}{3}D^{(3)}_{72}+\frac{\sqrt{3}}{6}D^{(3)}_{74},\;
\tilde{C}^{(3)}_{292}=\frac{2\,\sqrt{3}}{3}D^{(3)}_{76}-\frac{\sqrt{3}}{3}D^{(3)}_{77},\;
\tilde{C}^{(3)}_{293}=-\frac{2\,\sqrt{3}}{3}D^{(3)}_{76}+\frac{\sqrt{3}}{3}D^{(3)}_{77},\;
\tilde{C}^{(3)}_{294}=\frac{2\,\sqrt{3}}{3}D^{(3)}_{76}-\frac{\sqrt{3}}{3}D^{(3)}_{77},\;
\tilde{C}^{(3)}_{295}=-\frac{2\,\sqrt{3}}{3}D^{(3)}_{72}-\frac{\sqrt{3}}{3}D^{(3)}_{74}+\frac{\sqrt{3}}{3}D^{(3)}_{75},\;
\tilde{C}^{(3)}_{296}=-\frac{2\,\sqrt{3}}{3}D^{(3)}_{72}-\frac{\sqrt{3}}{3}D^{(3)}_{74},\;
\tilde{C}^{(3)}_{297}=\frac{2\,\sqrt{3}}{3}D^{(3)}_{78}+\frac{\sqrt{3}}{3}D^{(3)}_{79},\;
\tilde{C}^{(3)}_{298}=-\frac{2\,\sqrt{3}}{3}D^{(3)}_{78}-\frac{\sqrt{3}}{3}D^{(3)}_{79},\;
\tilde{C}^{(3)}_{302}=\frac{2\,\sqrt{3}}{3}D^{(3)}_{78}+\frac{\sqrt{3}}{3}D^{(3)}_{79}.
\end{autobreak}
\end{align}

\begin{align}\label{relation-p3-4}
\begin{autobreak}
~
\tilde{C}^{(3)}_{52}=-\frac{2}{3}\tilde{C}^{(3)}_{45}-\frac{1}{3}\tilde{C}^{(3)}_{46},\;
\tilde{C}^{(3)}_{54}=-\tilde{C}^{(3)}_{45}-\frac{1}{2}\tilde{C}^{(3)}_{46},\;
\tilde{C}^{(3)}_{86}=-\frac{3}{2}\tilde{C}^{(3)}_{45}+\frac{3}{4}\tilde{C}^{(3)}_{46}+\frac{3}{2}\tilde{C}^{(3)}_{55}+\frac{1}{2}\tilde{C}^{(3)}_{85},\;
\tilde{C}^{(3)}_{87}=-\frac{3}{2}\tilde{C}^{(3)}_{45}-\frac{3}{4}\tilde{C}^{(3)}_{46}-\frac{3}{2}\tilde{C}^{(3)}_{55}-\frac{1}{2}\tilde{C}^{(3)}_{85},\;
\tilde{C}^{(3)}_{88}=-\frac{3}{2}\tilde{C}^{(3)}_{43}-\frac{3}{4}\tilde{C}^{(3)}_{46},\;
\tilde{C}^{(3)}_{89}=-3\tilde{C}^{(3)}_{46}-\tilde{C}^{(3)}_{85},\;
\tilde{C}^{(3)}_{90}=-\frac{3}{2}\tilde{C}^{(3)}_{44}-\frac{3}{2}\tilde{C}^{(3)}_{45}+\frac{3}{4}\tilde{C}^{(3)}_{46},\;
\tilde{C}^{(3)}_{93}=6\tilde{C}^{(3)}_{49}-3\tilde{C}^{(3)}_{56}+2\tilde{C}^{(3)}_{92},\;
\tilde{C}^{(3)}_{94}=-3\tilde{C}^{(3)}_{49}-\tilde{C}^{(3)}_{92},\;
\tilde{C}^{(3)}_{95}=3\tilde{C}^{(3)}_{56}-2\tilde{C}^{(3)}_{92},\;
\tilde{C}^{(3)}_{96}=-\frac{3}{2}\tilde{C}^{(3)}_{47}-3\tilde{C}^{(3)}_{49},\;
\tilde{C}^{(3)}_{97}=-\frac{3}{2}\tilde{C}^{(3)}_{48}+\frac{3}{2}\tilde{C}^{(3)}_{49},\;
\tilde{C}^{(3)}_{99}=\frac{1}{2}\tilde{C}^{(3)}_{45}+\frac{1}{4}\tilde{C}^{(3)}_{46}-\frac{3}{2}\tilde{C}^{(3)}_{51},\;
\tilde{C}^{(3)}_{100}=\tilde{C}^{(3)}_{45}+\frac{1}{2}\tilde{C}^{(3)}_{46}+3\tilde{C}^{(3)}_{57},\;
\tilde{C}^{(3)}_{101}=\tilde{C}^{(3)}_{45}+\frac{1}{2}\tilde{C}^{(3)}_{46}-3\tilde{C}^{(3)}_{57},\;
\tilde{C}^{(3)}_{103}=\frac{3}{4}\tilde{C}^{(3)}_{45}+\frac{3}{8}\tilde{C}^{(3)}_{46}-\frac{3}{2}\tilde{C}^{(3)}_{53},\;
\tilde{C}^{(3)}_{104}=\frac{3}{2}\tilde{C}^{(3)}_{45}+\frac{3}{4}\tilde{C}^{(3)}_{46},\;
\tilde{C}^{(3)}_{107}=-3\tilde{C}^{(3)}_{59}-3\tilde{C}^{(3)}_{60}-\tilde{C}^{(3)}_{105}+\tilde{C}^{(3)}_{106},\;
\tilde{C}^{(3)}_{109}=-3\tilde{C}^{(3)}_{60}+3\tilde{C}^{(3)}_{63}+\tilde{C}^{(3)}_{106},\;
\tilde{C}^{(3)}_{110}=-\tilde{C}^{(3)}_{58}+\tilde{C}^{(3)}_{59}+2\tilde{C}^{(3)}_{60}-\tilde{C}^{(3)}_{61}-\tilde{C}^{(3)}_{63}+\frac{2}{3}\tilde{C}^{(3)}_{105}-\frac{2}{3}\tilde{C}^{(3)}_{106}-\tilde{C}^{(3)}_{108},\;
\tilde{C}^{(3)}_{111}=\tilde{C}^{(3)}_{60}-\tilde{C}^{(3)}_{62}-\tilde{C}^{(3)}_{63}-\frac{1}{3}\tilde{C}^{(3)}_{106},\;
\tilde{C}^{(3)}_{112}=-\frac{3}{2}\tilde{C}^{(3)}_{66}-\frac{3}{2}\tilde{C}^{(3)}_{68},\;
\tilde{C}^{(3)}_{113}=3\tilde{C}^{(3)}_{67}-3\tilde{C}^{(3)}_{68},\;
\tilde{C}^{(3)}_{115}=-3\tilde{C}^{(3)}_{71}+6\tilde{C}^{(3)}_{72}+2\tilde{C}^{(3)}_{114},\;
\tilde{C}^{(3)}_{116}=-3\tilde{C}^{(3)}_{71}+\tilde{C}^{(3)}_{114},\;
\tilde{C}^{(3)}_{117}=-\frac{3}{2}\tilde{C}^{(3)}_{70}-\frac{3}{2}\tilde{C}^{(3)}_{71},\;
\tilde{C}^{(3)}_{119}=-6\tilde{C}^{(3)}_{73}+2\tilde{C}^{(3)}_{118},\;
\tilde{C}^{(3)}_{123}=-6\tilde{C}^{(3)}_{74}+2\tilde{C}^{(3)}_{122},\;
\tilde{C}^{(3)}_{130}=\tilde{C}^{(3)}_{58}-\tilde{C}^{(3)}_{59}-\tilde{C}^{(3)}_{60}-\tilde{C}^{(3)}_{82}-\frac{2}{3}\tilde{C}^{(3)}_{105}+\frac{1}{3}\tilde{C}^{(3)}_{106}+\tilde{C}^{(3)}_{108},\;
\tilde{C}^{(3)}_{201}=\sqrt{3}\tilde{C}^{(3)}_{138}+\sqrt{3}\tilde{C}^{(3)}_{148}-\frac{\sqrt{3}}{2}\tilde{C}^{(3)}_{174},\;
\tilde{C}^{(3)}_{202}=\sqrt{3}\tilde{C}^{(3)}_{137}+\sqrt{3}\tilde{C}^{(3)}_{148}+\sqrt{3}\tilde{C}^{(3)}_{174},\;
\tilde{C}^{(3)}_{203}=\sqrt{3}\tilde{C}^{(3)}_{136}+\sqrt{3}\tilde{C}^{(3)}_{148}-\frac{\sqrt{3}}{2}\tilde{C}^{(3)}_{174},\;
\tilde{C}^{(3)}_{215}=\sqrt{3}\tilde{C}^{(3)}_{146}-\sqrt{3}\tilde{C}^{(3)}_{148},\;
\tilde{C}^{(3)}_{220}=-2\,\sqrt{3}\tilde{C}^{(3)}_{154}+\frac{\sqrt{3}}{2}\tilde{C}^{(3)}_{163},\;
\tilde{C}^{(3)}_{221}=\sqrt{3}\tilde{C}^{(3)}_{153}-\frac{\sqrt{3}}{2}\tilde{C}^{(3)}_{163},\;
\tilde{C}^{(3)}_{224}=\sqrt{3}\tilde{C}^{(3)}_{157}-\frac{\sqrt{3}}{2}\tilde{C}^{(3)}_{163},\;
\tilde{C}^{(3)}_{229}=\sqrt{3}\tilde{C}^{(3)}_{165}+\sqrt{3}\tilde{C}^{(3)}_{167},\;
\tilde{C}^{(3)}_{231}=\sqrt{3}\tilde{C}^{(3)}_{165}-\sqrt{3}\tilde{C}^{(3)}_{167},\;
\tilde{C}^{(3)}_{235}=-\sqrt{3}\tilde{C}^{(3)}_{170}-\sqrt{3}\tilde{C}^{(3)}_{174},\;
\tilde{C}^{(3)}_{253}=-\frac{\sqrt{3}}{2}\tilde{C}^{(3)}_{43}-\frac{3\,\sqrt{3}}{4}\tilde{C}^{(3)}_{46}-\frac{\sqrt{3}}{2}\tilde{C}^{(3)}_{55}-\frac{\sqrt{3}}{2}\tilde{C}^{(3)}_{85},\;
\tilde{C}^{(3)}_{255}=-\frac{\sqrt{3}}{2}\tilde{C}^{(3)}_{43}+\frac{3\,\sqrt{3}}{4}\tilde{C}^{(3)}_{46}+\frac{\sqrt{3}}{2}\tilde{C}^{(3)}_{55}+\frac{\sqrt{3}}{2}\tilde{C}^{(3)}_{85},\;
\tilde{C}^{(3)}_{256}=-\frac{\sqrt{3}}{2}\tilde{C}^{(3)}_{46}-\frac{\sqrt{3}}{2}\tilde{C}^{(3)}_{55}+\frac{\sqrt{3}}{2}\tilde{C}^{(3)}_{71}-\sqrt{3}\tilde{C}^{(3)}_{72}-\frac{\sqrt{3}}{2}\tilde{C}^{(3)}_{85}-\frac{\sqrt{3}}{2}\tilde{C}^{(3)}_{114},\;
\tilde{C}^{(3)}_{257}=\frac{\sqrt{3}}{2}\tilde{C}^{(3)}_{45}-\frac{\sqrt{3}}{4}\tilde{C}^{(3)}_{46}-\sqrt{3}\tilde{C}^{(3)}_{71}+2\,\sqrt{3}\tilde{C}^{(3)}_{72}+\sqrt{3}\tilde{C}^{(3)}_{114},\;
\tilde{C}^{(3)}_{258}=\sqrt{3}\tilde{C}^{(3)}_{46}+\frac{\sqrt{3}}{2}\tilde{C}^{(3)}_{55}+\frac{\sqrt{3}}{2}\tilde{C}^{(3)}_{71}-\sqrt{3}\tilde{C}^{(3)}_{72}+\frac{\sqrt{3}}{2}\tilde{C}^{(3)}_{85}-\frac{\sqrt{3}}{2}\tilde{C}^{(3)}_{114},\;
\tilde{C}^{(3)}_{260}=-\frac{3\,\sqrt{3}}{2}\tilde{C}^{(3)}_{46}-\sqrt{3}\tilde{C}^{(3)}_{55}-\sqrt{3}\tilde{C}^{(3)}_{85},\;
\tilde{C}^{(3)}_{262}=-\frac{\sqrt{3}}{2}\tilde{C}^{(3)}_{48}+\frac{3\,\sqrt{3}}{2}\tilde{C}^{(3)}_{49}-\sqrt{3}\tilde{C}^{(3)}_{56}+\sqrt{3}\tilde{C}^{(3)}_{92},\;
\tilde{C}^{(3)}_{263}=-\frac{\sqrt{3}}{2}\tilde{C}^{(3)}_{48}-\frac{3\,\sqrt{3}}{2}\tilde{C}^{(3)}_{49}+\sqrt{3}\tilde{C}^{(3)}_{56}-\sqrt{3}\tilde{C}^{(3)}_{92},\;
\tilde{C}^{(3)}_{265}=-2\,\sqrt{3}\tilde{C}^{(3)}_{49}+\sqrt{3}\tilde{C}^{(3)}_{56}-\sqrt{3}\tilde{C}^{(3)}_{92},\;
\tilde{C}^{(3)}_{266}=\sqrt{3}\tilde{C}^{(3)}_{49}-\sqrt{3}\tilde{C}^{(3)}_{56}+\sqrt{3}\tilde{C}^{(3)}_{92},\;
\tilde{C}^{(3)}_{267}=-3\,\sqrt{3}\tilde{C}^{(3)}_{49}+2\,\sqrt{3}\tilde{C}^{(3)}_{56}-2\,\sqrt{3}\tilde{C}^{(3)}_{92},\;
\tilde{C}^{(3)}_{270}=-\frac{\sqrt{3}}{6}\tilde{C}^{(3)}_{45}-\frac{\sqrt{3}}{12}\tilde{C}^{(3)}_{46},\;
\tilde{C}^{(3)}_{272}=-\frac{\sqrt{3}}{4}\tilde{C}^{(3)}_{45}-\frac{\sqrt{3}}{8}\tilde{C}^{(3)}_{46},\;
\tilde{C}^{(3)}_{274}=-\sqrt{3}\tilde{C}^{(3)}_{58}+\sqrt{3}\tilde{C}^{(3)}_{59}+2\,\sqrt{3}\tilde{C}^{(3)}_{60}-\sqrt{3}\tilde{C}^{(3)}_{63}+\frac{\sqrt{3}}{4}\tilde{C}^{(3)}_{76}+\sqrt{3}\tilde{C}^{(3)}_{105}-\frac{2\,\sqrt{3}}{3}\tilde{C}^{(3)}_{106}-\sqrt{3}\tilde{C}^{(3)}_{108}+\frac{\sqrt{3}}{4}\tilde{C}^{(3)}_{124},\;
\tilde{C}^{(3)}_{275}=\frac{\sqrt{3}}{2}\tilde{C}^{(3)}_{58}+\frac{\sqrt{3}}{2}\tilde{C}^{(3)}_{60}-\frac{\sqrt{3}}{8}\tilde{C}^{(3)}_{76}-\frac{\sqrt{3}}{6}\tilde{C}^{(3)}_{105}+\frac{\sqrt{3}}{6}\tilde{C}^{(3)}_{106}+\frac{\sqrt{3}}{2}\tilde{C}^{(3)}_{108}-\frac{\sqrt{3}}{8}\tilde{C}^{(3)}_{124},\;
\tilde{C}^{(3)}_{276}=-\frac{\sqrt{3}}{2}\tilde{C}^{(3)}_{58}-\sqrt{3}\tilde{C}^{(3)}_{59}-\frac{\sqrt{3}}{2}\tilde{C}^{(3)}_{60}+\frac{\sqrt{3}}{8}\tilde{C}^{(3)}_{76}-\frac{\sqrt{3}}{6}\tilde{C}^{(3)}_{105}+\frac{\sqrt{3}}{6}\tilde{C}^{(3)}_{106}-\frac{\sqrt{3}}{2}\tilde{C}^{(3)}_{108}+\frac{\sqrt{3}}{8}\tilde{C}^{(3)}_{124},\;
\tilde{C}^{(3)}_{277}=-\sqrt{3}\tilde{C}^{(3)}_{60}+2\,\sqrt{3}\tilde{C}^{(3)}_{63}+\frac{\sqrt{3}}{3}\tilde{C}^{(3)}_{106},\;
\tilde{C}^{(3)}_{278}=-\frac{\sqrt{3}}{2}\tilde{C}^{(3)}_{58}+\sqrt{3}\tilde{C}^{(3)}_{59}+\frac{\sqrt{3}}{2}\tilde{C}^{(3)}_{60}+\frac{\sqrt{3}}{8}\tilde{C}^{(3)}_{76}+\frac{\sqrt{3}}{2}\tilde{C}^{(3)}_{105}-\frac{\sqrt{3}}{6}\tilde{C}^{(3)}_{106}-\frac{\sqrt{3}}{2}\tilde{C}^{(3)}_{108}+\frac{\sqrt{3}}{8}\tilde{C}^{(3)}_{124},\;
\tilde{C}^{(3)}_{279}=\frac{\sqrt{3}}{2}\tilde{C}^{(3)}_{58}-\sqrt{3}\tilde{C}^{(3)}_{59}-\frac{3\,\sqrt{3}}{2}\tilde{C}^{(3)}_{60}-\frac{\sqrt{3}}{8}\tilde{C}^{(3)}_{76}-\frac{\sqrt{3}}{2}\tilde{C}^{(3)}_{105}+\frac{\sqrt{3}}{2}\tilde{C}^{(3)}_{106}+\frac{\sqrt{3}}{2}\tilde{C}^{(3)}_{108}-\frac{\sqrt{3}}{8}\tilde{C}^{(3)}_{124},\;
\tilde{C}^{(3)}_{282}=\sqrt{3}\tilde{C}^{(3)}_{58}-\sqrt{3}\tilde{C}^{(3)}_{59}-\sqrt{3}\tilde{C}^{(3)}_{60}+\sqrt{3}\tilde{C}^{(3)}_{63}-\frac{\sqrt{3}}{4}\tilde{C}^{(3)}_{76}-\frac{\sqrt{3}}{3}\tilde{C}^{(3)}_{105}+\frac{\sqrt{3}}{3}\tilde{C}^{(3)}_{106}+\sqrt{3}\tilde{C}^{(3)}_{108}-\frac{\sqrt{3}}{4}\tilde{C}^{(3)}_{124},\;
\tilde{C}^{(3)}_{283}=-\sqrt{3}\tilde{C}^{(3)}_{58}+\sqrt{3}\tilde{C}^{(3)}_{59}+2\,\sqrt{3}\tilde{C}^{(3)}_{60}+\frac{\sqrt{3}}{4}\tilde{C}^{(3)}_{76}+\frac{2\,\sqrt{3}}{3}\tilde{C}^{(3)}_{105}-\frac{\sqrt{3}}{3}\tilde{C}^{(3)}_{106}-\sqrt{3}\tilde{C}^{(3)}_{108}+\frac{\sqrt{3}}{4}\tilde{C}^{(3)}_{124},\;
\tilde{C}^{(3)}_{284}=-\frac{\sqrt{3}}{2}\tilde{C}^{(3)}_{66}-\frac{\sqrt{3}}{2}\tilde{C}^{(3)}_{68},\;
\tilde{C}^{(3)}_{286}=-\frac{\sqrt{3}}{2}\tilde{C}^{(3)}_{66}+\frac{\sqrt{3}}{2}\tilde{C}^{(3)}_{68},\;
\tilde{C}^{(3)}_{289}=-\frac{\sqrt{3}}{2}\tilde{C}^{(3)}_{70}+\frac{3\,\sqrt{3}}{2}\tilde{C}^{(3)}_{71}-2\,\sqrt{3}\tilde{C}^{(3)}_{72}-\sqrt{3}\tilde{C}^{(3)}_{114},\;
\tilde{C}^{(3)}_{290}=3\,\sqrt{3}\tilde{C}^{(3)}_{71}-4\,\sqrt{3}\tilde{C}^{(3)}_{72}-2\,\sqrt{3}\tilde{C}^{(3)}_{114},\;
\tilde{C}^{(3)}_{291}=-\frac{3\,\sqrt{3}}{2}\tilde{C}^{(3)}_{71}+2\,\sqrt{3}\tilde{C}^{(3)}_{72}+\sqrt{3}\tilde{C}^{(3)}_{114},\;
\tilde{C}^{(3)}_{292}=-2\,\sqrt{3}\tilde{C}^{(3)}_{73}+\sqrt{3}\tilde{C}^{(3)}_{118},\;
\tilde{C}^{(3)}_{293}=2\,\sqrt{3}\tilde{C}^{(3)}_{73}-\sqrt{3}\tilde{C}^{(3)}_{118},\;
\tilde{C}^{(3)}_{294}=-2\,\sqrt{3}\tilde{C}^{(3)}_{73}+\sqrt{3}\tilde{C}^{(3)}_{118},\;
\tilde{C}^{(3)}_{295}=\frac{\sqrt{3}}{2}\tilde{C}^{(3)}_{70}+\frac{3\,\sqrt{3}}{2}\tilde{C}^{(3)}_{71}-2\,\sqrt{3}\tilde{C}^{(3)}_{72}-\sqrt{3}\tilde{C}^{(3)}_{114},\;
\tilde{C}^{(3)}_{296}=3\,\sqrt{3}\tilde{C}^{(3)}_{71}-4\,\sqrt{3}\tilde{C}^{(3)}_{72}-2\,\sqrt{3}\tilde{C}^{(3)}_{114},\;
\tilde{C}^{(3)}_{297}=-2\,\sqrt{3}\tilde{C}^{(3)}_{74}+\sqrt{3}\tilde{C}^{(3)}_{122},\;
\tilde{C}^{(3)}_{298}=2\,\sqrt{3}\tilde{C}^{(3)}_{74}-\sqrt{3}\tilde{C}^{(3)}_{122},\;
\tilde{C}^{(3)}_{302}=-2\,\sqrt{3}\tilde{C}^{(3)}_{74}+\sqrt{3}\tilde{C}^{(3)}_{122}
	\end{autobreak}
\end{align}

\begin{longtable}{cccccc}
\caption{\label{tab:table4}Independent Lagrangian terms in the heavy quark limit at the $\mathcal{O}(p^{3})$ order.}\\
\hline\hline $O_n$ & $SU(2)$ & $SU(3)$ & $O_n$  & $SU(2)$ & $SU(3)$ \\
\hline\endfirsthead

\hline\hline $O_n$ & $SU(2)$ & $SU(3)$ & $O_n$  & $SU(2)$ & $SU(3)$\\
\hline\endhead

\hline\hline 
\endfoot

\hline\endlastfoot
$\la\Btb u^{\mu}u_{\mu}u^{\nu}\psi_{\nu}\ra$ & 1 & 1 & $i\epsilon^{\mu\nu\lambda\rho}\la\psib_{\mu}u_{\nu}u^{\sigma}u_{\sigma}v_{\lambda}\psi_{\rho}\ra+\mathrm{H.c.}$ & 32 & 46  \\
$\la\Btb u^{\mu}u^{\nu}u_{\mu}\psi_{\nu}\ra$ & 2 & 2 & $i\epsilon^{\mu\nu\lambda\rho}\la\psib_{\mu}u^{\sigma}u_{\nu}u_{\lambda}v_{\rho}\psi_{\sigma}\ra+\mathrm{H.c.}$ &  & 47  \\
$\la\Btb u^{\mu}u^{\nu}u_{\nu}\psi_{\mu}\ra$ &  & 3 & $i\epsilon^{\mu\nu\lambda\rho}\la\psib_{\mu}u_{\nu}u_{\lambda}v_{\rho}\psi^{\sigma}{u^{T}}_{\sigma}\ra+\mathrm{H.c.}$ & 33 & 48  \\
$\la\Btb u^{\mu}u^{\nu}u^{\lambda}v_{\mu}v_{\nu}\psi_{\lambda}\ra$ & 3 & 4 & $i\epsilon^{\mu\nu\lambda\rho}\la\psib_{\mu}u_{\nu}u^{\sigma}v_{\lambda}\psi_{\rho}{u^{T}}_{\sigma}\ra+\mathrm{H.c.}$ &  & 49  \\
$\la\Btb u^{\mu}u^{\nu}u^{\lambda}v_{\mu}v_{\lambda}\psi_{\nu}\ra$ & 4 & 5 & $i\epsilon^{\mu\nu\lambda\rho}\la\psib_{\mu}u_{\nu}u^{\sigma}v_{\lambda}\psi_{\sigma}{u^{T}}_{\rho}\ra+\mathrm{H.c.}$ & 34 & 50  \\
$\la\Btb u^{\mu}u^{\nu}u^{\lambda}v_{\nu}v_{\lambda}\psi_{\mu}\ra$ &  & 6 & $i\epsilon^{\mu\nu\lambda\rho}\la\psib_{\mu}u_{\nu}u^{\sigma}v_{\sigma}\psi_{\lambda}{u^{T}}_{\rho}\ra+\mathrm{H.c.}$ &  & 51  \\
$\la\Btb u^{\mu}u_{\mu}\psi^{\nu}{u^{T}}_{\nu}\ra$ &  & 7 & $i\epsilon^{\mu\nu\lambda\rho}\la\psib_{\mu}u^{\sigma}u_{\nu}v_{\lambda}\psi_{\sigma}{u^{T}}_{\rho}\ra+\mathrm{H.c.}$ &  & 52  \\
$\la\Btb u^{\mu}u^{\nu}\psi_{\mu}{u^{T}}_{\nu}\ra$ & 5 & 8 & $i\epsilon^{\mu\nu\lambda\rho}\la\psib_{\mu}v_{\nu}\psi_{\lambda}\ra\la u_{\rho}u^{\sigma}u_{\sigma}\ra$ &  & 53  \\
\hline
\end{longtable}

\section{$\mathcal{O}(p^{4})$ order results}\label{app:B}
\begin{longtable}{lcrcrlcrcr}
\caption{\label{tab:table6}Independent terms in the relativistic Lagrangian at the $\mathcal{O}(p^{4})$ order. In the columns 3, 5, 8, and 10, ``I'' means that we choose the term as an independent one when employing the heavy quark symmetry. ``P.'' denotes the parity transformed part and ``P.C.H.'' denotes the part obtained after the parity transformation, charge conjugation, and Hermitian conjugation are performed simultaneously.}\\
\hline\hline $O_n$ & $SU(2)$ & $\tilde{e}^{(4)}_n$ & $SU(3)$ & $\tilde{E}^{(4)}_n$ & $O_n$ & $SU(2)$ & $\tilde{e}^{(4)}_n$ & $SU(3)$ & $\tilde{E}^{(4)}_n$\\
\hline\endfirsthead

\hline\hline $O_n$ & $SU(2)$ & $\tilde{e}^{(4)}_n$ & $SU(3)$ & $\tilde{E}^{(4)}_n$ & $O_n$ & $SU(2)$ & $\tilde{e}^{(4)}_n$ & $SU(3)$ & $\tilde{E}^{(4)}_n$\\
\hline\endhead

\hline\hline 
\endfoot

$\la\Btb u^{\mu}u_{\mu}u^{\nu}u_{\nu}\Bt\ra$ & 1&  & 1& & $\la\Btb u^{\mu}\nabla_{\mu}h^{\nu\lambda}D_{\nu\lambda}\Bs\ra+\mathrm{H.c.}$ & 621&  & 1226&  \\
$\la\Btb u^{\mu}u^{\nu}u_{\mu}u_{\nu}\Bt\ra$ & 2&  & 2& & $\la\Btb u^{\mu}\nabla^{\nu}h^{\lambda\rho}D_{\mu\nu\lambda\rho}\Bs\ra+\mathrm{H.c.}$ & 622&  & 1227&  \\
$\la\Btb u^{\mu}u^{\nu}u_{\nu}u_{\mu}\Bt\ra$ &  &  & 3& & $i\la\Btb u^{\mu}\nabla^{\nu}h^{\lambda\rho}\sigma_{\mu\nu}D_{\lambda\rho}\Bs\ra+\mathrm{H.c.}$ & 623& $\mathrm{I}\quad$ & 1228& $\mathrm{I}\quad$  \\
$\la\Btb u^{\mu}u_{\mu}u^{\nu}u^{\lambda}D_{\nu\lambda}\Bt\ra+\mathrm{H.c.}$ & 3&  & 4& & $\la\Btb u^{\mu}\Bs\nabla^{\nu}{f_{-}^{T}}_{\mu\nu}\ra+\mathrm{H.c.}$ & 624&  & 1229&  \\
$\la\Btb u^{\mu}u^{\nu}u_{\mu}u^{\lambda}D_{\nu\lambda}\Bt\ra+\mathrm{H.c.}$ & 4&  & 5& & $\la\Btb u^{\mu}{D_{\mu}}^{\nu}\Bs\nabla^{\lambda}{f_{-}^{T}}_{\nu\lambda}\ra+\mathrm{H.c.}$ & 625&  & 1230&  \\
$\la\Btb u^{\mu}u^{\nu}u_{\nu}u^{\lambda}D_{\mu\lambda}\Bt\ra$ &  &  & 6& & $i\la\Btb u^{\mu}{\sigma_{\mu}}^{\nu}\Bs\nabla^{\lambda}{f_{-}^{T}}_{\nu\lambda}\ra+\mathrm{H.c.}$ & 626& $\mathrm{I}\quad$ & 1231& $\mathrm{I}\quad$  \\
$\la\Btb u^{\mu}u^{\nu}u^{\lambda}u_{\mu}D_{\nu\lambda}\Bt\ra$ &  &  & 7& & $i\la\Btb u^{\mu}\sigma^{\nu\lambda}\Bs\nabla_{\mu}{f_{-}^{T}}_{\nu\lambda}\ra+\mathrm{H.c.}$ & 627& $\mathrm{I}\quad$ & 1232& $\mathrm{I}\quad$  \\
$\la\Btb u^{\mu}u^{\nu}u^{\lambda}u^{\rho}D_{\mu\nu\lambda\rho}\Bt\ra$ & 5&  & 8& & $\la\Btb{f_{-}}^{\mu\nu}f_{-\mu\nu}\Bs\ra+\mathrm{H.c.}$ &  &  & 1233&  \\
\end{longtable}

\begin{longtable}{lrrlrr}
\caption{\label{tab:table7}Independent Lagrangian terms in the heavy quark limit at the $\mathcal{O}(p^{4})$ order.}\\
\hline\hline $P_n$ & $SU(2)$ & $SU(3)$ & $P_n$ & $SU(2)$ & $SU(3)$\\
\hline\endfirsthead

\hline\hline $P_n$ & $SU(2)$ & $SU(3)$ & $P_n$ & $SU(2)$ & $SU(3)$\\
\hline\endhead

\hline\hline 
\endfoot

$i\epsilon^{\mu\nu\lambda\rho}\la\Btb u_{\mu}u_{\nu}u_{\lambda}u^{\sigma}v_{\rho}\psi_{\sigma}\ra$ & 1 & 1 & $\la\psib^{\mu}u_{\mu}\psi^{\nu}\ra\la u_{\nu}u^{\lambda}u_{\lambda}\ra+\mathrm{H.c.}$ &  & 359  \\
$i\epsilon^{\mu\nu\lambda\rho}\la\Btb u_{\mu}u_{\nu}u_{\lambda}u^{\sigma}v_{\sigma}\psi_{\rho}\ra$ & 2 & 2 & $\la\psib^{\mu}u^{\nu}\psi_{\mu}\ra\la u_{\nu}u^{\lambda}u_{\lambda}\ra$ &  & 360  \\
$i\epsilon^{\mu\nu\lambda\rho}\la\Btb u_{\mu}u_{\nu}u^{\sigma}u_{\lambda}v_{\rho}\psi_{\sigma}\ra$ &  & 3 & $\la\psib^{\mu}u^{\nu}\psi^{\lambda}\ra\la u_{\mu}u_{\nu}u_{\lambda}\ra$ &  & 361  \\
$i\epsilon^{\mu\nu\lambda\rho}\la\Btb u_{\mu}u_{\nu}u^{\sigma}u_{\lambda}v_{\sigma}\psi_{\rho}\ra$ &  & 4 & $\la\psib^{\mu}u^{\nu}\psi^{\lambda}\ra\la u_{\mu}u_{\lambda}u_{\nu}\ra$ &  & 362  \\
$i\epsilon^{\mu\nu\lambda\rho}\la\Btb u_{\mu}u_{\nu}u^{\sigma}u_{\sigma}v_{\lambda}\psi_{\rho}\ra$ & 3 & 5 & $\la\psib^{\mu}u_{\mu}v^{\nu}v^{\lambda}\psi^{\rho}\ra\la u_{\nu}u_{\lambda}u_{\rho}\ra+\mathrm{H.c.}$ &  & 363  \\
$i\epsilon^{\mu\nu\lambda\rho}\la\Btb u_{\mu}u^{\sigma}u_{\nu}u_{\lambda}v_{\rho}\psi_{\sigma}\ra$ &  & 6 & $\la\psib^{\mu}u^{\nu}v_{\nu}v^{\lambda}\psi_{\mu}\ra\la u_{\lambda}u^{\rho}u_{\rho}\ra$ &  & 364  \\
$i\epsilon^{\mu\nu\lambda\rho}\la\Btb u_{\mu}u^{\sigma}u_{\nu}u_{\lambda}v_{\sigma}\psi_{\rho}\ra$ &  & 7 & $\la\psib^{\mu}u^{\nu}v_{\nu}v^{\lambda}\psi^{\rho}\ra\la u_{\mu}u_{\lambda}u_{\rho}\ra$ &  & 365  \\
$i\epsilon^{\mu\nu\lambda\rho}\la\Btb u_{\mu}u^{\sigma}u_{\nu}u_{\sigma}v_{\lambda}\psi_{\rho}\ra$ &  & 8 & $\la\psib^{\mu}u^{\nu}v_{\nu}v^{\lambda}\psi^{\rho}\ra\la u_{\mu}u_{\rho}u_{\lambda}\ra$ &  & 366  \\
\end{longtable}


\bibliography{bibtex}

\begin{thebibliography}{76}%
\makeatletter
\providecommand \@ifxundefined [1]{%
 \@ifx{#1\undefined}
}%
\providecommand \@ifnum [1]{%
 \ifnum #1\expandafter \@firstoftwo
 \else \expandafter \@secondoftwo
 \fi
}%
\providecommand \@ifx [1]{%
 \ifx #1\expandafter \@firstoftwo
 \else \expandafter \@secondoftwo
 \fi
}%
\providecommand \natexlab [1]{#1}%
\providecommand \enquote  [1]{``#1''}%
\providecommand \bibnamefont  [1]{#1}%
\providecommand \bibfnamefont [1]{#1}%
\providecommand \citenamefont [1]{#1}%
\providecommand \href@noop [0]{\@secondoftwo}%
\providecommand \href [0]{\begingroup \@sanitize@url \@href}%
\providecommand \@href[1]{\@@startlink{#1}\@@href}%
\providecommand \@@href[1]{\endgroup#1\@@endlink}%
\providecommand \@sanitize@url [0]{\catcode `\\12\catcode `\$12\catcode
  `\&12\catcode `\#12\catcode `\^12\catcode `\_12\catcode `\%12\relax}%
\providecommand \@@startlink[1]{}%
\providecommand \@@endlink[0]{}%
\providecommand \url  [0]{\begingroup\@sanitize@url \@url }%
\providecommand \@url [1]{\endgroup\@href {#1}{\urlprefix }}%
\providecommand \urlprefix  [0]{URL }%
\providecommand \Eprint [0]{\href }%
\providecommand \doibase [0]{http://dx.doi.org/}%
\providecommand \selectlanguage [0]{\@gobble}%
\providecommand \bibinfo  [0]{\@secondoftwo}%
\providecommand \bibfield  [0]{\@secondoftwo}%
\providecommand \translation [1]{[#1]}%
\providecommand \BibitemOpen [0]{}%
\providecommand \bibitemStop [0]{}%
\providecommand \bibitemNoStop [0]{.\EOS\space}%
\providecommand \EOS [0]{\spacefactor3000\relax}%
\providecommand \BibitemShut  [1]{\csname bibitem#1\endcsname}%
\let\auto@bib@innerbib\@empty
\bibitem [{\citenamefont {Cazzoli}\ \emph {et~al.}(1975)\citenamefont
  {Cazzoli}, \citenamefont {Cnops}, \citenamefont {Connolly}, \citenamefont
  {Louttit}, \citenamefont {Murtagh}, \citenamefont {Palmer}, \citenamefont
  {Samios}, \citenamefont {Tso},\ and\ \citenamefont
  {Williams}}]{Cazzoli:1975et}%
  \BibitemOpen
  \bibfield  {author} {\bibinfo {author} {\bibfnamefont {E.~G.}\ \bibnamefont
  {Cazzoli}}, \bibinfo {author} {\bibfnamefont {A.~M.}\ \bibnamefont {Cnops}},
  \bibinfo {author} {\bibfnamefont {P.~L.}\ \bibnamefont {Connolly}}, \bibinfo
  {author} {\bibfnamefont {R.~I.}\ \bibnamefont {Louttit}}, \bibinfo {author}
  {\bibfnamefont {M.~J.}\ \bibnamefont {Murtagh}}, \bibinfo {author}
  {\bibfnamefont {R.~B.}\ \bibnamefont {Palmer}}, \bibinfo {author}
  {\bibfnamefont {N.~P.}\ \bibnamefont {Samios}}, \bibinfo {author}
  {\bibfnamefont {T.~T.}\ \bibnamefont {Tso}}, \ and\ \bibinfo {author}
  {\bibfnamefont {H.~H.}\ \bibnamefont {Williams}},\ }\bibfield  {title}
  {\enquote {\bibinfo {title} {{Evidence for $\Delta S = - \Delta Q$ Currents
  or Charmed Baryon Production by Neutrinos}},}\ }\href {\doibase
  10.1103/PhysRevLett.34.1125} {\bibfield  {journal} {\bibinfo  {journal}
  {Phys. Rev. Lett.}\ }\textbf {\bibinfo {volume} {34}},\ \bibinfo {pages}
  {1125--1128} (\bibinfo {year} {1975})}\BibitemShut {NoStop}%
\bibitem [{\citenamefont {Abrams}\ \emph {et~al.}(1980)\citenamefont {Abrams}
  \emph {et~al.}}]{Abrams:1979iu}%
  \BibitemOpen
  \bibfield  {author} {\bibinfo {author} {\bibfnamefont {G.~S.}\ \bibnamefont
  {Abrams}} \emph {et~al.},\ }\bibfield  {title} {\enquote {\bibinfo {title}
  {{Observation of Charmed Baryon Production in $e^+e^-$ Annihilation}},}\
  }\href {\doibase 10.1103/PhysRevLett.44.10} {\bibfield  {journal} {\bibinfo
  {journal} {Phys. Rev. Lett.}\ }\textbf {\bibinfo {volume} {44}},\ \bibinfo
  {pages} {10} (\bibinfo {year} {1980})}\BibitemShut {NoStop}%
\bibitem [{\citenamefont {Baltay}\ \emph {et~al.}(1979)\citenamefont {Baltay}
  \emph {et~al.}}]{Baltay:1979rn}%
  \BibitemOpen
  \bibfield  {author} {\bibinfo {author} {\bibfnamefont {C.}~\bibnamefont
  {Baltay}} \emph {et~al.},\ }\bibfield  {title} {\enquote {\bibinfo {title}
  {{Confirmation of the Existence of the $\Sigma_c^{++}$ and $\Lambda_c^+$
  Charmed Baryons}},}\ }\href {\doibase 10.1103/PhysRevLett.42.1721} {\bibfield
   {journal} {\bibinfo  {journal} {Phys. Rev. Lett.}\ }\textbf {\bibinfo
  {volume} {42}},\ \bibinfo {pages} {1721} (\bibinfo {year}
  {1979})}\BibitemShut {NoStop}%
\bibitem [{\citenamefont {Giboni}\ \emph {et~al.}(1979)\citenamefont {Giboni}
  \emph {et~al.}}]{Giboni:1979rm}%
  \BibitemOpen
  \bibfield  {author} {\bibinfo {author} {\bibfnamefont {K.~L.}\ \bibnamefont
  {Giboni}} \emph {et~al.},\ }\bibfield  {title} {\enquote {\bibinfo {title}
  {{Diffractive Production of the Charmed Baryon $\Lambda_c^+$ at the CERN
  ISR}},}\ }\href {\doibase 10.1016/0370-2693(79)91291-7} {\bibfield  {journal}
  {\bibinfo  {journal} {Phys. Lett. B}\ }\textbf {\bibinfo {volume} {85}},\
  \bibinfo {pages} {437--442} (\bibinfo {year} {1979})}\BibitemShut {NoStop}%
\bibitem [{\citenamefont {Workman}\ \emph {et~al.}(2022)\citenamefont {Workman}
  \emph {et~al.}}]{ParticleDataGroup:2022pth}%
  \BibitemOpen
  \bibfield  {author} {\bibinfo {author} {\bibfnamefont {R.~L.}\ \bibnamefont
  {Workman}} \emph {et~al.} (\bibinfo {collaboration} {Particle Data Group}),\
  }\bibfield  {title} {\enquote {\bibinfo {title} {{Review of Particle
  Physics}},}\ }\href {\doibase 10.1093/ptep/ptac097} {\bibfield  {journal}
  {\bibinfo  {journal} {PTEP}\ }\textbf {\bibinfo {volume} {2022}},\ \bibinfo
  {pages} {083C01} (\bibinfo {year} {2022})}\BibitemShut {NoStop}%
\bibitem [{\citenamefont {Chen}\ \emph {et~al.}(2017)\citenamefont {Chen},
  \citenamefont {Chen}, \citenamefont {Liu}, \citenamefont {Liu},\ and\
  \citenamefont {Zhu}}]{Chen:2016spr}%
  \BibitemOpen
  \bibfield  {author} {\bibinfo {author} {\bibfnamefont {Hua-Xing}\
  \bibnamefont {Chen}}, \bibinfo {author} {\bibfnamefont {Wei}\ \bibnamefont
  {Chen}}, \bibinfo {author} {\bibfnamefont {Xiang}\ \bibnamefont {Liu}},
  \bibinfo {author} {\bibfnamefont {Yan-Rui}\ \bibnamefont {Liu}}, \ and\
  \bibinfo {author} {\bibfnamefont {Shi-Lin}\ \bibnamefont {Zhu}},\ }\bibfield
  {title} {\enquote {\bibinfo {title} {{A review of the open charm and open
  bottom systems}},}\ }\href {\doibase 10.1088/1361-6633/aa6420} {\bibfield
  {journal} {\bibinfo  {journal} {Rept. Prog. Phys.}\ }\textbf {\bibinfo
  {volume} {80}},\ \bibinfo {pages} {076201} (\bibinfo {year} {2017})},\
  \Eprint {http://arxiv.org/abs/1609.08928} {arXiv:1609.08928 [hep-ph]}
  \BibitemShut {NoStop}%
\bibitem [{\citenamefont {Padmanath}(2019)}]{Padmanath:2019ybu}%
  \BibitemOpen
  \bibfield  {author} {\bibinfo {author} {\bibfnamefont {M.}~\bibnamefont
  {Padmanath}},\ }\href@noop {} {\enquote {\bibinfo {title} {{Heavy baryon
  spectroscopy from lattice QCD}},}\ } (\bibinfo {year} {2019}),\ \Eprint
  {http://arxiv.org/abs/1905.10168} {arXiv:1905.10168 [hep-lat]} \BibitemShut
  {NoStop}%
\bibitem [{\citenamefont {Na}\ and\ \citenamefont
  {Gottlieb}(2006)}]{Na:2006qz}%
  \BibitemOpen
  \bibfield  {author} {\bibinfo {author} {\bibfnamefont {Heechang}\
  \bibnamefont {Na}}\ and\ \bibinfo {author} {\bibfnamefont {Steven}\
  \bibnamefont {Gottlieb}},\ }\bibfield  {title} {\enquote {\bibinfo {title}
  {{Heavy baryon mass spectrum from lattice QCD with 2+1 flavors}},}\ }\href
  {\doibase 10.22323/1.032.0191} {\bibfield  {journal} {\bibinfo  {journal}
  {Proc. Sci. LAT2006}\ ,\ \bibinfo {pages} {191}} (\bibinfo {year} {2006})},\
  \Eprint {http://arxiv.org/abs/hep-lat/0610009} {arXiv:hep-lat/0610009}
  \BibitemShut {NoStop}%
\bibitem [{\citenamefont {Can}\ \emph {et~al.}(2014)\citenamefont {Can},
  \citenamefont {Erkol}, \citenamefont {Isildak}, \citenamefont {Oka},\ and\
  \citenamefont {Takahashi}}]{Can:2013tna}%
  \BibitemOpen
  \bibfield  {author} {\bibinfo {author} {\bibfnamefont {K.~U.}\ \bibnamefont
  {Can}}, \bibinfo {author} {\bibfnamefont {G.}~\bibnamefont {Erkol}}, \bibinfo
  {author} {\bibfnamefont {B.}~\bibnamefont {Isildak}}, \bibinfo {author}
  {\bibfnamefont {M.}~\bibnamefont {Oka}}, \ and\ \bibinfo {author}
  {\bibfnamefont {T.~T.}\ \bibnamefont {Takahashi}},\ }\bibfield  {title}
  {\enquote {\bibinfo {title} {{Electromagnetic structure of charmed baryons in
  Lattice QCD}},}\ }\href {\doibase 10.1007/JHEP05(2014)125} {\bibfield
  {journal} {\bibinfo  {journal} {JHEP}\ }\textbf {\bibinfo {volume} {05}},\
  \bibinfo {pages} {125} (\bibinfo {year} {2014})},\ \Eprint
  {http://arxiv.org/abs/1310.5915} {arXiv:1310.5915 [hep-lat]} \BibitemShut
  {NoStop}%
\bibitem [{\citenamefont {Can}(2021)}]{Can:2021ehb}%
  \BibitemOpen
  \bibfield  {author} {\bibinfo {author} {\bibfnamefont {Kadir~Utku}\
  \bibnamefont {Can}},\ }\bibfield  {title} {\enquote {\bibinfo {title}
  {{Lattice QCD study of the elastic and transition form factors of charmed
  baryons}},}\ }\href {\doibase 10.1142/S0217751X21300131} {\bibfield
  {journal} {\bibinfo  {journal} {Int. J. Mod. Phys. A}\ }\textbf {\bibinfo
  {volume} {36}},\ \bibinfo {pages} {2130013} (\bibinfo {year} {2021})},\
  \Eprint {http://arxiv.org/abs/2107.13159} {arXiv:2107.13159 [hep-lat]}
  \BibitemShut {NoStop}%
\bibitem [{\citenamefont {Wang}(2010)}]{Wang:2010fq}%
  \BibitemOpen
  \bibfield  {author} {\bibinfo {author} {\bibfnamefont {Zhi-Gang}\
  \bibnamefont {Wang}},\ }\bibfield  {title} {\enquote {\bibinfo {title}
  {{Analysis of the ${1\over 2}^{\pm}$ antitriplet heavy baryon states with QCD
  sum rules}},}\ }\href {\doibase 10.1140/epjc/s10052-010-1365-8} {\bibfield
  {journal} {\bibinfo  {journal} {Eur. Phys. J. C}\ }\textbf {\bibinfo {volume}
  {68}},\ \bibinfo {pages} {479--486} (\bibinfo {year} {2010})},\ \Eprint
  {http://arxiv.org/abs/1001.1652} {arXiv:1001.1652 [hep-ph]} \BibitemShut
  {NoStop}%
\bibitem [{\citenamefont {Zhang}\ and\ \citenamefont
  {Huang}(2009)}]{Zhang:2009iya}%
  \BibitemOpen
  \bibfield  {author} {\bibinfo {author} {\bibfnamefont {Jian-Rong}\
  \bibnamefont {Zhang}}\ and\ \bibinfo {author} {\bibfnamefont {Ming-Qiu}\
  \bibnamefont {Huang}},\ }\bibfield  {title} {\enquote {\bibinfo {title}
  {{Heavy flavor baryon spectra via QCD sum rules}},}\ }\href {\doibase
  10.1088/1674-1137/33/12/061} {\bibfield  {journal} {\bibinfo  {journal}
  {Chin. Phys. C}\ }\textbf {\bibinfo {volume} {33}},\ \bibinfo {pages}
  {1385--1388} (\bibinfo {year} {2009})},\ \Eprint
  {http://arxiv.org/abs/0904.3391} {arXiv:0904.3391 [hep-ph]} \BibitemShut
  {NoStop}%
\bibitem [{\citenamefont {Zhu}\ \emph {et~al.}(1997)\citenamefont {Zhu},
  \citenamefont {Hwang},\ and\ \citenamefont {Yang}}]{Zhu:1997as}%
  \BibitemOpen
  \bibfield  {author} {\bibinfo {author} {\bibfnamefont {Shi-Lin}\ \bibnamefont
  {Zhu}}, \bibinfo {author} {\bibfnamefont {W-Y.~P.}\ \bibnamefont {Hwang}}, \
  and\ \bibinfo {author} {\bibfnamefont {Ze-Sen}\ \bibnamefont {Yang}},\
  }\bibfield  {title} {\enquote {\bibinfo {title} {{The Sigma(c) and Lambda(c)
  magnetic moments from QCD spectral sum rules}},}\ }\href {\doibase
  10.1103/PhysRevD.56.7273} {\bibfield  {journal} {\bibinfo  {journal} {Phys.
  Rev. D}\ }\textbf {\bibinfo {volume} {56}},\ \bibinfo {pages} {7273--7275}
  (\bibinfo {year} {1997})},\ \Eprint {http://arxiv.org/abs/hep-ph/9708411}
  {arXiv:hep-ph/9708411} \BibitemShut {NoStop}%
\bibitem [{\citenamefont {Zhu}\ and\ \citenamefont {Dai}(1999)}]{Zhu:1998ih}%
  \BibitemOpen
  \bibfield  {author} {\bibinfo {author} {\bibfnamefont {Shi-Lin}\ \bibnamefont
  {Zhu}}\ and\ \bibinfo {author} {\bibfnamefont {Yuan-Ben}\ \bibnamefont
  {Dai}},\ }\bibfield  {title} {\enquote {\bibinfo {title} {{Radiative decays
  of heavy hadrons from light cone QCD sum rules in the leading order of
  HQET}},}\ }\href {\doibase 10.1103/PhysRevD.59.114015} {\bibfield  {journal}
  {\bibinfo  {journal} {Phys. Rev. D}\ }\textbf {\bibinfo {volume} {59}},\
  \bibinfo {pages} {114015} (\bibinfo {year} {1999})},\ \Eprint
  {http://arxiv.org/abs/hep-ph/9810243} {arXiv:hep-ph/9810243} \BibitemShut
  {NoStop}%
\bibitem [{\citenamefont {Agamaliev}\ \emph {et~al.}(2017)\citenamefont
  {Agamaliev}, \citenamefont {Aliev},\ and\ \citenamefont
  {Savc\i{}}}]{Agamaliev:2016fou}%
  \BibitemOpen
  \bibfield  {author} {\bibinfo {author} {\bibfnamefont {A.~K.}\ \bibnamefont
  {Agamaliev}}, \bibinfo {author} {\bibfnamefont {T.~M.}\ \bibnamefont
  {Aliev}}, \ and\ \bibinfo {author} {\bibfnamefont {M.}~\bibnamefont
  {Savc\i{}}},\ }\bibfield  {title} {\enquote {\bibinfo {title} {{Radiative
  decays of negative parity heavy baryons in the framework of the light cone
  QCD sum rules}},}\ }\href {\doibase 10.1016/j.nuclphysa.2016.11.005}
  {\bibfield  {journal} {\bibinfo  {journal} {Nucl. Phys. A}\ }\textbf
  {\bibinfo {volume} {958}},\ \bibinfo {pages} {38--47} (\bibinfo {year}
  {2017})},\ \Eprint {http://arxiv.org/abs/1606.07666} {arXiv:1606.07666
  [hep-ph]} \BibitemShut {NoStop}%
\bibitem [{\citenamefont {Bose}\ and\ \citenamefont
  {Singh}(1980)}]{Bose:1980vy}%
  \BibitemOpen
  \bibfield  {author} {\bibinfo {author} {\bibfnamefont {S.~K.}\ \bibnamefont
  {Bose}}\ and\ \bibinfo {author} {\bibfnamefont {L.~P.}\ \bibnamefont
  {Singh}},\ }\bibfield  {title} {\enquote {\bibinfo {title} {{Magnetic Moments
  of Charmed and $B$ Flavored Hadrons in {MIT} Bag Model}},}\ }\href {\doibase
  10.1103/PhysRevD.22.773} {\bibfield  {journal} {\bibinfo  {journal} {Phys.
  Rev. D}\ }\textbf {\bibinfo {volume} {22}},\ \bibinfo {pages} {773} (\bibinfo
  {year} {1980})}\BibitemShut {NoStop}%
\bibitem [{\citenamefont {Bernotas}\ and\ \citenamefont
  {\v{S}imonis}(2013)}]{Bernotas:2013eia}%
  \BibitemOpen
  \bibfield  {author} {\bibinfo {author} {\bibfnamefont {Andrius}\ \bibnamefont
  {Bernotas}}\ and\ \bibinfo {author} {\bibfnamefont {Vytautas}\ \bibnamefont
  {\v{S}imonis}},\ }\bibfield  {title} {\enquote {\bibinfo {title} {{Radiative
  M1 transitions of heavy baryons in the bag model}},}\ }\href {\doibase
  10.1103/PhysRevD.87.074016} {\bibfield  {journal} {\bibinfo  {journal} {Phys.
  Rev. D}\ }\textbf {\bibinfo {volume} {87}},\ \bibinfo {pages} {074016}
  (\bibinfo {year} {2013})},\ \Eprint {http://arxiv.org/abs/1302.5918}
  {arXiv:1302.5918 [hep-ph]} \BibitemShut {NoStop}%
\bibitem [{\citenamefont {Kim}\ \emph {et~al.}(2021)\citenamefont {Kim},
  \citenamefont {Kim}, \citenamefont {Yang},\ and\ \citenamefont
  {Oka}}]{Kim:2021xpp}%
  \BibitemOpen
  \bibfield  {author} {\bibinfo {author} {\bibfnamefont {June-Young}\
  \bibnamefont {Kim}}, \bibinfo {author} {\bibfnamefont {Hyun-Chul}\
  \bibnamefont {Kim}}, \bibinfo {author} {\bibfnamefont {Ghil-Seok}\
  \bibnamefont {Yang}}, \ and\ \bibinfo {author} {\bibfnamefont {Makoto}\
  \bibnamefont {Oka}},\ }\bibfield  {title} {\enquote {\bibinfo {title}
  {{Electromagnetic transitions of the singly charmed baryons with spin
  3/2}},}\ }\href {\doibase 10.1103/PhysRevD.103.074025} {\bibfield  {journal}
  {\bibinfo  {journal} {Phys. Rev. D}\ }\textbf {\bibinfo {volume} {103}},\
  \bibinfo {pages} {074025} (\bibinfo {year} {2021})},\ \Eprint
  {http://arxiv.org/abs/2101.10653} {arXiv:2101.10653 [hep-ph]} \BibitemShut
  {NoStop}%
\bibitem [{\citenamefont {Scholl}\ and\ \citenamefont
  {Weigel}(2004)}]{Scholl:2003ip}%
  \BibitemOpen
  \bibfield  {author} {\bibinfo {author} {\bibfnamefont {Stephan}\ \bibnamefont
  {Scholl}}\ and\ \bibinfo {author} {\bibfnamefont {Herbert}\ \bibnamefont
  {Weigel}},\ }\bibfield  {title} {\enquote {\bibinfo {title} {{Magnetic
  moments of baryons with a single heavy quark}},}\ }\href {\doibase
  10.1016/j.nuclphysa.2004.01.132} {\bibfield  {journal} {\bibinfo  {journal}
  {Nucl. Phys. A}\ }\textbf {\bibinfo {volume} {735}},\ \bibinfo {pages}
  {163--184} (\bibinfo {year} {2004})},\ \Eprint
  {http://arxiv.org/abs/hep-ph/0312282} {arXiv:hep-ph/0312282} \BibitemShut
  {NoStop}%
\bibitem [{\citenamefont {Wise}(1992)}]{Wise:1992hn}%
  \BibitemOpen
  \bibfield  {author} {\bibinfo {author} {\bibfnamefont {Mark~B.}\ \bibnamefont
  {Wise}},\ }\bibfield  {title} {\enquote {\bibinfo {title} {{Chiral
  perturbation theory for hadrons containing a heavy quark}},}\ }\href
  {\doibase 10.1103/PhysRevD.45.R2188} {\bibfield  {journal} {\bibinfo
  {journal} {Phys. Rev. D}\ }\textbf {\bibinfo {volume} {45}},\ \bibinfo
  {pages} {R2188} (\bibinfo {year} {1992})}\BibitemShut {NoStop}%
\bibitem [{\citenamefont {Yan}\ \emph {et~al.}(1992)\citenamefont {Yan},
  \citenamefont {Cheng}, \citenamefont {Cheung}, \citenamefont {Lin},
  \citenamefont {Lin},\ and\ \citenamefont {Yu}}]{Yan:1992gz}%
  \BibitemOpen
  \bibfield  {author} {\bibinfo {author} {\bibfnamefont {Tung-Mow}\
  \bibnamefont {Yan}}, \bibinfo {author} {\bibfnamefont {Hai-Yang}\
  \bibnamefont {Cheng}}, \bibinfo {author} {\bibfnamefont {Chi-Yee}\
  \bibnamefont {Cheung}}, \bibinfo {author} {\bibfnamefont {Guey-Lin}\
  \bibnamefont {Lin}}, \bibinfo {author} {\bibfnamefont {Y.~C.}\ \bibnamefont
  {Lin}}, \ and\ \bibinfo {author} {\bibfnamefont {Hoi-Lai}\ \bibnamefont
  {Yu}},\ }\bibfield  {title} {\enquote {\bibinfo {title} {{Heavy quark
  symmetry and chiral dynamics}},}\ }\href {\doibase 10.1103/PhysRevD.46.1148}
  {\bibfield  {journal} {\bibinfo  {journal} {Phys. Rev. D}\ }\textbf {\bibinfo
  {volume} {46}},\ \bibinfo {pages} {1148--1164} (\bibinfo {year} {1992})},\
  \bibinfo {note} {[Erratum: Phys.Rev.D 55, 5851 (1997)]}\BibitemShut {NoStop}%
\bibitem [{\citenamefont {Cho}(1994)}]{Cho:1994vg}%
  \BibitemOpen
  \bibfield  {author} {\bibinfo {author} {\bibfnamefont {Peter~L.}\
  \bibnamefont {Cho}},\ }\bibfield  {title} {\enquote {\bibinfo {title}
  {{Strong and electromagnetic decays of two new Lambda(c)* baryons}},}\ }\href
  {\doibase 10.1103/PhysRevD.50.3295} {\bibfield  {journal} {\bibinfo
  {journal} {Phys. Rev. D}\ }\textbf {\bibinfo {volume} {50}},\ \bibinfo
  {pages} {3295--3302} (\bibinfo {year} {1994})},\ \Eprint
  {http://arxiv.org/abs/hep-ph/9401276} {arXiv:hep-ph/9401276} \BibitemShut
  {NoStop}%
\bibitem [{\citenamefont {Savage}(1995{\natexlab{a}})}]{Savage:1994wa}%
  \BibitemOpen
  \bibfield  {author} {\bibinfo {author} {\bibfnamefont {Martin~J.}\
  \bibnamefont {Savage}},\ }\bibfield  {title} {\enquote {\bibinfo {title} {{E2
  strength in the radiative charmed baryon decay Sigma(c)* ---\ensuremath{>}
  Lambda(c) gamma}},}\ }\href {\doibase 10.1016/0370-2693(94)01597-6}
  {\bibfield  {journal} {\bibinfo  {journal} {Phys. Lett. B}\ }\textbf
  {\bibinfo {volume} {345}},\ \bibinfo {pages} {61--66} (\bibinfo {year}
  {1995}{\natexlab{a}})},\ \Eprint {http://arxiv.org/abs/hep-ph/9408294}
  {arXiv:hep-ph/9408294} \BibitemShut {NoStop}%
\bibitem [{\citenamefont {Banuls}\ \emph {et~al.}(2000)\citenamefont {Banuls},
  \citenamefont {Pich},\ and\ \citenamefont {Scimemi}}]{Banuls:1999br}%
  \BibitemOpen
  \bibfield  {author} {\bibinfo {author} {\bibfnamefont {M.~C.}\ \bibnamefont
  {Banuls}}, \bibinfo {author} {\bibfnamefont {A.}~\bibnamefont {Pich}}, \ and\
  \bibinfo {author} {\bibfnamefont {I.}~\bibnamefont {Scimemi}},\ }\bibfield
  {title} {\enquote {\bibinfo {title} {{Electromagnetic decays of heavy
  baryons}},}\ }\href {\doibase 10.1103/PhysRevD.61.094009} {\bibfield
  {journal} {\bibinfo  {journal} {Phys. Rev. D}\ }\textbf {\bibinfo {volume}
  {61}},\ \bibinfo {pages} {094009} (\bibinfo {year} {2000})},\ \Eprint
  {http://arxiv.org/abs/hep-ph/9911502} {arXiv:hep-ph/9911502} \BibitemShut
  {NoStop}%
\bibitem [{\citenamefont {Tiburzi}(2005)}]{Tiburzi:2004mv}%
  \BibitemOpen
  \bibfield  {author} {\bibinfo {author} {\bibfnamefont {Brian~C.}\
  \bibnamefont {Tiburzi}},\ }\bibfield  {title} {\enquote {\bibinfo {title}
  {{Baryon electromagnetic properties in partially quenched heavy hadron chiral
  perturbation theory}},}\ }\href {\doibase 10.1103/PhysRevD.71.054504}
  {\bibfield  {journal} {\bibinfo  {journal} {Phys. Rev. D}\ }\textbf {\bibinfo
  {volume} {71}},\ \bibinfo {pages} {054504} (\bibinfo {year} {2005})},\
  \Eprint {http://arxiv.org/abs/hep-lat/0412025} {arXiv:hep-lat/0412025}
  \BibitemShut {NoStop}%
\bibitem [{\citenamefont {Jiang}\ \emph
  {et~al.}(2015{\natexlab{a}})\citenamefont {Jiang}, \citenamefont {Chen},\
  and\ \citenamefont {Zhu}}]{Jiang:2015xqa}%
  \BibitemOpen
  \bibfield  {author} {\bibinfo {author} {\bibfnamefont {Nan}\ \bibnamefont
  {Jiang}}, \bibinfo {author} {\bibfnamefont {Xiao-Lin}\ \bibnamefont {Chen}},
  \ and\ \bibinfo {author} {\bibfnamefont {Shi-Lin}\ \bibnamefont {Zhu}},\
  }\bibfield  {title} {\enquote {\bibinfo {title} {{Electromagnetic decays of
  the charmed and bottom baryons in chiral perturbation theory}},}\ }\href
  {\doibase 10.1103/PhysRevD.92.054017} {\bibfield  {journal} {\bibinfo
  {journal} {Phys. Rev. D}\ }\textbf {\bibinfo {volume} {92}},\ \bibinfo
  {pages} {054017} (\bibinfo {year} {2015}{\natexlab{a}})},\ \Eprint
  {http://arxiv.org/abs/1505.02999} {arXiv:1505.02999 [hep-ph]} \BibitemShut
  {NoStop}%
\bibitem [{\citenamefont {Wang}\ \emph {et~al.}(2018)\citenamefont {Wang},
  \citenamefont {Meng}, \citenamefont {Li}, \citenamefont {Liu},\ and\
  \citenamefont {Zhu}}]{Wang:2018gpl}%
  \BibitemOpen
  \bibfield  {author} {\bibinfo {author} {\bibfnamefont {Guang-Juan}\
  \bibnamefont {Wang}}, \bibinfo {author} {\bibfnamefont {Lu}~\bibnamefont
  {Meng}}, \bibinfo {author} {\bibfnamefont {Hao-Song}\ \bibnamefont {Li}},
  \bibinfo {author} {\bibfnamefont {Zhan-Wei}\ \bibnamefont {Liu}}, \ and\
  \bibinfo {author} {\bibfnamefont {Shi-Lin}\ \bibnamefont {Zhu}},\ }\bibfield
  {title} {\enquote {\bibinfo {title} {{Magnetic moments of the
  spin-$\frac{1}{2}$ singly charmed baryons in chiral perturbation theory}},}\
  }\href {\doibase 10.1103/PhysRevD.98.054026} {\bibfield  {journal} {\bibinfo
  {journal} {Phys. Rev. D}\ }\textbf {\bibinfo {volume} {98}},\ \bibinfo
  {pages} {054026} (\bibinfo {year} {2018})},\ \Eprint
  {http://arxiv.org/abs/1803.00229} {arXiv:1803.00229 [hep-ph]} \BibitemShut
  {NoStop}%
\bibitem [{\citenamefont {Meng}\ \emph {et~al.}(2018)\citenamefont {Meng},
  \citenamefont {Wang}, \citenamefont {Leng}, \citenamefont {Liu},\ and\
  \citenamefont {Zhu}}]{Meng:2018gan}%
  \BibitemOpen
  \bibfield  {author} {\bibinfo {author} {\bibfnamefont {Lu}~\bibnamefont
  {Meng}}, \bibinfo {author} {\bibfnamefont {Guang-Juan}\ \bibnamefont {Wang}},
  \bibinfo {author} {\bibfnamefont {Chang-Zhi}\ \bibnamefont {Leng}}, \bibinfo
  {author} {\bibfnamefont {Zhan-Wei}\ \bibnamefont {Liu}}, \ and\ \bibinfo
  {author} {\bibfnamefont {Shi-Lin}\ \bibnamefont {Zhu}},\ }\bibfield  {title}
  {\enquote {\bibinfo {title} {{Magnetic moments of the spin-${3\over 2}$
  singly heavy baryons}},}\ }\href {\doibase 10.1103/PhysRevD.98.094013}
  {\bibfield  {journal} {\bibinfo  {journal} {Phys. Rev. D}\ }\textbf {\bibinfo
  {volume} {98}},\ \bibinfo {pages} {094013} (\bibinfo {year} {2018})},\
  \Eprint {http://arxiv.org/abs/1805.09580} {arXiv:1805.09580 [hep-ph]}
  \BibitemShut {NoStop}%
\bibitem [{\citenamefont {Weinberg}(1979)}]{Weinberg:1978kz}%
  \BibitemOpen
  \bibfield  {author} {\bibinfo {author} {\bibfnamefont {Steven}\ \bibnamefont
  {Weinberg}},\ }\bibfield  {title} {\enquote {\bibinfo {title}
  {{Phenomenological Lagrangians}},}\ }\href {\doibase
  10.1016/0378-4371(79)90223-1} {\bibfield  {journal} {\bibinfo  {journal}
  {Physica A}\ }\textbf {\bibinfo {volume} {96}},\ \bibinfo {pages} {327--340}
  (\bibinfo {year} {1979})}\BibitemShut {NoStop}%
\bibitem [{\citenamefont {Gasser}\ and\ \citenamefont
  {Leutwyler}(1985)}]{Gasser:1984gg}%
  \BibitemOpen
  \bibfield  {author} {\bibinfo {author} {\bibfnamefont {J.}~\bibnamefont
  {Gasser}}\ and\ \bibinfo {author} {\bibfnamefont {H.}~\bibnamefont
  {Leutwyler}},\ }\bibfield  {title} {\enquote {\bibinfo {title} {{Chiral
  Perturbation Theory: Expansions in the Mass of the Strange Quark}},}\ }\href
  {\doibase 10.1016/0550-3213(85)90492-4} {\bibfield  {journal} {\bibinfo
  {journal} {Nucl. Phys. B}\ }\textbf {\bibinfo {volume} {250}},\ \bibinfo
  {pages} {465--516} (\bibinfo {year} {1985})}\BibitemShut {NoStop}%
\bibitem [{\citenamefont {Gasser}\ and\ \citenamefont
  {Leutwyler}(1984)}]{Gasser:1983yg}%
  \BibitemOpen
  \bibfield  {author} {\bibinfo {author} {\bibfnamefont {J.}~\bibnamefont
  {Gasser}}\ and\ \bibinfo {author} {\bibfnamefont {H.}~\bibnamefont
  {Leutwyler}},\ }\bibfield  {title} {\enquote {\bibinfo {title} {{Chiral
  Perturbation Theory to One Loop}},}\ }\href {\doibase
  10.1016/0003-4916(84)90242-2} {\bibfield  {journal} {\bibinfo  {journal}
  {Annals Phys.}\ }\textbf {\bibinfo {volume} {158}},\ \bibinfo {pages} {142}
  (\bibinfo {year} {1984})}\BibitemShut {NoStop}%
\bibitem [{\citenamefont {\"Unal}\ and\ \citenamefont
  {Mei\ss{}ner}(2021)}]{Unal:2020ezc}%
  \BibitemOpen
  \bibfield  {author} {\bibinfo {author} {\bibfnamefont {Y.}~\bibnamefont
  {\"Unal}}\ and\ \bibinfo {author} {\bibfnamefont {Ulf-G.}\ \bibnamefont
  {Mei\ss{}ner}},\ }\bibfield  {title} {\enquote {\bibinfo {title} {{Strong CP
  violation in spin-1/2 singly charmed baryons}},}\ }\href {\doibase
  10.1007/JHEP01(2021)115} {\bibfield  {journal} {\bibinfo  {journal} {JHEP}\
  }\textbf {\bibinfo {volume} {01}},\ \bibinfo {pages} {115} (\bibinfo {year}
  {2021})},\ \Eprint {http://arxiv.org/abs/2008.01371} {arXiv:2008.01371
  [hep-ph]} \BibitemShut {NoStop}%
\bibitem [{\citenamefont {Guo}\ \emph {et~al.}(2008)\citenamefont {Guo},
  \citenamefont {Hanhart},\ and\ \citenamefont {Meissner}}]{Guo:2008ns}%
  \BibitemOpen
  \bibfield  {author} {\bibinfo {author} {\bibfnamefont {Feng-Kun}\
  \bibnamefont {Guo}}, \bibinfo {author} {\bibfnamefont {Christoph}\
  \bibnamefont {Hanhart}}, \ and\ \bibinfo {author} {\bibfnamefont {Ulf-G.}\
  \bibnamefont {Meissner}},\ }\bibfield  {title} {\enquote {\bibinfo {title}
  {{Mass splittings within heavy baryon isospin multiplets in chiral
  perturbation theory}},}\ }\href {\doibase 10.1088/1126-6708/2008/09/136}
  {\bibfield  {journal} {\bibinfo  {journal} {JHEP}\ }\textbf {\bibinfo
  {volume} {09}},\ \bibinfo {pages} {136} (\bibinfo {year} {2008})},\ \Eprint
  {http://arxiv.org/abs/0809.2359} {arXiv:0809.2359 [hep-ph]} \BibitemShut
  {NoStop}%
\bibitem [{\citenamefont {Savage}(1995{\natexlab{b}})}]{Savage:1995dw}%
  \BibitemOpen
  \bibfield  {author} {\bibinfo {author} {\bibfnamefont {Martin~J.}\
  \bibnamefont {Savage}},\ }\bibfield  {title} {\enquote {\bibinfo {title}
  {{Charmed baryon masses in chiral perturbation theory}},}\ }\href {\doibase
  10.1016/0370-2693(95)01060-4} {\bibfield  {journal} {\bibinfo  {journal}
  {Phys. Lett. B}\ }\textbf {\bibinfo {volume} {359}},\ \bibinfo {pages}
  {189--193} (\bibinfo {year} {1995}{\natexlab{b}})},\ \Eprint
  {http://arxiv.org/abs/hep-ph/9508268} {arXiv:hep-ph/9508268} \BibitemShut
  {NoStop}%
\bibitem [{\citenamefont {Guo}\ and\ \citenamefont
  {Thomas}(2003)}]{Guo:2002tg}%
  \BibitemOpen
  \bibfield  {author} {\bibinfo {author} {\bibfnamefont {Xin-Heng}\
  \bibnamefont {Guo}}\ and\ \bibinfo {author} {\bibfnamefont {Anthony~William}\
  \bibnamefont {Thomas}},\ }\bibfield  {title} {\enquote {\bibinfo {title}
  {{Chiral extrapolation of lattice data for heavy baryons}},}\ }\href
  {\doibase 10.1103/PhysRevD.67.074005} {\bibfield  {journal} {\bibinfo
  {journal} {Phys. Rev. D}\ }\textbf {\bibinfo {volume} {67}},\ \bibinfo
  {pages} {074005} (\bibinfo {year} {2003})},\ \Eprint
  {http://arxiv.org/abs/hep-ph/0210394} {arXiv:hep-ph/0210394} \BibitemShut
  {NoStop}%
\bibitem [{\citenamefont {Jiang}\ \emph
  {et~al.}(2014{\natexlab{a}})\citenamefont {Jiang}, \citenamefont {Chen},\
  and\ \citenamefont {Zhu}}]{Jiang:2014ena}%
  \BibitemOpen
  \bibfield  {author} {\bibinfo {author} {\bibfnamefont {Nan}\ \bibnamefont
  {Jiang}}, \bibinfo {author} {\bibfnamefont {Xiao-Lin}\ \bibnamefont {Chen}},
  \ and\ \bibinfo {author} {\bibfnamefont {Shi-Lin}\ \bibnamefont {Zhu}},\
  }\bibfield  {title} {\enquote {\bibinfo {title} {{Mass and axial charge of
  heavy baryons}},}\ }\href {\doibase 10.1103/PhysRevD.90.074011} {\bibfield
  {journal} {\bibinfo  {journal} {Phys. Rev. D}\ }\textbf {\bibinfo {volume}
  {90}},\ \bibinfo {pages} {074011} (\bibinfo {year} {2014}{\natexlab{a}})},\
  \Eprint {http://arxiv.org/abs/1403.5404} {arXiv:1403.5404 [hep-ph]}
  \BibitemShut {NoStop}%
\bibitem [{\citenamefont {Groote}\ and\ \citenamefont
  {Korner}(2000)}]{Groote:2000ma}%
  \BibitemOpen
  \bibfield  {author} {\bibinfo {author} {\bibfnamefont {S.}~\bibnamefont
  {Groote}}\ and\ \bibinfo {author} {\bibfnamefont {J.~G.}\ \bibnamefont
  {Korner}},\ }\bibfield  {title} {\enquote {\bibinfo {title} {{Theory of heavy
  baryon decay}},}\ }in\ \href {\doibase 10.1142/9789812791870_0028} {\emph
  {\bibinfo {booktitle} {{Proceedings of the 3rd International Conference on B
  Physics and CP Violation (BCONF99)}}}}\ (\bibinfo  {publisher} {World
  Scientific Publishing},\ \bibinfo {address} {Singapore},\ \bibinfo {year}
  {2000})\ pp.\ \bibinfo {pages} {185--191},\ \Eprint
  {http://arxiv.org/abs/hep-ph/0003116} {arXiv:hep-ph/0003116} \BibitemShut
  {NoStop}%
\bibitem [{\citenamefont {Shi}\ \emph {et~al.}(2022)\citenamefont {Shi},
  \citenamefont {Li}, \citenamefont {Lu},\ and\ \citenamefont
  {Geng}}]{Shi:2022dhw}%
  \BibitemOpen
  \bibfield  {author} {\bibinfo {author} {\bibfnamefont {Rui-Xiang}\
  \bibnamefont {Shi}}, \bibinfo {author} {\bibfnamefont {Shuang-Yi}\
  \bibnamefont {Li}}, \bibinfo {author} {\bibfnamefont {Jun-Xu}\ \bibnamefont
  {Lu}}, \ and\ \bibinfo {author} {\bibfnamefont {Li-Sheng}\ \bibnamefont
  {Geng}},\ }\bibfield  {title} {\enquote {\bibinfo {title} {{Weak radiative
  hyperon decays in covariant baryon chiral perturbation theory}},}\ }\href
  {\doibase 10.1016/j.scib.2022.10.026} {\bibfield  {journal} {\bibinfo
  {journal} {Sci. Bull.}\ }\textbf {\bibinfo {volume} {67}},\ \bibinfo {pages}
  {2298--2304} (\bibinfo {year} {2022})},\ \Eprint
  {http://arxiv.org/abs/2206.11773} {arXiv:2206.11773 [hep-ph]} \BibitemShut
  {NoStop}%
\bibitem [{\citenamefont {Wang}\ \emph {et~al.}(2019)\citenamefont {Wang},
  \citenamefont {Meng},\ and\ \citenamefont {Zhu}}]{Wang:2018cre}%
  \BibitemOpen
  \bibfield  {author} {\bibinfo {author} {\bibfnamefont {Guang-Juan}\
  \bibnamefont {Wang}}, \bibinfo {author} {\bibfnamefont {Lu}~\bibnamefont
  {Meng}}, \ and\ \bibinfo {author} {\bibfnamefont {Shi-Lin}\ \bibnamefont
  {Zhu}},\ }\bibfield  {title} {\enquote {\bibinfo {title} {{Radiative decays
  of the singly heavy baryons in chiral perturbation theory}},}\ }\href
  {\doibase 10.1103/PhysRevD.99.034021} {\bibfield  {journal} {\bibinfo
  {journal} {Phys. Rev. D}\ }\textbf {\bibinfo {volume} {99}},\ \bibinfo
  {pages} {034021} (\bibinfo {year} {2019})},\ \Eprint
  {http://arxiv.org/abs/1811.06208} {arXiv:1811.06208 [hep-ph]} \BibitemShut
  {NoStop}%
\bibitem [{\citenamefont {Isgur}\ and\ \citenamefont
  {Wise}(1989)}]{Isgur:1989vq}%
  \BibitemOpen
  \bibfield  {author} {\bibinfo {author} {\bibfnamefont {Nathan}\ \bibnamefont
  {Isgur}}\ and\ \bibinfo {author} {\bibfnamefont {Mark~B.}\ \bibnamefont
  {Wise}},\ }\bibfield  {title} {\enquote {\bibinfo {title} {{Weak Decays of
  Heavy Mesons in the Static Quark Approximation}},}\ }\href {\doibase
  10.1016/0370-2693(89)90566-2} {\bibfield  {journal} {\bibinfo  {journal}
  {Phys. Lett. B}\ }\textbf {\bibinfo {volume} {232}},\ \bibinfo {pages}
  {113--117} (\bibinfo {year} {1989})}\BibitemShut {NoStop}%
\bibitem [{\citenamefont {Isgur}\ and\ \citenamefont
  {Wise}(1990)}]{Isgur:1990yhj}%
  \BibitemOpen
  \bibfield  {author} {\bibinfo {author} {\bibfnamefont {Nathan}\ \bibnamefont
  {Isgur}}\ and\ \bibinfo {author} {\bibfnamefont {Mark~B.}\ \bibnamefont
  {Wise}},\ }\bibfield  {title} {\enquote {\bibinfo {title} {{Weak transition
  form-factors between heavy mesons}},}\ }\href {\doibase
  10.1016/0370-2693(90)91219-2} {\bibfield  {journal} {\bibinfo  {journal}
  {Phys. Lett. B}\ }\textbf {\bibinfo {volume} {237}},\ \bibinfo {pages}
  {527--530} (\bibinfo {year} {1990})}\BibitemShut {NoStop}%
\bibitem [{\citenamefont {Georgi}(1990)}]{Georgi:1990um}%
  \BibitemOpen
  \bibfield  {author} {\bibinfo {author} {\bibfnamefont {Howard}\ \bibnamefont
  {Georgi}},\ }\bibfield  {title} {\enquote {\bibinfo {title} {{An Effective
  Field Theory for Heavy Quarks at Low-energies}},}\ }\href {\doibase
  10.1016/0370-2693(90)91128-X} {\bibfield  {journal} {\bibinfo  {journal}
  {Phys. Lett. B}\ }\textbf {\bibinfo {volume} {240}},\ \bibinfo {pages}
  {447--450} (\bibinfo {year} {1990})}\BibitemShut {NoStop}%
\bibitem [{\citenamefont {Cheng}\ \emph {et~al.}(1994)\citenamefont {Cheng},
  \citenamefont {Cheung}, \citenamefont {Lin}, \citenamefont {Lin},
  \citenamefont {Yan},\ and\ \citenamefont {Yu}}]{Cheng:1993kp}%
  \BibitemOpen
  \bibfield  {author} {\bibinfo {author} {\bibfnamefont {Hai-Yang}\
  \bibnamefont {Cheng}}, \bibinfo {author} {\bibfnamefont {Chi-Yee}\
  \bibnamefont {Cheung}}, \bibinfo {author} {\bibfnamefont {Guey-Lin}\
  \bibnamefont {Lin}}, \bibinfo {author} {\bibfnamefont {Y.~C.}\ \bibnamefont
  {Lin}}, \bibinfo {author} {\bibfnamefont {Tung-Mow}\ \bibnamefont {Yan}}, \
  and\ \bibinfo {author} {\bibfnamefont {Hoi-Lai}\ \bibnamefont {Yu}},\
  }\bibfield  {title} {\enquote {\bibinfo {title} {{Corrections to chiral
  dynamics of heavy hadrons: SU(3) symmetry breaking}},}\ }\href {\doibase
  10.1103/PhysRevD.49.5857} {\bibfield  {journal} {\bibinfo  {journal} {Phys.
  Rev. D}\ }\textbf {\bibinfo {volume} {49}},\ \bibinfo {pages} {5857--5881}
  (\bibinfo {year} {1994})},\ \bibinfo {note} {[Erratum: Phys.Rev.D 55,
  5851--5852 (1997)]},\ \Eprint {http://arxiv.org/abs/hep-ph/9312304}
  {arXiv:hep-ph/9312304} \BibitemShut {NoStop}%
\bibitem [{\citenamefont {Fettes}\ \emph {et~al.}(2000)\citenamefont {Fettes},
  \citenamefont {Meissner}, \citenamefont {Mojzis},\ and\ \citenamefont
  {Steininger}}]{Fettes:2000gb}%
  \BibitemOpen
  \bibfield  {author} {\bibinfo {author} {\bibfnamefont {Nadia}\ \bibnamefont
  {Fettes}}, \bibinfo {author} {\bibfnamefont {Ulf-G.}\ \bibnamefont
  {Meissner}}, \bibinfo {author} {\bibfnamefont {Martin}\ \bibnamefont
  {Mojzis}}, \ and\ \bibinfo {author} {\bibfnamefont {Sven}\ \bibnamefont
  {Steininger}},\ }\bibfield  {title} {\enquote {\bibinfo {title} {{The Chiral
  effective pion nucleon Lagrangian of order p**4}},}\ }\href {\doibase
  10.1006/aphy.2000.6059} {\bibfield  {journal} {\bibinfo  {journal} {Annals
  Phys.}\ }\textbf {\bibinfo {volume} {283}},\ \bibinfo {pages} {273--302}
  (\bibinfo {year} {2000})},\ \bibinfo {note} {[Erratum: Annals Phys. 288,
  249--250 (2001)]},\ \Eprint {http://arxiv.org/abs/hep-ph/0001308}
  {arXiv:hep-ph/0001308} \BibitemShut {NoStop}%
\bibitem [{\citenamefont {Oller}\ \emph {et~al.}(2006)\citenamefont {Oller},
  \citenamefont {Verbeni},\ and\ \citenamefont {Prades}}]{Oller:2006yh}%
  \BibitemOpen
  \bibfield  {author} {\bibinfo {author} {\bibfnamefont {Jose~Antonio}\
  \bibnamefont {Oller}}, \bibinfo {author} {\bibfnamefont {Michela}\
  \bibnamefont {Verbeni}}, \ and\ \bibinfo {author} {\bibfnamefont {Joaquim}\
  \bibnamefont {Prades}},\ }\bibfield  {title} {\enquote {\bibinfo {title}
  {{Meson-baryon effective chiral lagrangians to O(q**3)}},}\ }\href {\doibase
  10.1088/1126-6708/2006/09/079} {\bibfield  {journal} {\bibinfo  {journal}
  {JHEP}\ }\textbf {\bibinfo {volume} {09}},\ \bibinfo {pages} {079} (\bibinfo
  {year} {2006})},\ \Eprint {http://arxiv.org/abs/hep-ph/0608204}
  {arXiv:hep-ph/0608204} \BibitemShut {NoStop}%
\bibitem [{\citenamefont {Frink}\ and\ \citenamefont
  {Meissner}(2006)}]{Frink:2006hx}%
  \BibitemOpen
  \bibfield  {author} {\bibinfo {author} {\bibfnamefont {Matthias}\
  \bibnamefont {Frink}}\ and\ \bibinfo {author} {\bibfnamefont {Ulf-G.}\
  \bibnamefont {Meissner}},\ }\bibfield  {title} {\enquote {\bibinfo {title}
  {{On the chiral effective meson-baryon Lagrangian at third order}},}\ }\href
  {\doibase 10.1140/epja/i2006-10105-x} {\bibfield  {journal} {\bibinfo
  {journal} {Eur. Phys. J. A}\ }\textbf {\bibinfo {volume} {29}},\ \bibinfo
  {pages} {255--260} (\bibinfo {year} {2006})},\ \Eprint
  {http://arxiv.org/abs/hep-ph/0609256} {arXiv:hep-ph/0609256} \BibitemShut
  {NoStop}%
\bibitem [{\citenamefont {Jiang}\ \emph {et~al.}(2017)\citenamefont {Jiang},
  \citenamefont {Chen},\ and\ \citenamefont {Liu}}]{Jiang:2016vax}%
  \BibitemOpen
  \bibfield  {author} {\bibinfo {author} {\bibfnamefont {Shao-Zhou}\
  \bibnamefont {Jiang}}, \bibinfo {author} {\bibfnamefont {Qing-Sen}\
  \bibnamefont {Chen}}, \ and\ \bibinfo {author} {\bibfnamefont {Yan-Rui}\
  \bibnamefont {Liu}},\ }\bibfield  {title} {\enquote {\bibinfo {title}
  {{Meson-baryon effective chiral Lagrangians at order $p^4$}},}\ }\href
  {\doibase 10.1103/PhysRevD.95.014012} {\bibfield  {journal} {\bibinfo
  {journal} {Phys. Rev. D}\ }\textbf {\bibinfo {volume} {95}},\ \bibinfo
  {pages} {014012} (\bibinfo {year} {2017})},\ \Eprint
  {http://arxiv.org/abs/1608.06104} {arXiv:1608.06104 [hep-ph]} \BibitemShut
  {NoStop}%
\bibitem [{\citenamefont {Jiang}\ \emph
  {et~al.}(2018{\natexlab{a}})\citenamefont {Jiang}, \citenamefont {Liu},\ and\
  \citenamefont {Wang}}]{Jiang:2017yda}%
  \BibitemOpen
  \bibfield  {author} {\bibinfo {author} {\bibfnamefont {Shao-Zhou}\
  \bibnamefont {Jiang}}, \bibinfo {author} {\bibfnamefont {Yan-Rui}\
  \bibnamefont {Liu}}, \ and\ \bibinfo {author} {\bibfnamefont {Hong-Qian}\
  \bibnamefont {Wang}},\ }\bibfield  {title} {\enquote {\bibinfo {title}
  {{Chiral Lagrangians with $\Delta(1232)$ to one loop}},}\ }\href {\doibase
  10.1103/PhysRevD.97.014002} {\bibfield  {journal} {\bibinfo  {journal} {Phys.
  Rev. D}\ }\textbf {\bibinfo {volume} {97}},\ \bibinfo {pages} {014002}
  (\bibinfo {year} {2018}{\natexlab{a}})},\ \Eprint
  {http://arxiv.org/abs/1707.09690} {arXiv:1707.09690 [hep-ph]} \BibitemShut
  {NoStop}%
\bibitem [{\citenamefont {Jiang}\ \emph
  {et~al.}(2018{\natexlab{b}})\citenamefont {Jiang}, \citenamefont {Liu},
  \citenamefont {Wang},\ and\ \citenamefont {Yang}}]{Jiang:2018mzd}%
  \BibitemOpen
  \bibfield  {author} {\bibinfo {author} {\bibfnamefont {Shao-Zhou}\
  \bibnamefont {Jiang}}, \bibinfo {author} {\bibfnamefont {Yan-Rui}\
  \bibnamefont {Liu}}, \bibinfo {author} {\bibfnamefont {Hong-Qian}\
  \bibnamefont {Wang}}, \ and\ \bibinfo {author} {\bibfnamefont {Qin-He}\
  \bibnamefont {Yang}},\ }\bibfield  {title} {\enquote {\bibinfo {title}
  {{Chiral Lagrangians with decuplet baryons to one loop}},}\ }\href {\doibase
  10.1103/PhysRevD.97.054031} {\bibfield  {journal} {\bibinfo  {journal} {Phys.
  Rev. D}\ }\textbf {\bibinfo {volume} {97}},\ \bibinfo {pages} {054031}
  (\bibinfo {year} {2018}{\natexlab{b}})},\ \Eprint
  {http://arxiv.org/abs/1801.09879} {arXiv:1801.09879 [hep-ph]} \BibitemShut
  {NoStop}%
\bibitem [{\citenamefont {Holmberg}\ and\ \citenamefont
  {Leupold}(2018)}]{Holmberg:2018dtv}%
  \BibitemOpen
  \bibfield  {author} {\bibinfo {author} {\bibfnamefont {M\r{a}ns}\
  \bibnamefont {Holmberg}}\ and\ \bibinfo {author} {\bibfnamefont {Stefan}\
  \bibnamefont {Leupold}},\ }\bibfield  {title} {\enquote {\bibinfo {title}
  {{The relativistic chiral Lagrangian for decuplet and octet baryons at
  next-to-leading order}},}\ }\href {\doibase 10.1140/epja/i2018-12533-3}
  {\bibfield  {journal} {\bibinfo  {journal} {Eur. Phys. J. A}\ }\textbf
  {\bibinfo {volume} {54}},\ \bibinfo {pages} {103} (\bibinfo {year} {2018})},\
  \Eprint {http://arxiv.org/abs/1802.05168} {arXiv:1802.05168 [hep-ph]}
  \BibitemShut {NoStop}%
\bibitem [{\citenamefont {Jiang}\ \emph {et~al.}(2019)\citenamefont {Jiang},
  \citenamefont {Liu},\ and\ \citenamefont {Yang}}]{Jiang:2019hgs}%
  \BibitemOpen
  \bibfield  {author} {\bibinfo {author} {\bibfnamefont {Shao-Zhou}\
  \bibnamefont {Jiang}}, \bibinfo {author} {\bibfnamefont {Yan-Rui}\
  \bibnamefont {Liu}}, \ and\ \bibinfo {author} {\bibfnamefont {Qin-He}\
  \bibnamefont {Yang}},\ }\bibfield  {title} {\enquote {\bibinfo {title}
  {{Chiral Lagrangians for mesons with a single heavy quark}},}\ }\href
  {\doibase 10.1103/PhysRevD.99.074018} {\bibfield  {journal} {\bibinfo
  {journal} {Phys. Rev. D}\ }\textbf {\bibinfo {volume} {99}},\ \bibinfo
  {pages} {074018} (\bibinfo {year} {2019})},\ \Eprint
  {http://arxiv.org/abs/1901.09479} {arXiv:1901.09479 [hep-ph]} \BibitemShut
  {NoStop}%
\bibitem [{\citenamefont {Qiu}\ and\ \citenamefont {Yao}(2021)}]{Qiu:2020omj}%
  \BibitemOpen
  \bibfield  {author} {\bibinfo {author} {\bibfnamefont {Peng-Cheng}\
  \bibnamefont {Qiu}}\ and\ \bibinfo {author} {\bibfnamefont {De-Liang}\
  \bibnamefont {Yao}},\ }\bibfield  {title} {\enquote {\bibinfo {title}
  {{Chiral effective Lagrangian for doubly charmed baryons up to
  $\mathcal{O}(q^4)$}},}\ }\href {\doibase 10.1103/PhysRevD.103.034006}
  {\bibfield  {journal} {\bibinfo  {journal} {Phys. Rev. D}\ }\textbf {\bibinfo
  {volume} {103}},\ \bibinfo {pages} {034006} (\bibinfo {year} {2021})},\
  \Eprint {http://arxiv.org/abs/2012.11117} {arXiv:2012.11117 [hep-ph]}
  \BibitemShut {NoStop}%
\bibitem [{\citenamefont {Liu}\ \emph {et~al.}(2023)\citenamefont {Liu},
  \citenamefont {Zou}, \citenamefont {Liu},\ and\ \citenamefont
  {Jiang}}]{Liu:2023lsg}%
  \BibitemOpen
  \bibfield  {author} {\bibinfo {author} {\bibfnamefont {Hao}\ \bibnamefont
  {Liu}}, \bibinfo {author} {\bibfnamefont {Yuan-He}\ \bibnamefont {Zou}},
  \bibinfo {author} {\bibfnamefont {Yan-Rui}\ \bibnamefont {Liu}}, \ and\
  \bibinfo {author} {\bibfnamefont {Shao-Zhou}\ \bibnamefont {Jiang}},\
  }\bibfield  {title} {\enquote {\bibinfo {title} {{Chiral Lagrangians for
  spin-$\frac{1}{2}$ and spin-$\frac{3}{2}$ doubly charmed baryons}},}\
  }\href@noop {} {\bibfield  {journal} {\bibinfo  {journal} {Phys. Rev. D}\
  }\textbf {\bibinfo {volume} {108}},\ \bibinfo {pages} {014032} (\bibinfo
  {year} {2023})},\ \Eprint {http://arxiv.org/abs/2304.04575} {arXiv:2304.04575
  [hep-ph]} \BibitemShut {NoStop}%
\bibitem [{\citenamefont {Fearing}\ and\ \citenamefont
  {Scherer}(1996)}]{Fearing:1994ga}%
  \BibitemOpen
  \bibfield  {author} {\bibinfo {author} {\bibfnamefont {H.~W.}\ \bibnamefont
  {Fearing}}\ and\ \bibinfo {author} {\bibfnamefont {S.}~\bibnamefont
  {Scherer}},\ }\bibfield  {title} {\enquote {\bibinfo {title} {{Extension of
  the chiral perturbation theory meson Lagrangian to order p(6)}},}\ }\href
  {\doibase 10.1103/PhysRevD.53.315} {\bibfield  {journal} {\bibinfo  {journal}
  {Phys. Rev. D}\ }\textbf {\bibinfo {volume} {53}},\ \bibinfo {pages}
  {315--348} (\bibinfo {year} {1996})},\ \Eprint
  {http://arxiv.org/abs/hep-ph/9408346} {arXiv:hep-ph/9408346} \BibitemShut
  {NoStop}%
\bibitem [{\citenamefont {Bijnens}\ \emph {et~al.}(1999)\citenamefont
  {Bijnens}, \citenamefont {Colangelo},\ and\ \citenamefont
  {Ecker}}]{Bijnens:1999sh}%
  \BibitemOpen
  \bibfield  {author} {\bibinfo {author} {\bibfnamefont {Johan}\ \bibnamefont
  {Bijnens}}, \bibinfo {author} {\bibfnamefont {Gilberto}\ \bibnamefont
  {Colangelo}}, \ and\ \bibinfo {author} {\bibfnamefont {Gerhard}\ \bibnamefont
  {Ecker}},\ }\bibfield  {title} {\enquote {\bibinfo {title} {{The Mesonic
  chiral Lagrangian of order p**6}},}\ }\href {\doibase
  10.1088/1126-6708/1999/02/020} {\bibfield  {journal} {\bibinfo  {journal}
  {JHEP}\ }\textbf {\bibinfo {volume} {02}},\ \bibinfo {pages} {020} (\bibinfo
  {year} {1999})},\ \Eprint {http://arxiv.org/abs/hep-ph/9902437}
  {arXiv:hep-ph/9902437} \BibitemShut {NoStop}%
\bibitem [{\citenamefont {Bijnens}\ \emph {et~al.}(2002)\citenamefont
  {Bijnens}, \citenamefont {Girlanda},\ and\ \citenamefont
  {Talavera}}]{Bijnens:2001bb}%
  \BibitemOpen
  \bibfield  {author} {\bibinfo {author} {\bibfnamefont {J.}~\bibnamefont
  {Bijnens}}, \bibinfo {author} {\bibfnamefont {L.}~\bibnamefont {Girlanda}}, \
  and\ \bibinfo {author} {\bibfnamefont {P.}~\bibnamefont {Talavera}},\
  }\bibfield  {title} {\enquote {\bibinfo {title} {{The Anomalous chiral
  Lagrangian of order p**6}},}\ }\href {\doibase 10.1007/s100520100887}
  {\bibfield  {journal} {\bibinfo  {journal} {Eur. Phys. J. C}\ }\textbf
  {\bibinfo {volume} {23}},\ \bibinfo {pages} {539--544} (\bibinfo {year}
  {2002})},\ \Eprint {http://arxiv.org/abs/hep-ph/0110400}
  {arXiv:hep-ph/0110400} \BibitemShut {NoStop}%
\bibitem [{\citenamefont {Bijnens}\ \emph {et~al.}(2019)\citenamefont
  {Bijnens}, \citenamefont {Hermansson-Truedsson},\ and\ \citenamefont
  {Wang}}]{Bijnens:2018lez}%
  \BibitemOpen
  \bibfield  {author} {\bibinfo {author} {\bibfnamefont {Johan}\ \bibnamefont
  {Bijnens}}, \bibinfo {author} {\bibfnamefont {Nils}\ \bibnamefont
  {Hermansson-Truedsson}}, \ and\ \bibinfo {author} {\bibfnamefont
  {Si}~\bibnamefont {Wang}},\ }\bibfield  {title} {\enquote {\bibinfo {title}
  {{The order p$^{8}$ mesonic chiral Lagrangian}},}\ }\href {\doibase
  10.1007/JHEP01(2019)102} {\bibfield  {journal} {\bibinfo  {journal} {JHEP}\
  }\textbf {\bibinfo {volume} {01}},\ \bibinfo {pages} {102} (\bibinfo {year}
  {2019})},\ \Eprint {http://arxiv.org/abs/1810.06834} {arXiv:1810.06834
  [hep-ph]} \BibitemShut {NoStop}%
\bibitem [{\citenamefont {Gasser}\ \emph {et~al.}(1988)\citenamefont {Gasser},
  \citenamefont {Sainio},\ and\ \citenamefont {Svarc}}]{Gasser:1987rb}%
  \BibitemOpen
  \bibfield  {author} {\bibinfo {author} {\bibfnamefont {J.}~\bibnamefont
  {Gasser}}, \bibinfo {author} {\bibfnamefont {M.~E.}\ \bibnamefont {Sainio}},
  \ and\ \bibinfo {author} {\bibfnamefont {A.}~\bibnamefont {Svarc}},\
  }\bibfield  {title} {\enquote {\bibinfo {title} {{Nucleons with Chiral
  Loops}},}\ }\href {\doibase 10.1016/0550-3213(88)90108-3} {\bibfield
  {journal} {\bibinfo  {journal} {Nucl. Phys. B}\ }\textbf {\bibinfo {volume}
  {307}},\ \bibinfo {pages} {779--853} (\bibinfo {year} {1988})}\BibitemShut
  {NoStop}%
\bibitem [{\citenamefont {Wise}(1993)}]{Wise:1993wa}%
  \BibitemOpen
  \bibfield  {author} {\bibinfo {author} {\bibfnamefont {Mark~B.}\ \bibnamefont
  {Wise}},\ }\bibfield  {title} {\enquote {\bibinfo {title} {{Combining chiral
  and heavy quark symmetry}},}\ }in\ \href@noop {} {\emph {\bibinfo {booktitle}
  {{Proceedings of the CCAST Symposium on Particle Physics at the Fermi
  Scale}}}}\ (\bibinfo  {publisher} {Gordon \& Breach},\ \bibinfo {address}
  {Yverdon},\ \bibinfo {year} {1993})\ pp.\ \bibinfo {pages} {71--114},\
  \Eprint {http://arxiv.org/abs/hep-ph/9306277} {arXiv:hep-ph/9306277}
  \BibitemShut {NoStop}%
\bibitem [{\citenamefont {Jenkins}\ \emph {et~al.}(1993)\citenamefont
  {Jenkins}, \citenamefont {Manohar},\ and\ \citenamefont
  {Wise}}]{Jenkins:1992zx}%
  \BibitemOpen
  \bibfield  {author} {\bibinfo {author} {\bibfnamefont {Elizabeth~Ellen}\
  \bibnamefont {Jenkins}}, \bibinfo {author} {\bibfnamefont {Aneesh~V.}\
  \bibnamefont {Manohar}}, \ and\ \bibinfo {author} {\bibfnamefont {Mark~B.}\
  \bibnamefont {Wise}},\ }\bibfield  {title} {\enquote {\bibinfo {title}
  {{Baryons containing a heavy quark as solitons}},}\ }\href {\doibase
  10.1016/0550-3213(93)90256-O} {\bibfield  {journal} {\bibinfo  {journal}
  {Nucl. Phys. B}\ }\textbf {\bibinfo {volume} {396}},\ \bibinfo {pages}
  {27--37} (\bibinfo {year} {1993})},\ \Eprint
  {http://arxiv.org/abs/hep-ph/9205243} {arXiv:hep-ph/9205243} \BibitemShut
  {NoStop}%
\bibitem [{\citenamefont {Casalbuoni}\ \emph {et~al.}(1997)\citenamefont
  {Casalbuoni}, \citenamefont {Deandrea}, \citenamefont {Di~Bartolomeo},
  \citenamefont {Gatto}, \citenamefont {Feruglio},\ and\ \citenamefont
  {Nardulli}}]{Casalbuoni:1996pg}%
  \BibitemOpen
  \bibfield  {author} {\bibinfo {author} {\bibfnamefont {R.}~\bibnamefont
  {Casalbuoni}}, \bibinfo {author} {\bibfnamefont {A.}~\bibnamefont
  {Deandrea}}, \bibinfo {author} {\bibfnamefont {N.}~\bibnamefont
  {Di~Bartolomeo}}, \bibinfo {author} {\bibfnamefont {Raoul}\ \bibnamefont
  {Gatto}}, \bibinfo {author} {\bibfnamefont {F.}~\bibnamefont {Feruglio}}, \
  and\ \bibinfo {author} {\bibfnamefont {G.}~\bibnamefont {Nardulli}},\
  }\bibfield  {title} {\enquote {\bibinfo {title} {{Phenomenology of heavy
  meson chiral Lagrangians}},}\ }\href {\doibase 10.1016/S0370-1573(96)00027-0}
  {\bibfield  {journal} {\bibinfo  {journal} {Phys. Rept.}\ }\textbf {\bibinfo
  {volume} {281}},\ \bibinfo {pages} {145--238} (\bibinfo {year} {1997})},\
  \Eprint {http://arxiv.org/abs/hep-ph/9605342} {arXiv:hep-ph/9605342}
  \BibitemShut {NoStop}%
\bibitem [{\citenamefont {Cat\`a}\ and\ \citenamefont
  {Mateu}(2007)}]{Cata:2007ns}%
  \BibitemOpen
  \bibfield  {author} {\bibinfo {author} {\bibfnamefont {Oscar}\ \bibnamefont
  {Cat\`a}}\ and\ \bibinfo {author} {\bibfnamefont {Vicent}\ \bibnamefont
  {Mateu}},\ }\bibfield  {title} {\enquote {\bibinfo {title} {{Chiral
  perturbation theory with tensor sources}},}\ }\href {\doibase
  10.1088/1126-6708/2007/09/078} {\bibfield  {journal} {\bibinfo  {journal}
  {Journal of High Energy Physics}\ }\textbf {\bibinfo {volume} {0709}},\
  \bibinfo {pages} {078} (\bibinfo {year} {2007})},\ \Eprint
  {http://arxiv.org/abs/0705.2948} {arXiv:0705.2948 [hep-ph]} \BibitemShut
  {NoStop}%
\bibitem [{\citenamefont {Detmold}\ \emph {et~al.}(2011)\citenamefont
  {Detmold}, \citenamefont {Lin},\ and\ \citenamefont
  {Meinel}}]{Detmold:2011rb}%
  \BibitemOpen
  \bibfield  {author} {\bibinfo {author} {\bibfnamefont {William}\ \bibnamefont
  {Detmold}}, \bibinfo {author} {\bibfnamefont {C.~J.~David}\ \bibnamefont
  {Lin}}, \ and\ \bibinfo {author} {\bibfnamefont {Stefan}\ \bibnamefont
  {Meinel}},\ }\bibfield  {title} {\enquote {\bibinfo {title} {{Axial couplings
  in heavy hadron chiral perturbation theory at the next-to-leading order}},}\
  }\href {\doibase 10.1103/PhysRevD.84.094502} {\bibfield  {journal} {\bibinfo
  {journal} {Phys. Rev. D}\ }\textbf {\bibinfo {volume} {84}},\ \bibinfo
  {pages} {094502} (\bibinfo {year} {2011})},\ \Eprint
  {http://arxiv.org/abs/1108.5594} {arXiv:1108.5594 [hep-lat]} \BibitemShut
  {NoStop}%
\bibitem [{\citenamefont {Kawakami}\ and\ \citenamefont
  {Harada}(2018)}]{Kawakami:2018olq}%
  \BibitemOpen
  \bibfield  {author} {\bibinfo {author} {\bibfnamefont {Yohei}\ \bibnamefont
  {Kawakami}}\ and\ \bibinfo {author} {\bibfnamefont {Masayasu}\ \bibnamefont
  {Harada}},\ }\bibfield  {title} {\enquote {\bibinfo {title} {{Analysis of
  $\Lambda_c(2595)$, $\Lambda_c(2625)$, $\Lambda_b(5912)$, $\Lambda_b(5920)$
  based on a chiral partner structure}},}\ }\href {\doibase
  10.1103/PhysRevD.97.114024} {\bibfield  {journal} {\bibinfo  {journal} {Phys.
  Rev. D}\ }\textbf {\bibinfo {volume} {97}},\ \bibinfo {pages} {114024}
  (\bibinfo {year} {2018})},\ \Eprint {http://arxiv.org/abs/1804.04872}
  {arXiv:1804.04872 [hep-ph]} \BibitemShut {NoStop}%
\bibitem [{Note1()}]{Note1}%
  \BibitemOpen
  \bibinfo {note} {From now on, $v^\mu $ will no longer denote vector external
  source}\BibitemShut {NoStop}%
\bibitem [{\citenamefont {Hemmert}\ \emph {et~al.}(1998)\citenamefont
  {Hemmert}, \citenamefont {Holstein},\ and\ \citenamefont
  {Kambor}}]{Hemmert:1997ye}%
  \BibitemOpen
  \bibfield  {author} {\bibinfo {author} {\bibfnamefont {Thomas~R.}\
  \bibnamefont {Hemmert}}, \bibinfo {author} {\bibfnamefont {Barry~R.}\
  \bibnamefont {Holstein}}, \ and\ \bibinfo {author} {\bibfnamefont {Joachim}\
  \bibnamefont {Kambor}},\ }\bibfield  {title} {\enquote {\bibinfo {title}
  {{Chiral Lagrangians and delta(1232) interactions: Formalism}},}\ }\href
  {\doibase 10.1088/0954-3899/24/10/003} {\bibfield  {journal} {\bibinfo
  {journal} {J. Phys. G}\ }\textbf {\bibinfo {volume} {24}},\ \bibinfo {pages}
  {1831--1859} (\bibinfo {year} {1998})},\ \Eprint
  {http://arxiv.org/abs/hep-ph/9712496} {arXiv:hep-ph/9712496} \BibitemShut
  {NoStop}%
\bibitem [{\citenamefont {Ebertshauser}\ \emph {et~al.}(2002)\citenamefont
  {Ebertshauser}, \citenamefont {Fearing},\ and\ \citenamefont
  {Scherer}}]{Ebertshauser:2001nj}%
  \BibitemOpen
  \bibfield  {author} {\bibinfo {author} {\bibfnamefont {T.}~\bibnamefont
  {Ebertshauser}}, \bibinfo {author} {\bibfnamefont {H.~W.}\ \bibnamefont
  {Fearing}}, \ and\ \bibinfo {author} {\bibfnamefont {S.}~\bibnamefont
  {Scherer}},\ }\bibfield  {title} {\enquote {\bibinfo {title} {{The Anomalous
  chiral perturbation theory meson Lagrangian to order p**6 revisited}},}\
  }\href {\doibase 10.1103/PhysRevD.65.054033} {\bibfield  {journal} {\bibinfo
  {journal} {Phys. Rev. D}\ }\textbf {\bibinfo {volume} {65}},\ \bibinfo
  {pages} {054033} (\bibinfo {year} {2002})},\ \Eprint
  {http://arxiv.org/abs/hep-ph/0110261} {arXiv:hep-ph/0110261} \BibitemShut
  {NoStop}%
\bibitem [{\citenamefont {Jiang}\ \emph
  {et~al.}(2014{\natexlab{b}})\citenamefont {Jiang}, \citenamefont {Ge},\ and\
  \citenamefont {Wang}}]{Jiang:2014via}%
  \BibitemOpen
  \bibfield  {author} {\bibinfo {author} {\bibfnamefont {Shao-Zhou}\
  \bibnamefont {Jiang}}, \bibinfo {author} {\bibfnamefont {Feng-Jun}\
  \bibnamefont {Ge}}, \ and\ \bibinfo {author} {\bibfnamefont {Qing}\
  \bibnamefont {Wang}},\ }\bibfield  {title} {\enquote {\bibinfo {title} {{Full
  pseudoscalar mesonic chiral Lagrangian at $p^6$ order under the unitary
  group}},}\ }\href {\doibase 10.1103/PhysRevD.89.074048} {\bibfield  {journal}
  {\bibinfo  {journal} {Phys. Rev. D}\ }\textbf {\bibinfo {volume} {89}},\
  \bibinfo {pages} {074048} (\bibinfo {year} {2014}{\natexlab{b}})},\ \Eprint
  {http://arxiv.org/abs/1401.0317} {arXiv:1401.0317 [hep-ph]} \BibitemShut
  {NoStop}%
\bibitem [{\citenamefont {Cho}\ and\ \citenamefont
  {Georgi}(1992)}]{Cho:1992nt}%
  \BibitemOpen
  \bibfield  {author} {\bibinfo {author} {\bibfnamefont {Peter~L.}\
  \bibnamefont {Cho}}\ and\ \bibinfo {author} {\bibfnamefont {Howard}\
  \bibnamefont {Georgi}},\ }\bibfield  {title} {\enquote {\bibinfo {title}
  {{Electromagnetic interactions in heavy hadron chiral theory}},}\ }\href
  {\doibase 10.1016/0370-2693(92)91340-F} {\bibfield  {journal} {\bibinfo
  {journal} {Phys. Lett. B}\ }\textbf {\bibinfo {volume} {296}},\ \bibinfo
  {pages} {408--414} (\bibinfo {year} {1992})},\ \bibinfo {note} {[Erratum:
  Phys.Lett.B 300, 410 (1993)]},\ \Eprint {http://arxiv.org/abs/hep-ph/9209239}
  {arXiv:hep-ph/9209239} \BibitemShut {NoStop}%
\bibitem [{\citenamefont {Meng}\ \emph {et~al.}(2023)\citenamefont {Meng},
  \citenamefont {Wang}, \citenamefont {Wang},\ and\ \citenamefont
  {Zhu}}]{Meng:2022ozq}%
  \BibitemOpen
  \bibfield  {author} {\bibinfo {author} {\bibfnamefont {Lu}~\bibnamefont
  {Meng}}, \bibinfo {author} {\bibfnamefont {Bo}~\bibnamefont {Wang}}, \bibinfo
  {author} {\bibfnamefont {Guang-Juan}\ \bibnamefont {Wang}}, \ and\ \bibinfo
  {author} {\bibfnamefont {Shi-Lin}\ \bibnamefont {Zhu}},\ }\bibfield  {title}
  {\enquote {\bibinfo {title} {{Chiral perturbation theory for heavy hadrons
  and chiral effective field theory for heavy hadronic molecules}},}\ }\href
  {\doibase 10.1016/j.physrep.2023.04.003} {\bibfield  {journal} {\bibinfo
  {journal} {Phys. Rept.}\ }\textbf {\bibinfo {volume} {1019}},\ \bibinfo
  {pages} {1--149} (\bibinfo {year} {2023})},\ \Eprint
  {http://arxiv.org/abs/2204.08716} {arXiv:2204.08716 [hep-ph]} \BibitemShut
  {NoStop}%
\bibitem [{\citenamefont {Liu}\ and\ \citenamefont {Oka}(2012)}]{Liu:2011xc}%
  \BibitemOpen
  \bibfield  {author} {\bibinfo {author} {\bibfnamefont {Yan-Rui}\ \bibnamefont
  {Liu}}\ and\ \bibinfo {author} {\bibfnamefont {Makoto}\ \bibnamefont {Oka}},\
  }\bibfield  {title} {\enquote {\bibinfo {title} {{$\Lambda_c N$ bound states
  revisited}},}\ }\href {\doibase 10.1103/PhysRevD.85.014015} {\bibfield
  {journal} {\bibinfo  {journal} {Phys. Rev. D}\ }\textbf {\bibinfo {volume}
  {85}},\ \bibinfo {pages} {014015} (\bibinfo {year} {2012})},\ \Eprint
  {http://arxiv.org/abs/1103.4624} {arXiv:1103.4624 [hep-ph]} \BibitemShut
  {NoStop}%
\bibitem [{\citenamefont {Meguro}\ \emph {et~al.}(2011)\citenamefont {Meguro},
  \citenamefont {Liu},\ and\ \citenamefont {Oka}}]{Meguro:2011nr}%
  \BibitemOpen
  \bibfield  {author} {\bibinfo {author} {\bibfnamefont {Wakafumi}\
  \bibnamefont {Meguro}}, \bibinfo {author} {\bibfnamefont {Yan-Rui}\
  \bibnamefont {Liu}}, \ and\ \bibinfo {author} {\bibfnamefont {Makoto}\
  \bibnamefont {Oka}},\ }\bibfield  {title} {\enquote {\bibinfo {title}
  {{Possible $\Lambda_c\Lambda_c$ molecular bound state}},}\ }\href {\doibase
  10.1016/j.physletb.2011.09.088} {\bibfield  {journal} {\bibinfo  {journal}
  {Phys. Lett. B}\ }\textbf {\bibinfo {volume} {704}},\ \bibinfo {pages}
  {547--550} (\bibinfo {year} {2011})},\ \Eprint
  {http://arxiv.org/abs/1105.3693} {arXiv:1105.3693 [hep-ph]} \BibitemShut
  {NoStop}%
\bibitem [{\citenamefont {Jiang}\ \emph
  {et~al.}(2015{\natexlab{b}})\citenamefont {Jiang}, \citenamefont {Wei},
  \citenamefont {Chen},\ and\ \citenamefont {Wang}}]{Jiang:2015dba}%
  \BibitemOpen
  \bibfield  {author} {\bibinfo {author} {\bibfnamefont {Shao-Zhou}\
  \bibnamefont {Jiang}}, \bibinfo {author} {\bibfnamefont {Zhen-Long}\
  \bibnamefont {Wei}}, \bibinfo {author} {\bibfnamefont {Qing-Sen}\
  \bibnamefont {Chen}}, \ and\ \bibinfo {author} {\bibfnamefont {Qing}\
  \bibnamefont {Wang}},\ }\bibfield  {title} {\enquote {\bibinfo {title}
  {{Computation of the $O(p^6)$ order low-energy constants: An update}},}\
  }\href {\doibase 10.1103/PhysRevD.92.025014} {\bibfield  {journal} {\bibinfo
  {journal} {Phys. Rev. D}\ }\textbf {\bibinfo {volume} {92}},\ \bibinfo
  {pages} {025014} (\bibinfo {year} {2015}{\natexlab{b}})},\ \Eprint
  {http://arxiv.org/abs/1502.05087} {arXiv:1502.05087 [hep-ph]} \BibitemShut
  {NoStop}%
\bibitem [{\citenamefont {Jiang}\ \emph {et~al.}(2022)\citenamefont {Jiang},
  \citenamefont {Jiang}, \citenamefont {Li}, \citenamefont {Liu}, \citenamefont
  {Si},\ and\ \citenamefont {Wang}}]{Jiang:2022gjy}%
  \BibitemOpen
  \bibfield  {author} {\bibinfo {author} {\bibfnamefont {Jun}\ \bibnamefont
  {Jiang}}, \bibinfo {author} {\bibfnamefont {Shao-Zhou}\ \bibnamefont
  {Jiang}}, \bibinfo {author} {\bibfnamefont {Shi-Yuan}\ \bibnamefont {Li}},
  \bibinfo {author} {\bibfnamefont {Yan-Rui}\ \bibnamefont {Liu}}, \bibinfo
  {author} {\bibfnamefont {Zong-Guo}\ \bibnamefont {Si}}, \ and\ \bibinfo
  {author} {\bibfnamefont {Hong-Qian}\ \bibnamefont {Wang}},\ }\bibfield
  {title} {\enquote {\bibinfo {title} {{Relations for low-energy coupling
  constants in baryon chiral perturbation theory derived from the chiral quark
  model}},}\ }\href {\doibase 10.1103/PhysRevD.106.054023} {\bibfield
  {journal} {\bibinfo  {journal} {Phys. Rev. D}\ }\textbf {\bibinfo {volume}
  {106}},\ \bibinfo {pages} {054023} (\bibinfo {year} {2022})},\ \Eprint
  {http://arxiv.org/abs/2206.06570} {arXiv:2206.06570 [hep-ph]} \BibitemShut
  {NoStop}%
\bibitem [{\citenamefont {Jiang}\ \emph {et~al.}(2023)\citenamefont {Jiang},
  \citenamefont {Jiang}, \citenamefont {Li}, \citenamefont {Liu}, \citenamefont
  {Si},\ and\ \citenamefont {Wang}}]{Jiang:2023zqq}%
  \BibitemOpen
  \bibfield  {author} {\bibinfo {author} {\bibfnamefont {Jun}\ \bibnamefont
  {Jiang}}, \bibinfo {author} {\bibfnamefont {Shao-Zhou}\ \bibnamefont
  {Jiang}}, \bibinfo {author} {\bibfnamefont {Shi-Yuan}\ \bibnamefont {Li}},
  \bibinfo {author} {\bibfnamefont {Yan-Rui}\ \bibnamefont {Liu}}, \bibinfo
  {author} {\bibfnamefont {Zong-Guo}\ \bibnamefont {Si}}, \ and\ \bibinfo
  {author} {\bibfnamefont {Hong-Qian}\ \bibnamefont {Wang}},\ }\bibfield
  {title} {\enquote {\bibinfo {title} {{Relations for low-energy constants in
  baryon chiral perturbation theory with explicit $\Delta(1232)$ derived from
  the chiral quark model}},}\ }\href {\doibase 10.1140/epjc/s10052-023-11446-6}
  {\bibfield  {journal} {\bibinfo  {journal} {Eur. Phys. J. C}\ }\textbf
  {\bibinfo {volume} {83}},\ \bibinfo {pages} {296} (\bibinfo {year} {2023})},\
  \Eprint {http://arxiv.org/abs/2301.08059} {arXiv:2301.08059 [hep-ph]}
  \BibitemShut {NoStop}%
\bibitem [{\citenamefont {Graf}\ \emph {et~al.}(2021)\citenamefont {Graf},
  \citenamefont {Henning}, \citenamefont {Lu}, \citenamefont {Melia},\ and\
  \citenamefont {Murayama}}]{Graf:2020yxt}%
  \BibitemOpen
  \bibfield  {author} {\bibinfo {author} {\bibfnamefont {Lukas}\ \bibnamefont
  {Graf}}, \bibinfo {author} {\bibfnamefont {Brian}\ \bibnamefont {Henning}},
  \bibinfo {author} {\bibfnamefont {Xiaochuan}\ \bibnamefont {Lu}}, \bibinfo
  {author} {\bibfnamefont {Tom}\ \bibnamefont {Melia}}, \ and\ \bibinfo
  {author} {\bibfnamefont {Hitoshi}\ \bibnamefont {Murayama}},\ }\bibfield
  {title} {\enquote {\bibinfo {title} {{2, 12, 117, 1959, 45171, 1170086,
  \textellipsis{}: a Hilbert series for the QCD chiral Lagrangian}},}\ }\href
  {\doibase 10.1007/JHEP01(2021)142} {\bibfield  {journal} {\bibinfo  {journal}
  {JHEP}\ }\textbf {\bibinfo {volume} {01}},\ \bibinfo {pages} {142} (\bibinfo
  {year} {2021})},\ \Eprint {http://arxiv.org/abs/2009.01239} {arXiv:2009.01239
  [hep-ph]} \BibitemShut {NoStop}%
\end{thebibliography}%
\end{document}